\documentclass[%
reprint,
amsmath,amssymb,
aps,
prb,
]{revtex4-2}
\usepackage{graphicx}
\usepackage{dcolumn}
\usepackage{bm}
\usepackage[colorlinks=true,allcolors=blue]{hyperref}

\begin{document}
	
	\preprint{APS/123-QED}

	\title{Electric field and Strain-induced Band-gap Engineering and Manipulation of the Rashba Spin Splitting in Janus van der Waals Heterostructures}

	\author{Shubham Patel$^1$}\email{spatelphy@iitkgp.ac.in} 
	\author{Urmimala Dey$^2$, Narayan Prasad Adhikari$^3$} 
	\author{A. Taraphder$^{1}$}\email{arghya@phy.iitkgp.ac.in}
	
	\affiliation{$^1$ Department of Physics, Indian Institute of Technology,
		Kharagpur-721302, India}
	\affiliation{$^2$ Department of Physics, Durham University, South Road, Durham DH1 3LE, United Kingdom}
	\affiliation{$^3$ Central Department of Physics, Tribhuvan University, Kirtipur, Kathmandu, Nepal}

	\date{\today}
	
	\begin{abstract}
		The compositional as well as structural asymmetries in Janus transition metal dichalcogenides (J-TMDs) and their van der Waals heterostructures (vdW HSs) induce an intrinsic Rashba spin-splitting. We investigate the variation of band-gaps and the Rashba parameter in three different Janus heterostructures having AB-stacked Mo$XY$/W$XY$ ($X$, $Y$ = S, Se, Te; $X\neq Y$) geometry with a $Y-Y$ interface, using first-principles calculations. We consider the effect of external electric field and in-plane biaxial strain in tuning the strength of the intrinsic electric field, which leads to remarkable modifications of the band-gap and the Rashba spin-splitting. In particular, it is found that the positive applied field and compressive in-plane biaxial strain can lead to a notable increase in the Rashba spin-splitting of the valence bands about the $\Gamma$-point. Moreover, our \textit{ab-initio} density functional theory (DFT) calculations reveal the existence of a type-II band alignment in these heterostructures, which remains robust under positive external field and biaxial strain. These suggest novel ways of engineering the electronic, optical, and spin properties of J-TMD van der Waals heterostructures holding a huge promise in spintronic and optoelectronic devices.  Detailed  $\mathbf{k\cdot p}$ model analyses have been performed to investigate the electronic and spin properties near the $\Gamma$ and K points of the Brillouin zone.
	\end{abstract}

	\maketitle

	\section{\label{sec:level1}Introduction \protect}
	
	Two-dimensional (2D) transition metal dichalcogenides (TMDs) have drawn a considerable attention since the discovery of graphene \cite{castro2009electronic}. Endowed with several exotic and intriguing properties, such as higher stability, large spin-orbit coupling (SOC) and tunable band-gap, make these materials stand out in the fields of semiconductor physics, electronics, spintronics, and valleytronics. TMD monolayers (MLs) such as \textit{M}$X_2$, where \textit{M} is the transition metal atom and $X$ is a chalcogen atom, and numerous derivatives of TMDs have been successfully synthesized in experiments using chemical vapor deposition \cite{lee2012synthesis,shi2015recent}, mechanical exfoliation \cite{li2013mechanical,zeng2012effective}, physical vapor decomposition \cite{feng2015growth} and liquid exfoliation \cite{coleman2011two}.
	
	Breaking symmetries leads to many novel phenomena such as phase transitions, magnetism, superconductivity, and so on. Symmetries can be broken by applying electric field, strain, or stacking in different orders. In contrast to conventional TMDs, in which out-of-plane symmetry is preserved, a new class of 2D materials, Janus transition metal dichalcogenides (J-TMDs), having mirror asymmetry, are recently acquiring huge interest due to their distinct properties \cite{li2018recent}. Janus structure (\textit{M}$XY$; \textit{M}= Mo, W; $X$, $Y$ = S, Se, and Te; $X\neq Y$), based on group-VI chalcogens was introduced by Y.C. Cheng \textit{et al.} \cite{cheng2013spin}. Janus monolayers (J-MLs) of MoSSe and WSSe have already been synthesized experimentally. Janus MoSSe ML was fabricated by controlled selenization of MoS$_2$ or sulfurization of MoSe$_2$ ML \cite{lu2017janus,zhang2017janus}, while Janus WSSe ML was synthesized by implanting Se species into WS$_2$ ML with pulsed laser ablation plasmas \cite{lin2020low}. Contrary to usual TMDs, Janus materials display many novel properties such as Rashba effect \cite{dresselhaus1955spin,bychkov1984properties,manchon2015new,bihlmayer2015focus,riis2019classifying}, piezoelectric polarization \cite{ li2015piezoelectricity,dong2017large,yagmurcukardes2019electronic}, topological effects \cite{sato2009topological,maghirang2019predicting}, magnetic anisotropy \cite{dey2020structural}, which make them one of the most suitable candidates for applications in spintronics, optoelectronic devices and quantum computing. 
	
	Since 2D TMDs occupied much of the attention in the last few decades, researchers have started playing with the stacking of these materials in different layers. Efforts have been made to synthesize lateral and vertical van der Waals (vdW) heterostructures (HSs) experimentally \cite{georgiou2013vertical,anto2014two,zheng2018band}. Theoretically, vdW-HSs of TMDs have been investigated intensively as well. Enhanced SOC and interlayer hybridization can induce a direct to indirect band-gap transition when moving from MLs to multilayer phase \cite{kou2013nanoscale,komsa2013electronic,terrones2013novel} leading to novel applications \cite{long2016quantum,chhowalla2013chemistry,kou2014strain}. In addition, the stacking sequence strongly influences the electronic properties and interlayer excitation within vdW HSs \cite{arora2017interlayer,liu2020excitons,suzuki2014valley}. Based on J-ML TMDs with broken out-of-plane symmetry, the electronic, structural and optical properties can be modified substantially by stacking them in different layers and constructing Janus vdW HSs \cite{li2017electronic,zhou2019geometry,rezavand2021stacking,guo2020strain,wang2019mirror}.
	
	\begin{figure*}[!htb]
		\centering
		\begin{tabular}{lccccc}
			\includegraphics[scale=0.2]{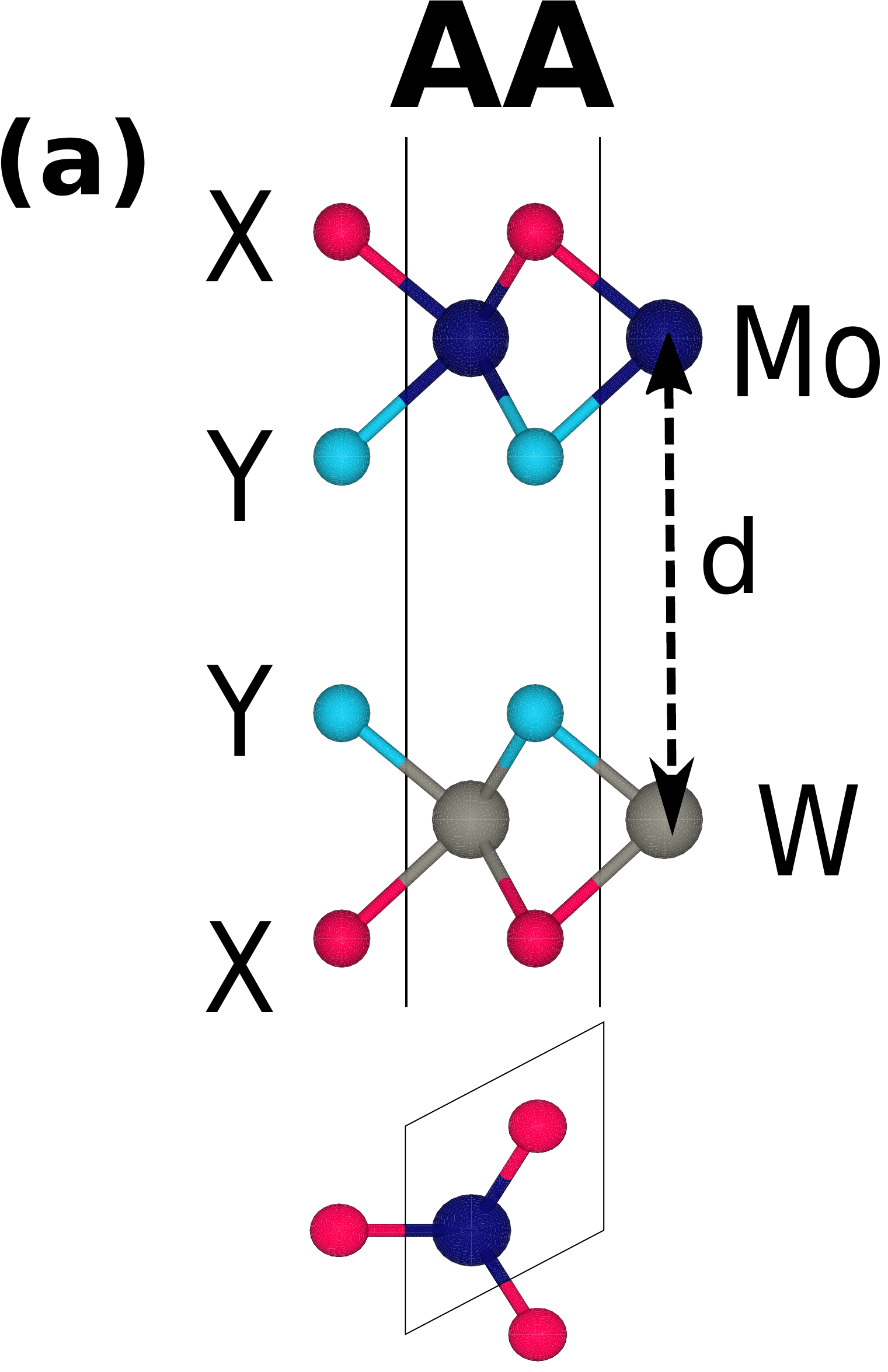}&
			\includegraphics[scale=0.2]{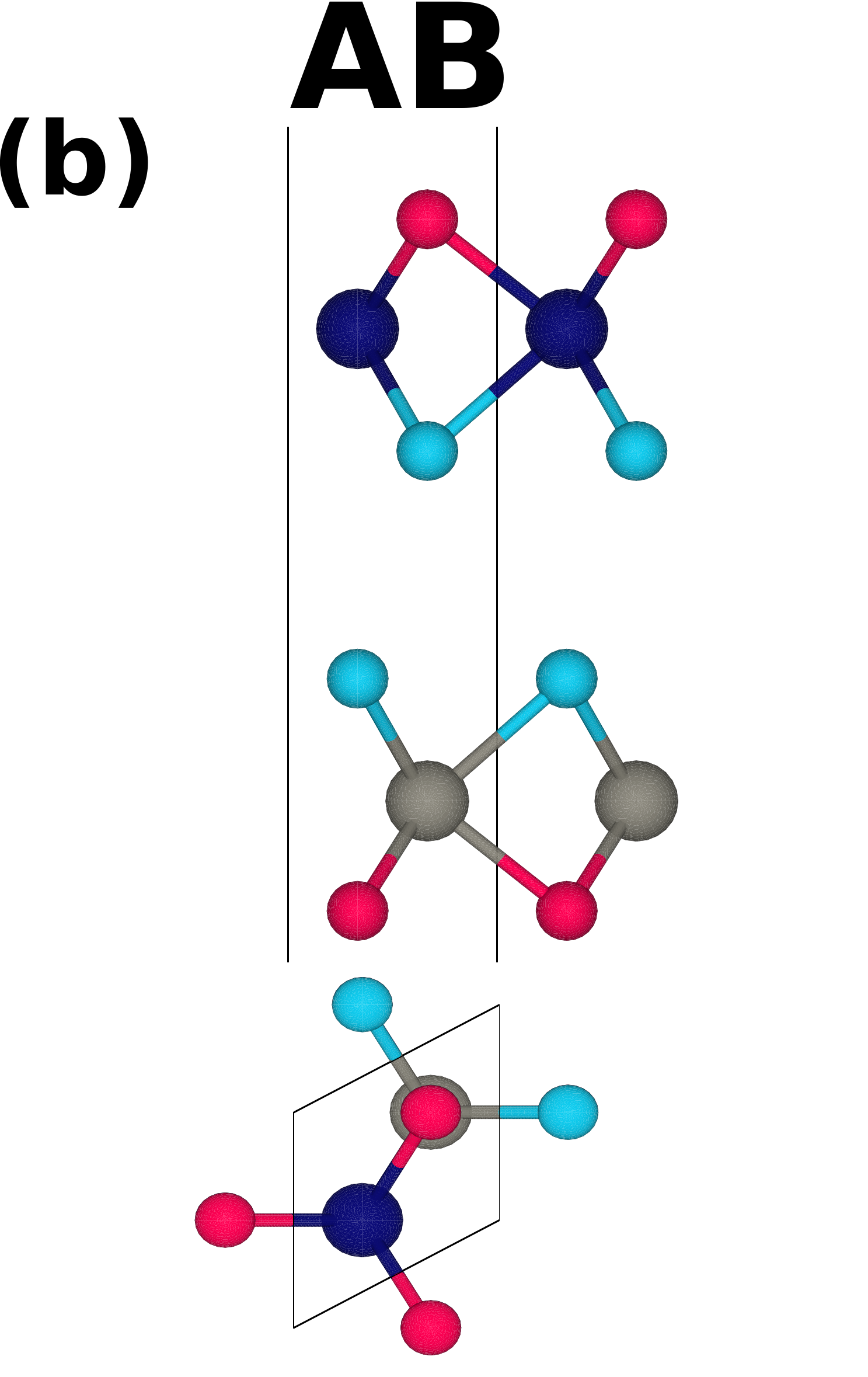}&
			\includegraphics[scale=0.2]{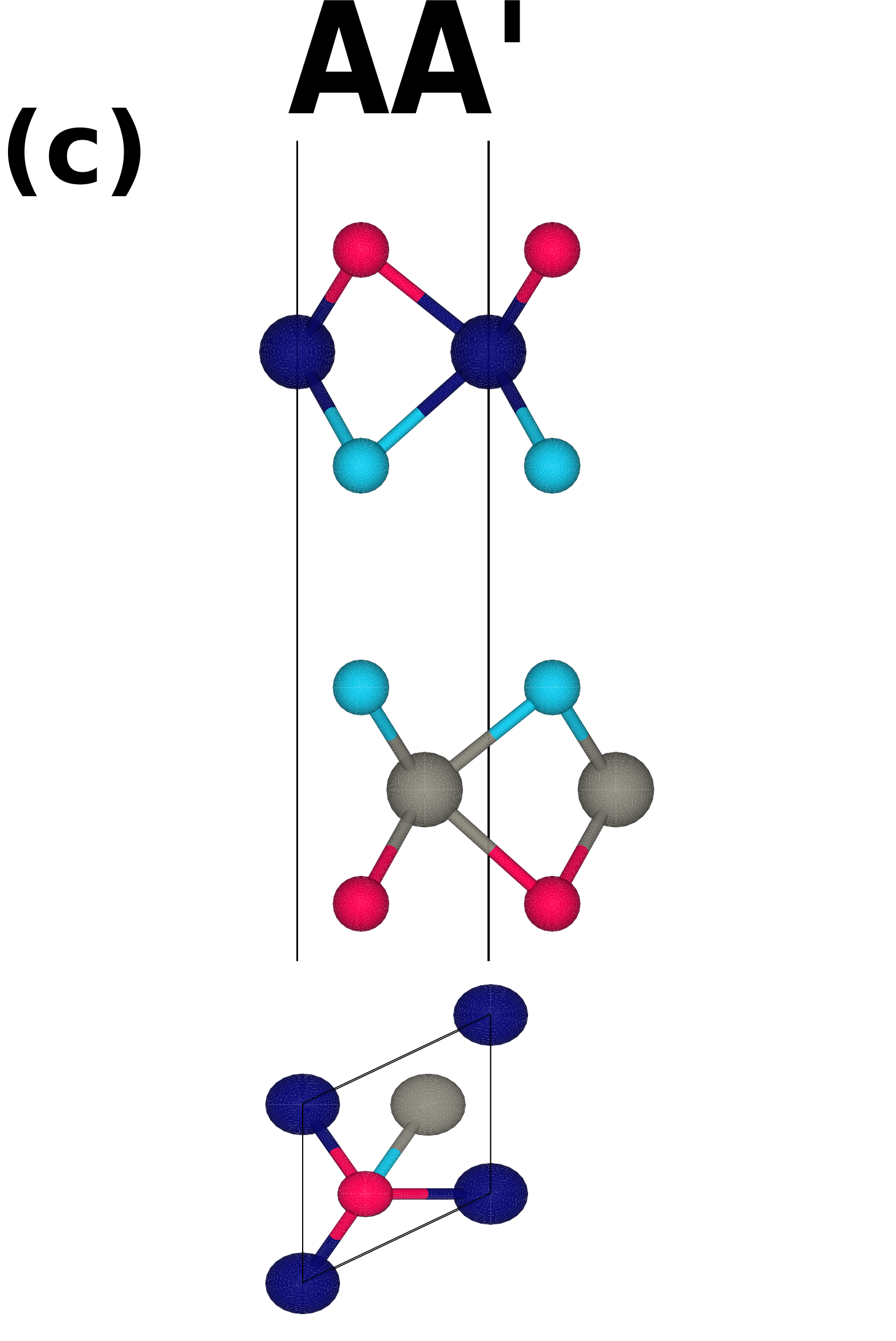}&
			\includegraphics[scale=0.2]{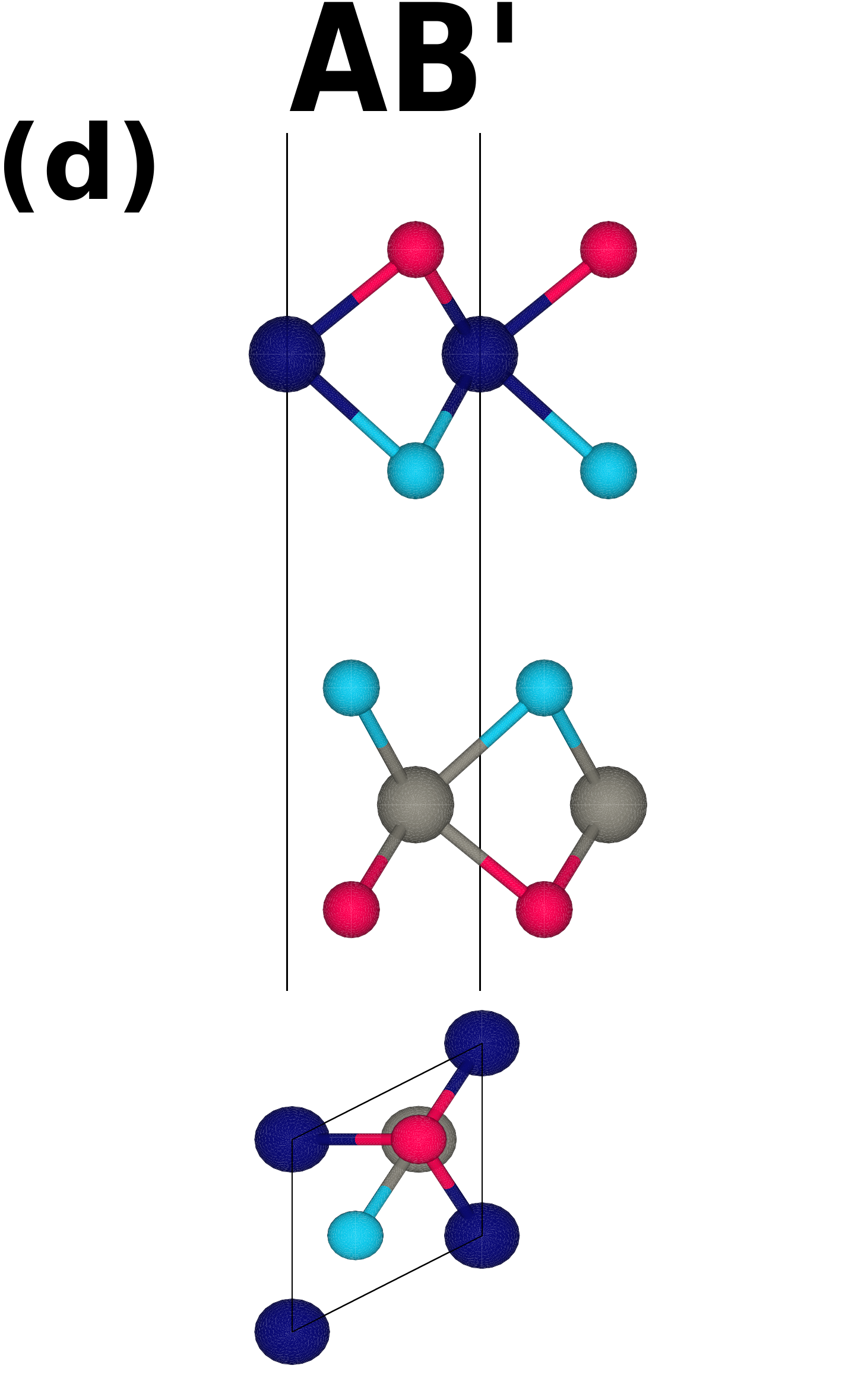}&
			\includegraphics[scale=0.2]{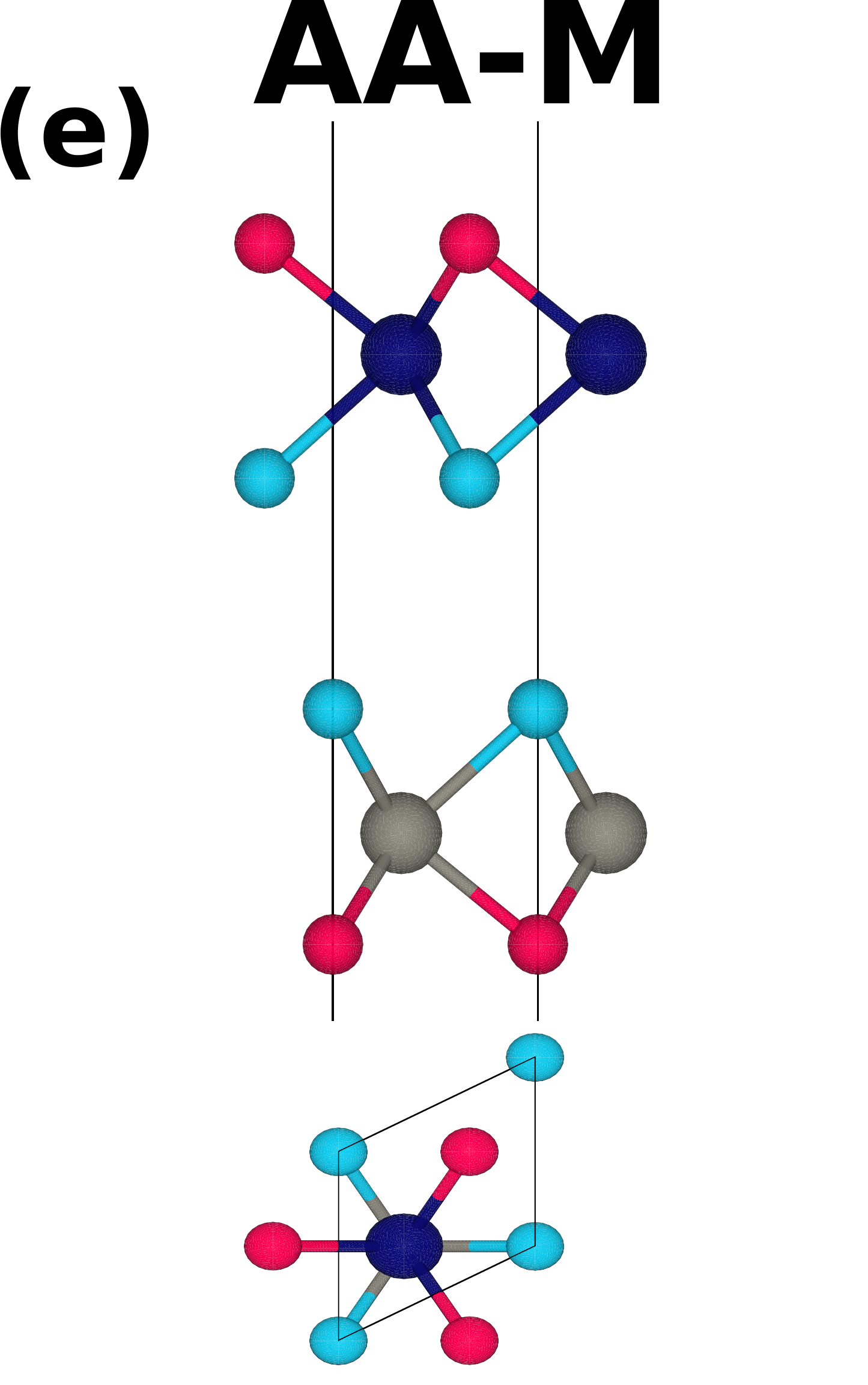} &
			\includegraphics[scale=0.2]{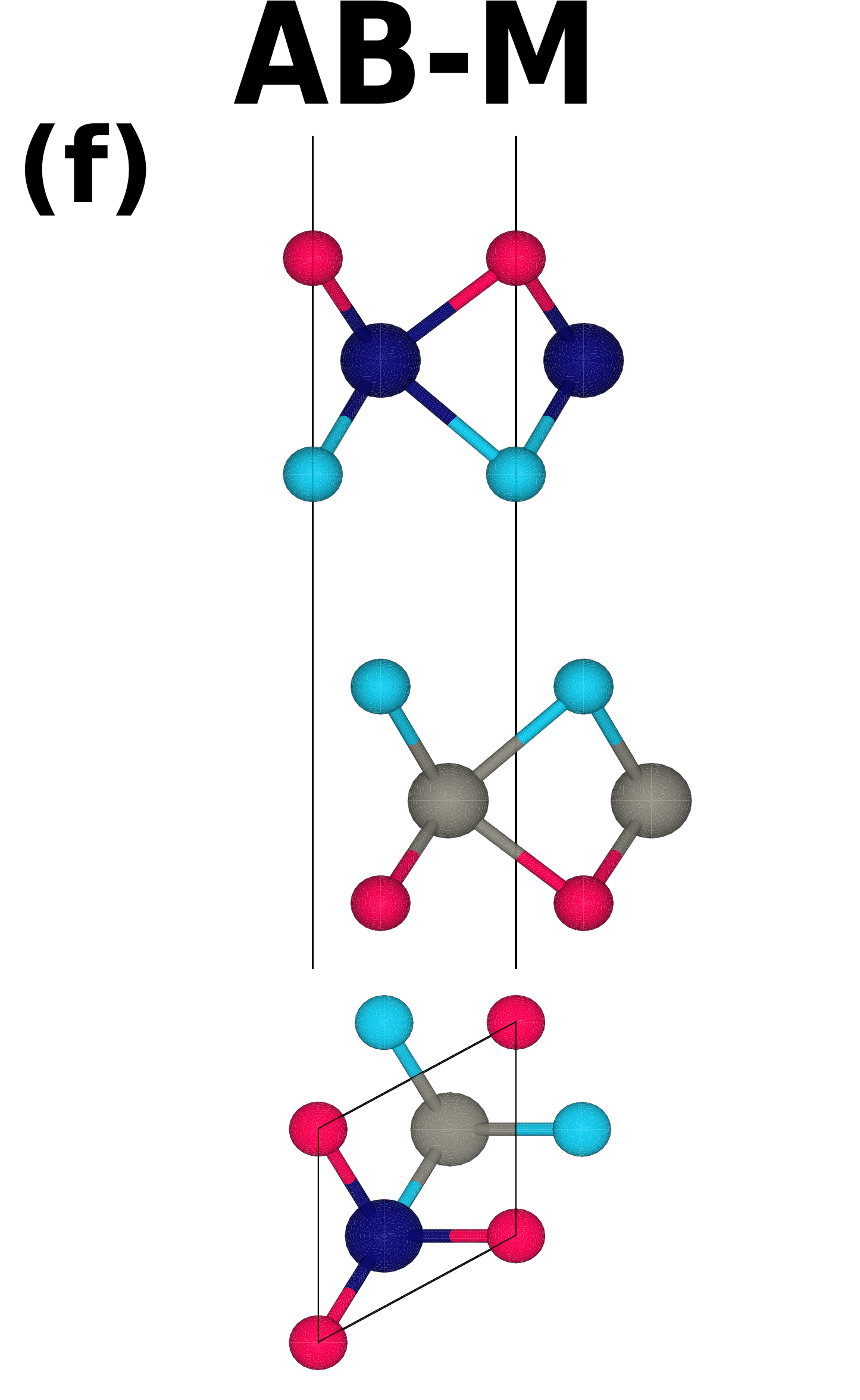}\\
		\end{tabular}
		\caption{Side and top views of all the stacking orders with $Y-Y$ interface. Gray and blue spheres represent W and Mo, while magenta and cyan spheres denote $X$ and $Y$ chalcogen atoms, respectively. $Y$ atom is the chalcogen atom with larger atomic number. The interlayer distance is denoted by d.}
		\label{fig:stackings}
	\end{figure*}
	
	In the search for materials with strong spin-orbit coupling (SOC), the Rashba effect appears as an important tool, resulting from the lack of out-of-plane mirror symmetry in the direction perpendicular to the two-dimensional plane. It is considered the origin of the spin Hall effect \cite{hirsch1999spin,sinova2004universal,wang2020spin, yu2021spin}. The capability of manipulating the Rashba effect can be exciting and essential for spintronic applications and making devices such as Spin-FETs \cite{datta1990electronic}. 
	
	The Rashba effect (a momentum-dependent splitting of spin bands) for a 2D electron gas can be described by the Bychkov-Rashba Hamiltonian \cite{bychkov1984properties}:
	
	\begin{center}
		\begin{equation}
		\label{eq:rashba}
		H_R = \alpha_R(\vec{E}_{int})\,(\vec{k}\times \vec{\sigma})\cdot\vec{z},
		\end{equation}
	\end{center}
	where, the factor $\alpha_R$ is the Rashba parameter. $\vec{\sigma}$ denotes the Pauli matrices, $\vec{k} = (k_x, k_y,0)$ is the in-plane momentum of electrons, and $\vec{z}=(0,0,1)$ is the out-of-plane direction. Therefore, spin degeneracy is lifted and the energy-momentum dispersion relation takes the following form: 
	
	\begin{center}
		\begin{equation}
		E_{\pm}(\vec{k}) = \frac{\hbar^2\vec{k}^2}{2m^*} \pm \alpha_R|\Delta k|,
		\end{equation}
	\end{center}
	
	This relation represents two parabolas which are shifted by $\pm \Delta k$ in the reciprocal space as depicted in Fig. \ref{fig:efield_direction}(a). Here, $m^*$ is the effective mass of the electrons.
	
	A significant reason for the success of semiconductor spintronics is the tuning ability of Rashba spin-orbit coupling via external gating. There are several ways to tune the Rashba parameter ($\alpha_R$), such as external electric field (EEF), strain, doping, etc. \cite{shanavas2014electric,liu2021tuning, singh2017giant,absor2018strong,chen2020tunable}. We investigate here the modification of electronic and spintronic properties of Janus vdW Mo$XY$/W$XY$ HSs by EEF and an in-plane biaxial strain using first-principles calculations. Our findings also shed light on the band-gap engineering and charge transfer phenomena via EEF. We outline the band-alignment facet for these HSs. Also, we present the transitions to various types of band gaps and tuning of $\alpha_R$ under strained conditions. 
	
	The paper is organized in the following manner. We narrate the methodology used in Sec.~\ref{sec:levelcomp}. Sec.~\ref{sec:levelcryst} describes the crystal structure of J-TMD HSs and insights for choosing the AB-stacking. In Sec.~\ref{sec:levelelec}, we study the electronic properties of three Mo$XY$/W$XY$ HSs, where we choose $X$, $Y$ = S, Se, Te; $X\neq Y$. Our \textit{ab initio} results are supplemented with $\mathbf{k\cdot p}$ analyses in Secs.~\ref{sec:GammapointHam} and \ref{sec:KpointHam}. This is followed by a detailed discussion of the effects of EEF and biaxial strains on band-gaps and the Rashba parameter in Secs.~\ref{sec:levelEEF} and \ref{sec:levelstrain}, respectively. Finally, we summarize our results in Sec.~\ref{sec:levelconclusion}.
	
	\section{\label{sec:levelcomp}Computational Details}

	\begin{table*}
		\centering
		\setlength{\tabcolsep}{15pt} 
		\renewcommand{\arraystretch}{1.2} 
		\caption{\label{table:optimized} Optimized parameters, formation energies and band-gaps (both PBE and HSE) for all three AB-stacked heterostructures with $Y-Y$ interface. The negative formation energies indicate the stability of these HSs. Symbol $^*$ represents a direct band-gap.}
		\begin{tabular}{cccc}
			\hline
			\hline
			Optimized parameters &   MoSSe/WSSe & MoSTe/WSTe &  MoSeTe/WSeTe \\
			\hline
			Lattice constant, a (\AA)  & 3.25 & 3.36 & 3.43 \\
			Bond length Mo-X/Y (\AA)  &  2.42/2.53 & 2.43/2.71 & 2.55/2.72 \\
			Bond length W-X/Y (\AA)  & 2.42/2.53 & 2.44/2.72 & 2.56/2.72 \\
			Interlayer distance, d (\AA)  &  6.57 & 7.20 & 7.09\\
			Angle, X-Mo-Y   & 81.29$^{\circ}$  & 81.58$^{\circ}$ &82.34$^{\circ}$ \\
			Angle, X-W-Y  & 81.46$^{\circ}$  & 81.68$^{\circ}$ & 82.49$^{\circ}$\\
			Formation Energy, $E_f$ (eV) & -1.90 & -2.25 & -2.46 \\
			$E_g^{\text{PBE+SOC}}$(eV) 			& 1.154$^*$   & 0.805 & 0.841 \\
			$E_g^{\text{HSE06+SOC}}$(eV)		 	& 1.492$^*$   & 1.207 & 1.149 \\ 
			\hline
			\hline
		\end{tabular}		
	\end{table*}
	
	First-principles calculations are performed using density functional theory (DFT) as implemented in the Vienna Ab initio Simulation Package (VASP) \cite{kresse1996efficiency,kresse1996efficient}. All the calculations are carried out with Perdew–Burke–Ernzerhof (PBE) \cite{perdew1996generalized} exchange-correlation (XC) functional in the framework of generalized gradient approximation (GGA) and projector augmented wave (PAW) method \cite{blochl1994projector,kresse1999ultrasoft} adopted for ion-electron interactions. The cut-off energy for the plane-wave basis is set to be 500 eV, and we choose an energy tolerance criterion of $10^{-6} $ eV. All the structures are fully relaxed by the conjugate gradient (CG) algorithm until the force on each atom is less than 0.01 eV/\AA. A Monkhorst-pack of 13 $\times$ 13 $\times$1 is used to sample the whole Brillouin zone for both geometry optimizations and static electronic structure calculations. Spin-orbit coupling (SOC) is considered in all the calculations. It is well known that the semi-local XC functionals like GGA-PBE underestimate the band-gaps of semiconductors and insulators. Therefore, in order to calculate the band-gaps of the Janus HSs with better precision, we use the screened hybrid Heyd-Scuseria-Ernzerhof (HSE06) exchange-correlation functional \cite{heyd2003hybrid, heyd2004efficient}, in which a fraction of the exact Hartree-Fock exchange is used. Our HSE06+SOC calculations are found to improve the band-gaps of the HSs significantly without a noticeable change in the overall band profile. That is why in the foregoing, we present all results calculated using PBE functional unless stated explicitly.
	
	To avoid the interactions between periodic layers, a vacuum of 20 \AA\ is added. DFT-D2 correction scheme developed by Grimme \cite{grimme2006semiempirical} is employed to include the van der Waals (vdW) interactions in our calculations. Phonon dispersion analysis is performed through the finite displacement method implemented in the PHONOPY code \cite{togo2015first}. The external electric field (EEF) is applied perpendicular to the heterostructure plane by introducing a dipolar sheet at the center of the simulation cell. This method is implemented in VASP and contributed by J. Neugebauer and M. Schffler \cite{neugebauer1992adsorbate}. The EEF is applied to the optimized structures to avoid field-induced modifications. The structures and the charge densities are plotted in the visualization software VESTA \cite{momma2011vesta}. 

	\section{\label{sec:level3}Results and Discussion}
	
	\subsection{\label{sec:levelcryst}Crystal structure}
	In the present study, we work on three vertical vdW Janus heterostructures (J-HSs), MoSSe/WSSe, MoSTe/WSTe, and MoSeTe/WSeTe. All three J-HSs consist of two different J-MLs with 2H polymorph (or trigonal prismatic phase). Although the TMDs and the J-TMDs also come with another symmetry, called 1T polymorph (or octahedral symmetric phase), the 2H phase is found to be more stable than the 1T phase \cite{wypych1998scanning,tang2018distorted,yang2019emerging} which has encouraged us to select the 2H phase for our investigations. The J-MLs possess $C_{3v}$ point-group symmetry with space group $P3m1$ (no. 156), unlike the conventional TMDs with $D_{3h}$ point-group symmetry and space group $P\bar{6}m2$ (no. 187). In this trigonal prismatic (2H) phase, the $X$ and $Y$ atoms are aligned vertically on top of each other along the $z$-direction. The mirror symmetry in J-TMDs is broken due to different chalcogen atoms on the opposite sides of the transition metal. Stacking the MLs of J-TMDs in different layers and different orders also breaks the structural symmetry.
	
	We first consider all possible stacking orders of J-HSs, in which the bottom layer is W$XY$ and the top layer is Mo$XY$ (where $X \neq Y$ = S, Se, Te). The six possible stacking patterns are shown in Fig.~\ref{fig:stackings}. In order to have a small lattice mismatch between two monolayers, we select J-TMDs with the same $X$ and $Y$ chalcogen atoms. The values of the lattice mismatch are 0.03, 0.08 and 0.09\% for MoSSe/WSSe, MoSTe/WSTe and MoSeTe/WSeTe HSs, respectively. For all the stacking orders, similar chalcogen atoms with larger atomic numbers face each other at the interface, making the $Y-Y$ interface, where $Y$ is the chalcogen atom with a larger atomic number. The stacking pattern at the interface has a measurable effect on the electronic properties of the system \cite{rezavand2021stacking}. Formation or binding energy is the difference between the total ground state energy of HS and the sum of ground-state energies of the corresponding MLs, $E_f = E_{MoXY/WXY} - E_{MoXY} - E_{WXY}$. The negative values of the formation energies are listed in Table \ref{table:optimized}, which confirm the suitability of the formation of the HSs. From our calculations and previous studies, it is found that the AB-stacking with the $Y-Y$ interface corresponds to the lowest ground state energy \cite{guo2020strain,wang2019mirror}, making it the most stable vertical stacking. So we choose the AB-stacking geometry for HSs in our work. In the AB-stacking, the chalcogen atom at the interface in the top layer is directly on top of the transition metal atom of the bottom layer, as shown in Fig.~\ref{fig:stackings}(b). The optimized lattice constants, interlayer distances, bond lengths, and bond angles of all three HSs with AB-stacking are presented in Table~\ref{table:optimized}. We find that the MoSTe/WSTe J-HS has the largest interlayer distance while MoSSe/WSSe has the lowest interlayer distance for this particular AB-stacking with the $Y-Y$ interface. Calculating the electronegativity differences ($\Delta_{en}$) for all three combinations show that the difference varies as $\Delta_{en}^{S-Te} > \Delta_{en}^{Se-Te} > \Delta_{en}^{S-Se}$, which is in accordance with their interlayer distances. $\Delta_{en}$ dictates the direction and strength of the intrinsic electric field, as described in a later section.
	
	\subsection{\label{sec:levelelec}Electronic properties}
	
	\begin{figure*}[!htb]
		\centering
		$\begin{array}{cc}
		\includegraphics[scale=0.9]{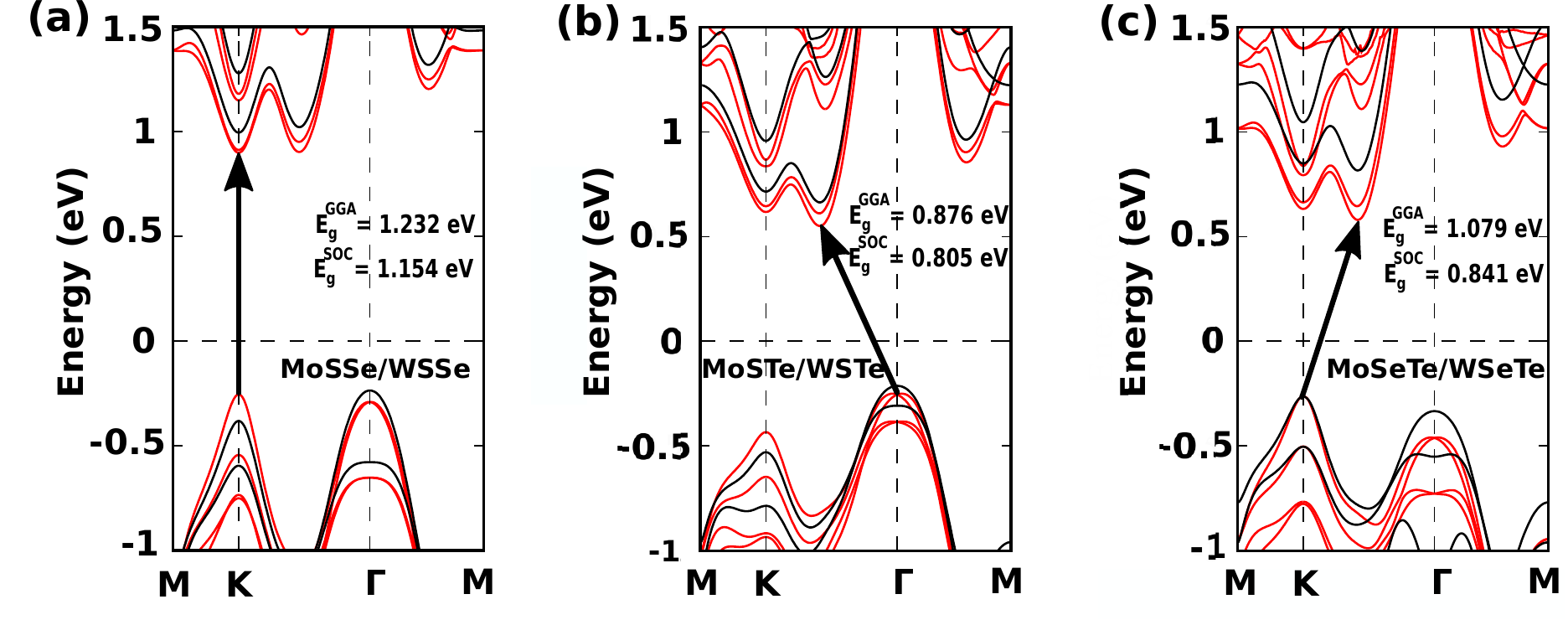}  \\
		\includegraphics[scale=0.9]{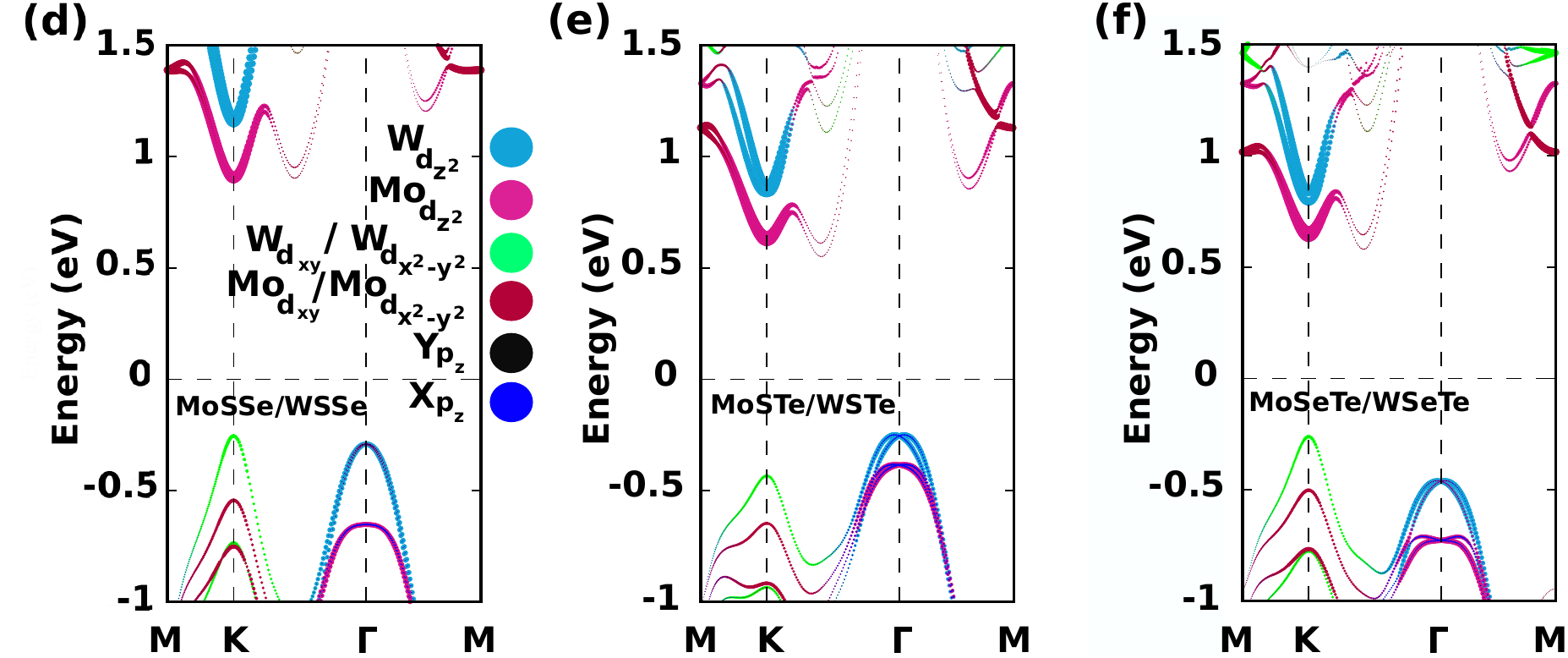} \\
		\end{array}$
		\caption{(a-c) Band structures of MoSSe/WSSe, MoSTe/WSTe and MoSeTe/WSeTe J-HSs with (red) and without SOC (black) using PBE XC functional. Arrows in the figures represent the system band-gaps (E$_g^{SOC}$) with SOC. As seen, MoSTe/WSTe and MoSeTe/WSeTe have indirect band gaps with and without SOC, while it changes from indirect to direct with the inclusion of SOC in the case of MoSSe/WSSe. M (1/2,0,0), K (1/3,1/3,0) and $\Gamma$(0,0,0) are the high symmetry points in the Brillouin zone. (d-f) Orbital-projected band structures of all three J-HSs. The size of the circles is proportional to the corresponding orbital weight. The orbitals weights at the VBM and CBM show that all the three HSs have type-II band alignments.}
		\label{fig:bands_hs}
	\end{figure*}
	
	\begin{figure*}[!htb]
		\centering
		\includegraphics[scale=0.26]{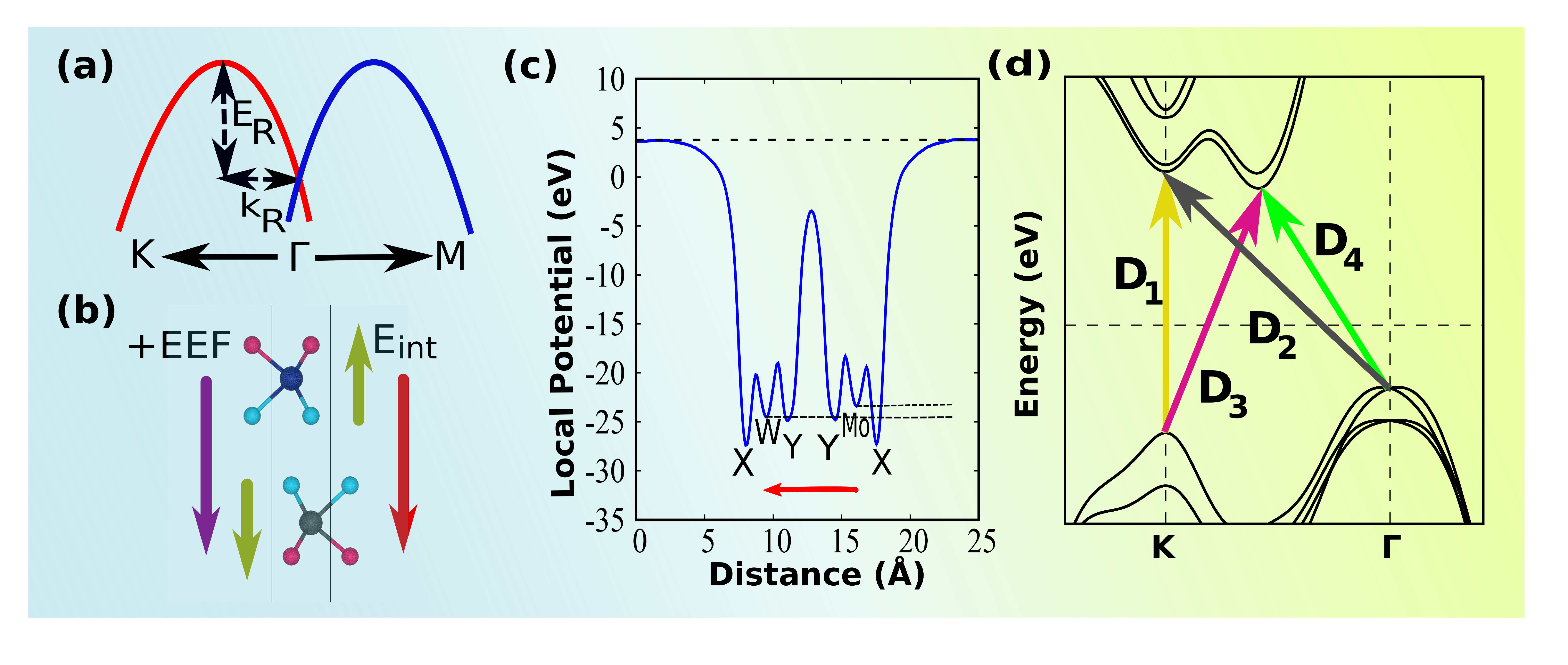}
		\caption{(a) Schematics of calculating the Rashba parameter $\alpha_R$, where $E_R$ is the Rashba energy and $k_R$ is the momentum offset. (b) Directions of the intrinsic and external electric fields. E$_\text{int}$, with the red arrow, represents the net intrinsic electric field, and olive green arrows denote the intrinsic fields due to the chalcogen atoms in each layer. +EEF is the positive external electric field applied in the downward direction from the Mo$XY$ layer to the W$XY$ layer. (c) Local electrostatic potential profile for Mo$XY$/W$XY$ HS without any external effect. The red arrow represents the direction of E$_\text{int}$. Here, the dashed lines indicate the potential drop across the interface, confirming the existence of an intrinsic electric field in the direction from Mo to W atom. (d) A general band structure showing different types of gaps between the valence band and the conduction band. D$_1$ is the direct gap at the K point, D$_2$ is the gap between VBM at $\Gamma$ and CBM at the K point, and D$_3$(D$_4$) is the gap between VBM at K ($\Gamma$) and CBM in the $\Gamma-$K direction. Out of these four gaps, one of them turns out to be the energy band-gap of the system.}
		\label{fig:efield_direction}
	\end{figure*}
	
	In order to determine the electronic properties of AB-Mo$XY$/W$XY$ J-HSs, we calculate the band structures of the three vdW HSs, which are displayed in Fig.~\ref{fig:bands_hs}. The MoSSe/WSSe HS has a direct band-gap at the K point of the Brillouin zone, while MoSTe/WSTe and MoSeTe/WSeTe HSs are indirect band-gap semiconductors. In MoSSe/WSSe and MoSeTe/WSeTe heterostructures, the valence band maximum (VBM) is located at the K point while it is located at the $\Gamma$ point in the MoSTe/WSTe heterostructure. The conduction band minimum (CBM) in MoSTe/WSTe and MoSeTe/WSeTe HSs are positioned between K and $\Gamma$ points. We also perform the band structure calculations for J-MLs and match our results with the previous studies. We find that \textit{M}SSe and \textit{M}SeTe (\textit{M} = Mo and W) are direct band-gap semiconducting MLs, while \textit{M}STe MLs have indirect band-gaps \cite{hu2018intrinsic, xia2018universality}. Among them, MoSSe ML has the largest band-gap of 1.45 eV. The reduced band-gap in MoSSe/WSSe and transition from direct band-gap in MLs to indirect in MoSeTe/WSeTe is the result of vdW interactions between both layers of the HS as already reported in previous studies. \cite{wang2017tuning,han2011band, sahoo2021electric}. Since GGA-PBE functional is known to underestimate the electronic band-gaps of semiconductors and insulators, we also perform HSE06+SOC calculations for the three HSs. The band-gaps of all the HSs are found to be significantly enhanced with the introduction of hybrid exchange-correlation functional, as seen from Table \ref{table:optimized}. The MoSSe/WSSe HS remains a direct band-gap, and the other two HSs remain indirect band-gap semiconductors with HSE functional (see Fig. S1 of the Supplemental Material (SM) \cite{patel2022rashba}). Fig. S1 also shows that the band-gap value of the MoSTe/WSTe HS becomes larger than that of the MoSeTe/WSeTe HS, keeping the nature of band-gaps unchanged, i.e. MoSTe/WSTe and MoSeTe/WSeTe HSs still remain D$_4$ and D$_3$ type band-gap semiconductors, respectively.
	\begin{center}
		\begin{figure*}[!htb]
			\centering
			\hspace{-0.3cm}\includegraphics[scale=0.425]{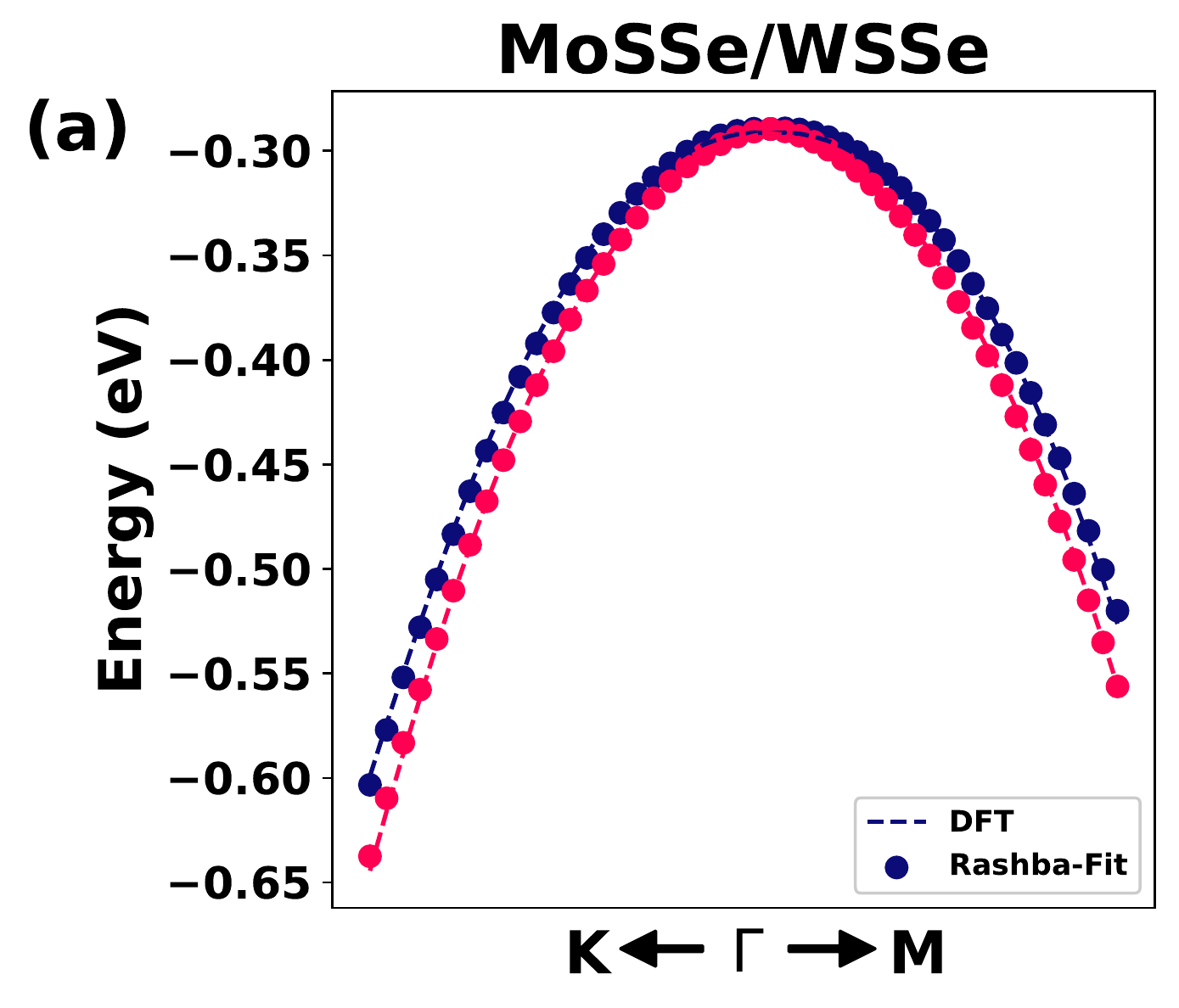}
			\hspace{-0.2cm}\includegraphics[scale=0.425]{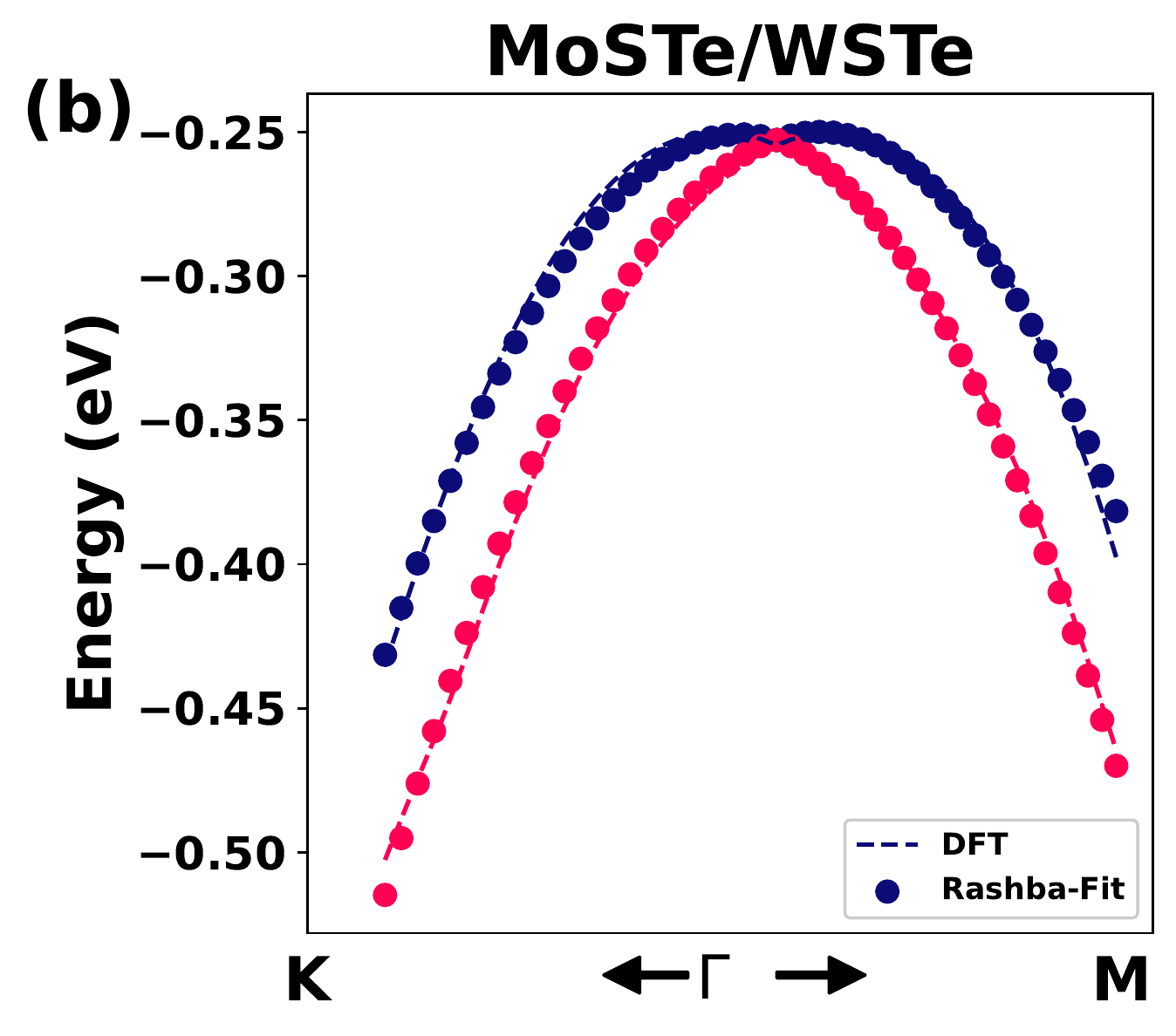}
			\includegraphics[scale=0.425]{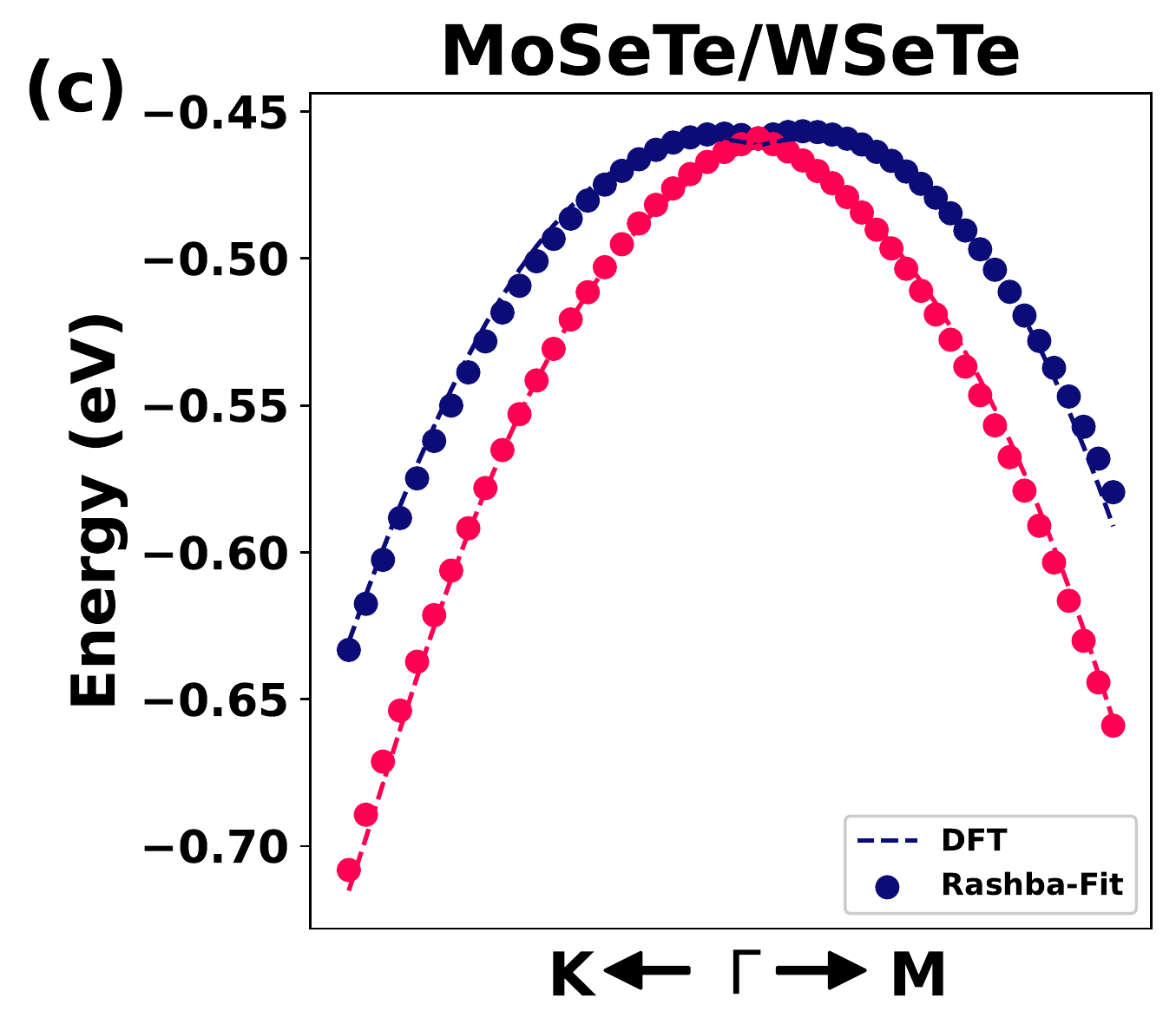}
			\includegraphics[scale=1.03]{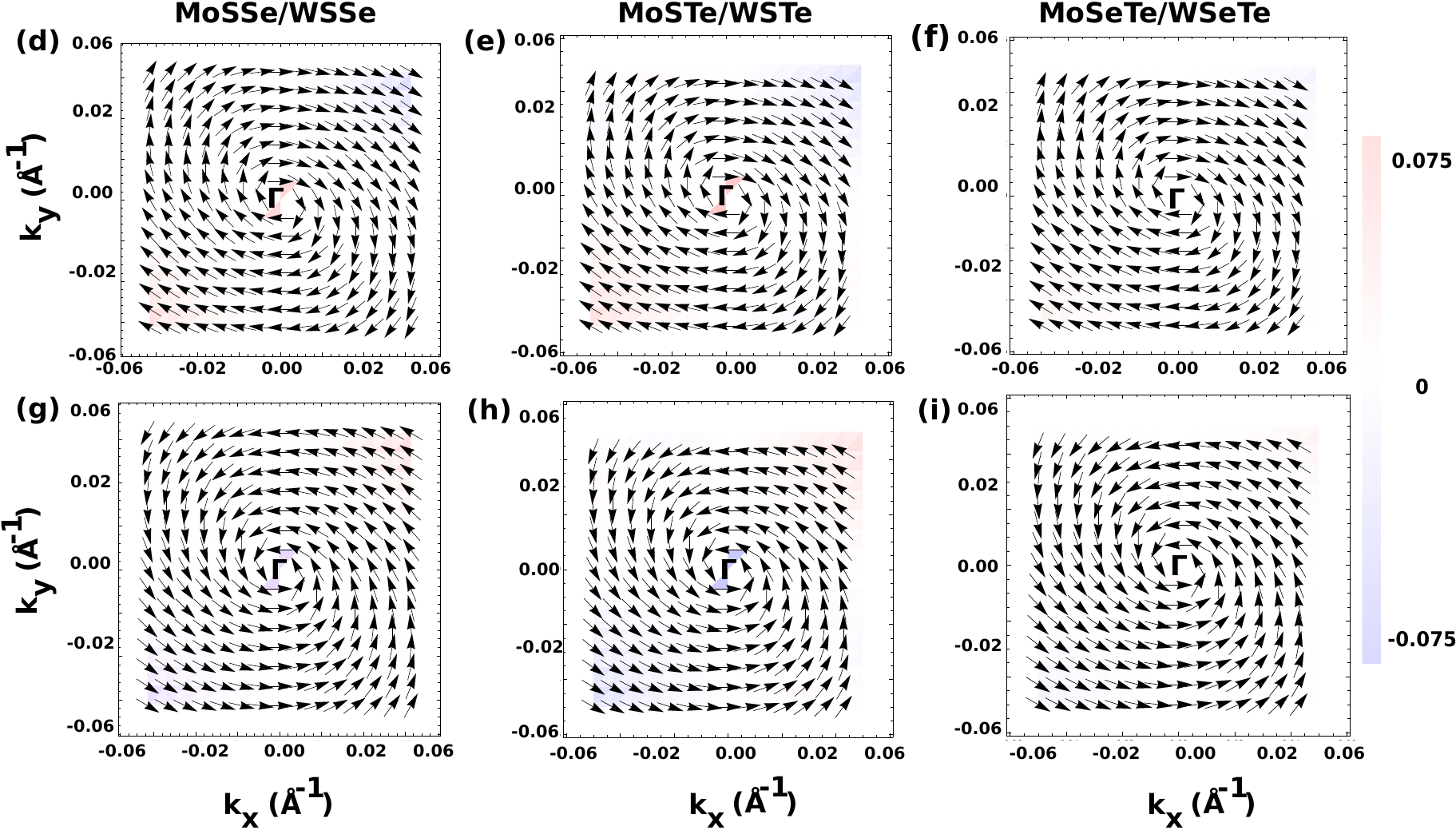}
			\caption{Upper panel(a-c): Bands around $\Gamma$ point fitted along $\frac{2\pi}{a}(0.08,0.08,0)-(0,0,0)-\frac{2\pi}{a}(0.12,0,0)$ direction (along K $-\Gamma-$ M) of the Brillouin zone, where $a$ is the in-plane lattice constant. The dashed lines and scattered points represent the DFT bands and the $k \cdot p$ bands, respectively. Middle panel(d-f): Spin textures for the inner branch (red bands in the upper panel). Bottom panel(g-i): Spin textures for the outer branch (blue bands in the upper panel). The colorbar in the right shows the magnitude of the out-of-plane spin projection ($S_z$). The almost colorless backgrounds in the middle and bottom panels depict the 2D nature of Rashba spin-splitting around the $\Gamma$ point. }
			\label{fig:fit_Rashba_ST}
		\end{figure*}
	\end{center}
	
	\begin{center}
		\begin{figure*}
			\centering
			\includegraphics[scale=0.63]{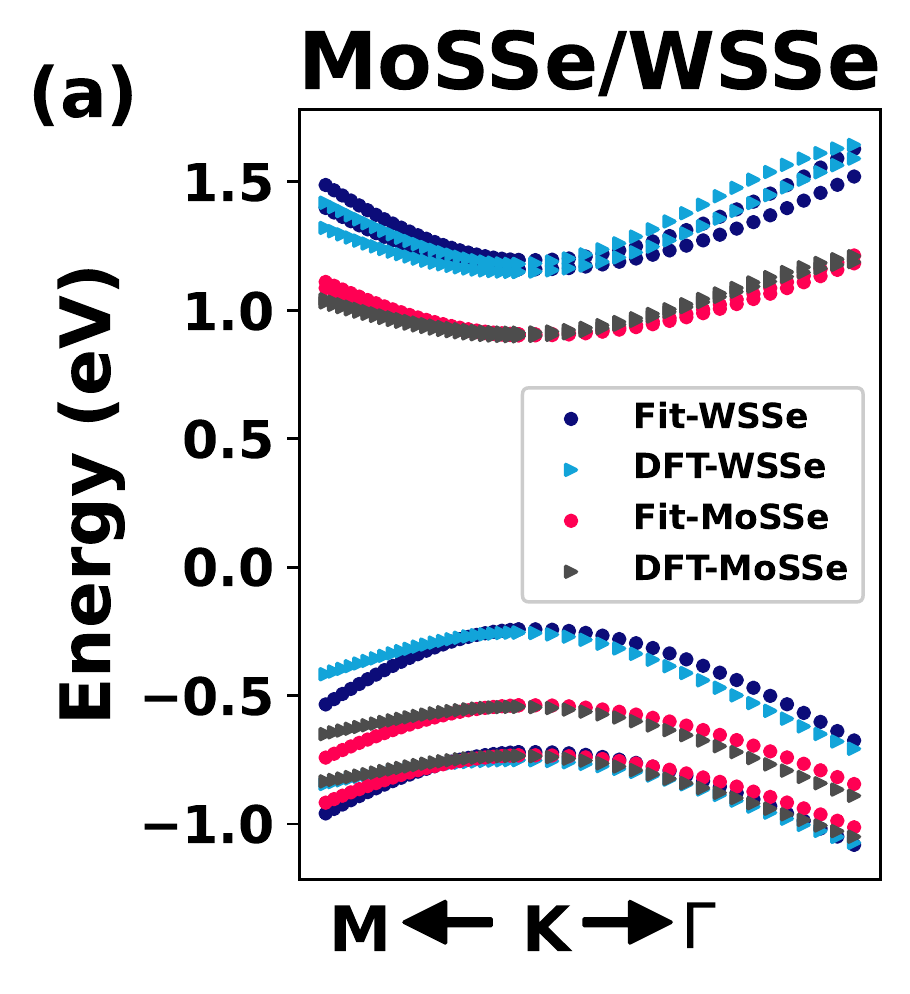}
			\includegraphics[scale=0.63]{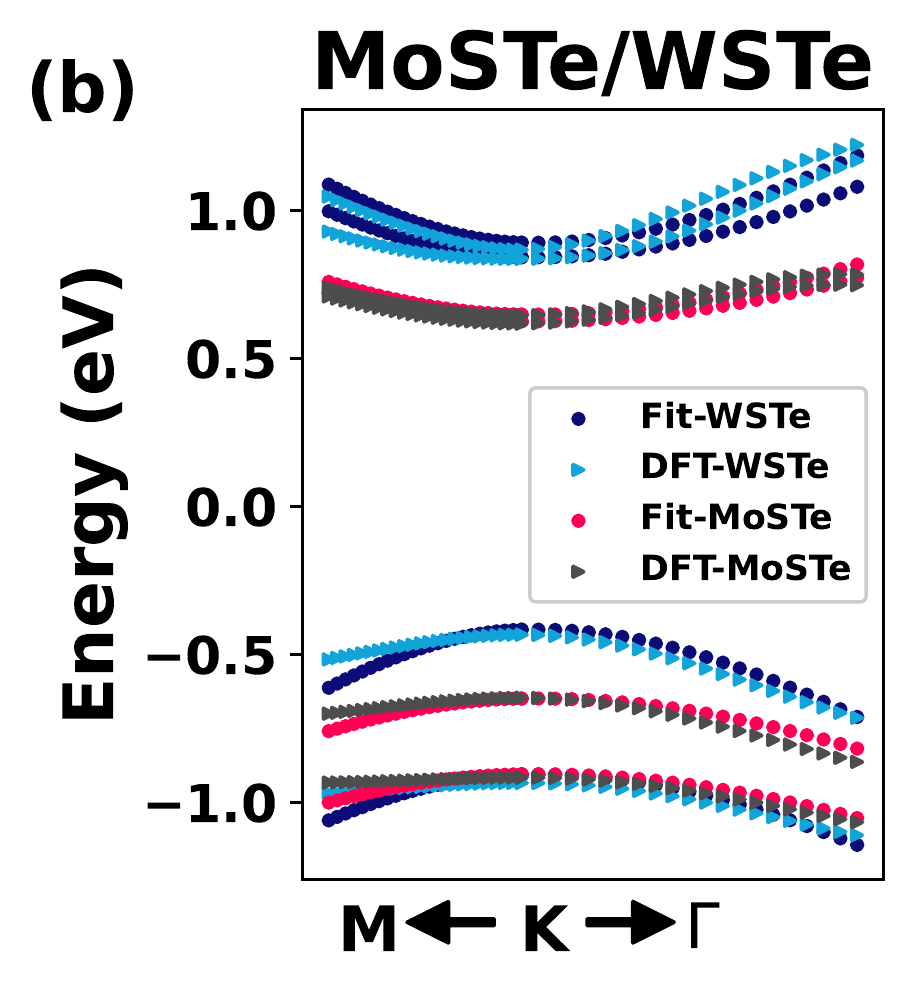}
			\includegraphics[scale=0.63]{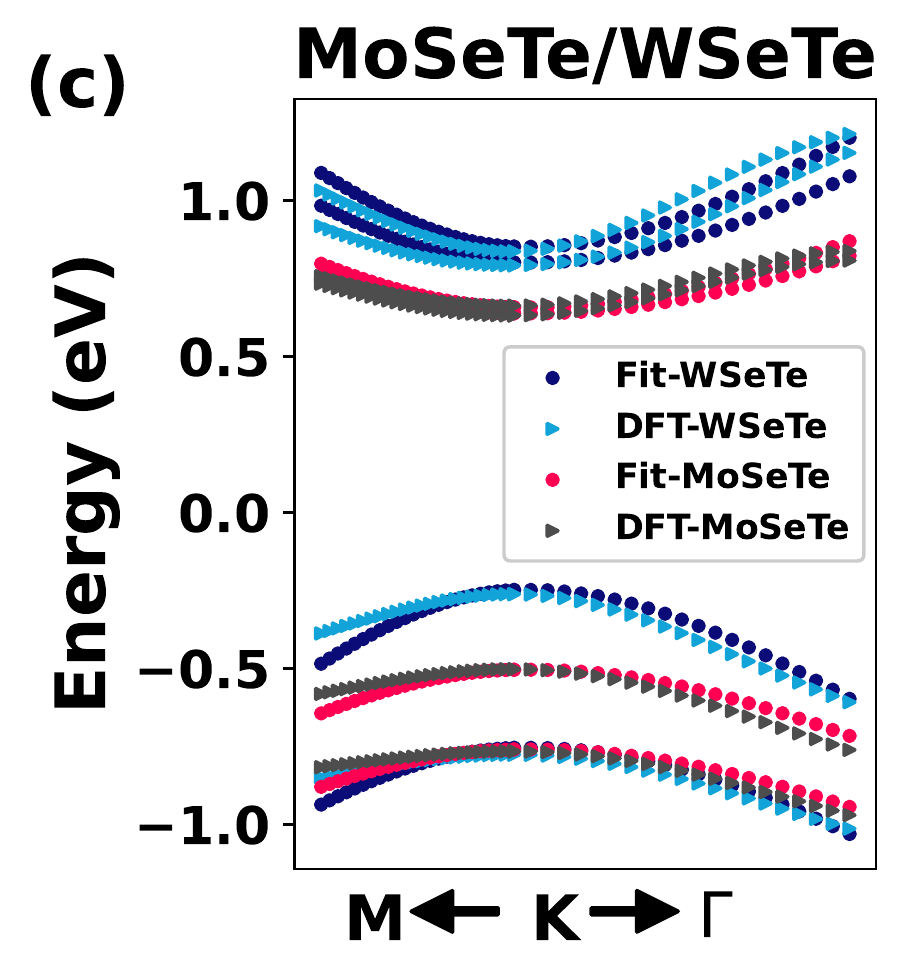}
			\caption{Band structures for (a) MoSSe/WSSe, (b) MosTe/WSTe, and (c) MoSeTe/WSeTe around the K point are fitted along the $\frac{2\pi}{a}(0.38,0.25,0)-\frac{2\pi}{a}(0.33,0.33,0)-\frac{2\pi}{a}(0.25,0.25,0)$ direction (along M$-$K$-\Gamma$) of the Brillouin zone, where $a$ is the in-plane lattice constant. Right triangles represent the DFT bands and circles denote the $k \cdot p$ bands in Eq. (\ref{eqn4}).} 
			\label{fig:fit_near_K}
		\end{figure*}
	\end{center}
	
	We also plot the orbital projected band structures for all three J-HSs in Fig.~\ref{fig:bands_hs}(d-f). One can see that the valence band and conduction band are mainly contributed by the $d$-orbitals of W and Mo, respectively. Moreover, the valence band maxima at the $\Gamma$ and K points are primarily occupied by W$_{d_{z^2}}$ and W$_{d_{x^2-y^2}}$ (equivalently, Mo$_{d_{xy}}$) orbitals, respectively. Similarly, the conduction band minima are dominated by the Mo$_{d_{z^2}}$ orbitals. There is also a small contribution coming from the $p_z$ orbitals of $X$ and $Y$ atoms, mostly away from the Fermi level. If we see the overall effect, The VBM is located at the W$XY$ layer while the CBM is found to be at the Mo$XY$ layer, resulting in a type-II band alignment for all three HSs.  The orbital symmetries of the bands and the nature of band alignment remain unchanged when we use HSE06 functional, as shown in Fig. S1.  Type-II band alignment in these HSs can lead to the generation of spatially separated photo-excited electrons and holes with increasing carrier lifetimes and thereby have immense technological importance in optoelectronics, valleytronic, photocatalytic and photovoltaic applications \cite{hong2014ultrafast,yu2015valley,guan2018tunable,yu2013highly}. 
	
	Janus materials with inversion asymmetry inherit a large spin-orbit coupling and thus show a large out-of-plane spin-splitting at the K point of the Brillouin zone (BZ) \cite{xiao2012coupled, kosmider2013large,yuan2013zeeman}. On the other hand, as a consequence of replacing one $X$ by $Y$ in \textit{M}$X_2$ MLs and further with vertical stacking, additional inversion symmetry breaking takes place. Due to this inversion asymmetry, these J-MLs and J-HSs inherit an intrinsic electric field which is the main source of the emerging in-plane Rashba spin-splitting \cite{bychkov1984properties}. The net transverse electric field mainly controls the strength of this Rashba spin-splitting. The schematics to calculate the $\alpha_R$ is presented in Fig.~\ref{fig:efield_direction}(a), which is defined as $\alpha_R = 2E_R/k_R$. Here, $E_R$ is the Rashba energy, and $k_R$ is the momentum offset around the $\Gamma$ point, also called the Rashba wavevector. The calculated values of $\alpha_R$ for three HSs, MoSSe/WSSe, MoSTe/WSTe and MoSeTe/WSeTe are 30.74, 178.8 and 129.8 meV \AA, respectively. 
	
	We list down the values of $\alpha_R$ for different Janus systems in Table S1 \cite{patel2022rashba}, which shows that Rashba spin-splittings depend upon the interfacial (or interlayer) distance, structural arrangements and the strength of the net intrinsic electric field. $\alpha_R$ is found to be highest for AA-stacking with the $Y-Y$ interface. This particular stacking with the said interface has the intra-layer dipoles in opposite directions, which results in an increase of the interlayer distance between the transition metals. As a result, the net intrinsic electric field due to the transition metals becomes larger, giving rise to a large spin-splitting. On the other hand, $\alpha_R$ has the lowest values for AB-stacked $Y-X$ interfaces having the smallest interfacial distance. As mentioned earlier, in the present work, we choose the Janus HSs, which possess the most stable interface structure ($Y-Y$) and stacking pattern (AB-stacking) and correspond to the correct configuration, which is most likely to be realized in experiments. Our chosen systems have lower $\alpha_R$ compared to the Janus monolayers, $X-Y$ interfaced bilayers and AA-stacked HSs due to the cancellation of intrinsic electric fields originating from the chalcogen atoms on each layer, leaving behind contributions from the transition metals only. We will discuss this in Sec.~\ref{sec:levelEEF} in more detail. Moreover, a detailed investigation (from Table S1) reveals that for a given interface, $\alpha_R$ varies as $\alpha_R^{\text{MoSTe/WSTe}} > \alpha_R^{\text{MoSeTe/WSeTe}} > \alpha_R^{\text{MoSSe/WSSe}}$ for AB-stacking and for AA-stacking it varies as $\alpha_R^{\text{MoSeTe/WSeTe}} > \alpha_R^{\text{MoSTe/WSTe}} > \alpha_R^{\text{MoSSe/WSSe}}$. Therefore, to have promising technological applications, it is essential to modify the strength of the Rashba spin-splitting in these HSs by external perturbations. In order to investigate the electronic properties at the $\Gamma$ and K points of the BZ in detail, we define $k \cdot p$ Hamiltonians in the following.

	\subsubsection{\label{sec:GammapointHam}k $\cdot$ p Hamiltonian at $\Gamma$ point}
	The Janus HSs possess $C_{3v}$ point group symmetry in which a center of inversion is absent. Absence of inversion symmetry in these 2D materials gives rise to a momentum dependent Rashba spin-splitting \cite{bychkov1984properties}. In our electronic structures, Rashba-like spin-splitting is observed only around the $\Gamma$ point, as seen from Fig. \ref{fig:efield_direction}(a). At this high symmetry point ($\Gamma$), the allowed linear-in-$k$ term in the $k\cdot p$ Hamiltonian is $(k_x\sigma_y - k_y\sigma_x)$ \cite{bychkov1984properties,vajna2012higher}. Other higher-order terms are also present \cite{vajna2012higher,yao2017manipulation,kormanyos2014spin}, which are cubic in $k$. The $k\cdot p$ Hamiltonian around $\Gamma$ subjected to spin-orbit coupling can be defined as follows:
	\begin{center}
		\begin{equation}
		\begin{split}
		H(k)^{\Gamma} &= H_0(k) + \alpha_R(k_x\sigma_y - k_y\sigma_x)+\beta_R[(k_x^3 + k_xk_y^2)\sigma_y \\
		&-(k_x^2k_y + k_y^3)\sigma_x] + \gamma_R (k_x^3 + 3k_xk_y^2)\sigma_z\\
		&\approx \left(Ak_x^2 + Bk_y^2\right) + \alpha_R(k_x\sigma_y - k_y\sigma_x)
		\end{split}	
		\label{eqn1}
		\end{equation}
	\end{center}
	where, $H_0(k)$ is the free electron Hamiltonian; $A = \frac{\hbar^2 }{2m_x^*}$ and $B = \frac{\hbar^2 }{2m_y^*}$ with $m_x^*$ and $m_y^*$ being the effective masses of electrons along different directions. $\alpha_R$ is the Rashba parameter in linear order, and $\beta_R$ and $\gamma_R$ are the coefficients in the cubic order.
	
	We ignore the small cubic Rashba terms in the Hamiltonian and numerically fit the first-principles bands around the $\Gamma$ point.
	One can see from Fig.~\ref{fig:fit_Rashba_ST}(a-c) that the eigenvalues of the Rashba Hamiltonian excellently fit the DFT bands around the $\Gamma$ point. From the numerical fit, we obtain the values of the $\alpha_R$ for the three Janus HSs, which are in agreement with the DFT values and listed in Table S2 \cite{patel2022rashba}. 
	
	In order to show the spin orientations for the two bands around $\Gamma$, we have plotted the spin textures in the momentum space, as shown in the middle and lower panels of Fig.~\ref{fig:fit_Rashba_ST}. The out-of-plane spin projection ($S_z$) is indicated by the background color. We observe circular spin textures for the outer (blue bands in the upper panel) and inner (red bands in the upper panel) branches which encircle in opposite directions. Moreover, the almost colorless backgrounds in the spin-textures indicate the presence of negligible out-of-plane spin components. Opposite spin orientations of the inner and outer branches and the presence of negligible out-of-plane spin components are the characteristic features of the pure Rashba effect \cite{tao2018persistent}. Therefore, from this analysis, it is evident that the bands around the $\Gamma$ point are best described by the Rashba-type Hamiltonian given in Eq. (\ref{eqn1}). Although very small, presence of a weak out-of-plane projection explains a slight mismatch in the numerical values of $\alpha_R$ obtained from DFT and numerical fittings.

	\begin{center}
		\begin{table*}
			\centering
			\caption{\label{table:fitted_K_bands} Values of the parameters in Eq. (\ref{eqn4}) obtained by fitting the first-principles bands around K.  $a$, $t$, $\Delta$, 2$\lambda_v$, and 2$\lambda_c$ are the lattice constant, hopping integral, band-gap, spin splittings at the VBM and CBM, respectively. Both the $MXY$ layers ($M$ = transition metal) are treated individually to perform the numerical fitting. $\Delta$ is the direct energy gap between the VBM and CBM of respective $MXY$ layers. }
			\begin{tabular}{ c c c c c c c}
				\hline
				\hline
				&				 &		Fitted 		&			&		&	DFT(GGA-PBE)	& \\
				\hline
				parameters	  	&   MoSSe (WSSe) & MoSTe (WSTe) & MoSeTe (WSeTe)	&	MoSSe (WSSe)	& MoSTe (WSTe)	&  MoSeTe (WSeTe)\\
				\hline
				$a$ 				& 3.25   & 3.36 & 3.43	&		&		&  \\ 
				$t$  				& 0.99 (1.21)& 0.67 (0.93) & 0.73 (0.96) 	&		&		&  \\
				$\Delta$ 		& 1.54 (1.66)& 1.41 (1.53) & 1.28 (1.33) 	&  1.40 (1.44)		& 1.26 (1.27)		& 1.13 (1.05)\\ 
				2$\lambda_v$ 	& 0.19 (0.48)   & 0.26 (0.49) & 0.26 (0.51) 	& 0.19 (0.49)	& 0.27 (0.50)		& 0.26 (0.51) \\
				2$\lambda_c$ 	& 0.01 (0.04)   & 0.02 (0.05) & 0.02 (0.05) 	& 0.01 (0.03)	& 0.03 (0.03)		& 0.03 (0.04) \\
				\hline
				\hline
			\end{tabular}
		\end{table*}
	\end{center}
	
	\subsubsection{\label{sec:KpointHam}k $\cdot$ p Hamiltonian at K point}

	While the electronic and spin properties of MoSTe/WSTe HS with VBM at $\Gamma$ are dictated by the Rashba Hamiltonian described above, the properties near the K point are also exciting and mainly crucial for MoSSe/WSSe and MoSeTe/WSeTe HSs for which the VBM lies at the K point. The valence band maxima and conduction band minima at the K point in both MoSSe/WSSe and MoSeTe/WSeTe HSs are contributed by $d_{xy}+d_{x^2-y^2}$ and $d_{z^2}$  orbitals of transition metals (both Mo and W), respectively.  Therefore, the symmetry adapted basis functions at the band edges are \cite{xiao2012coupled}: 
	\begin{center}
		\begin{equation}		
		| \phi_c > = | d_{z^2} >, \hspace{0.2cm} | \phi_v > = \frac{1}{\sqrt{2}} \left(| d_{x^2-y^2} > + i |d_{xy}> \right)
		\end{equation}
	\end{center}
	where, the subscripts $v$ and $c$ indicate the valence band and the conduction band, respectively. Spin-orbit coupling removes the spin degeneracy at VBM and CBM and results in two spin-split bands (total 4 bands) below and above the Fermi level for a single layer. However, the orbital symmetries of the bands remain preserved, as seen from Figs.~\ref{fig:bands_hs}(d-f). The $M$XY ($M$ = transition metal) layers have the similar orbital contributions at K and K$^\prime$ valleys except for the fact that the spin is flipped at another valley. It also emphasizes that the spin splitting does not depend on the model details and for the present study, we only focus on the spin-properties at K valley.
	
	A minimal four-band $\mathbf{k\cdot p}$ Hamiltonian with SOC can be constructed for each layer at the K point as \cite{xiao2012coupled}:
	\begin{center}
		\begin{equation}		
		H(k)^{K} = at(k_x\sigma_x + k_y\sigma_y) + \frac{\Delta}{2}\sigma_z + \lambda_v\frac{1-\sigma_z}{2}\hat{s}_z + \lambda_c\frac{1+\sigma_z}{2}\hat{s}_z
		\label{eqn4}
		\end{equation}
	\end{center}
	where a, t, 2$\lambda_v$, and 2$\lambda_c$ are the lattice constant, hopping integral, spin splittings at the valence band top and conduction band minimum, respectively. Here, $\hat{\sigma}_z$ are the Pauli matrices for the basis functions, while $\hat{s}_z$ are the spin Pauli matrices. $\Delta$ is the direct energy gap between the VBM and CBM of respective $MXY$ layers at the K point. The third and fourth terms in the Hamiltonian remove the degeneracy at the valence band and the conduction band, respectively. 
	
	We fit this Hamiltonian with our DFT bands at K (Fig.~\ref{fig:fit_near_K}(a-c)) and extract the parameters for each layer separately. The extracted values of the parameters are in excellent agreement with the DFT parameters, as listed in Table \ref{table:fitted_K_bands}. The spin-splittings at the top of valence bands ($\lambda_v$) are larger than the splittings at the conduction band minima, which is known for a while for normal TMDs \cite{xiao2012coupled,kosmider2013large}. Also, the splittings for W$XY$ layers in the valence band are almost double of that for the Mo$XY$ layers, which is related to the fact that the SOC is larger for the heavier atoms. It is important to note that MoSTe/WSTe and MoSeTe/WSeTe HSs are not direct band gap semiconductors. However, for the purpose of fitting, we have considered the conduction bands in the DFT band structures, which represent the D$_1$ type gap (direct band gap) at the K point. The extracted parameters, which are in very good agreement with the DFT bands, show that our band structure near the K point can be safely described by the $k\cdot p$ Hamiltonian defined by Xiao \textit{et al.} \cite{xiao2012coupled}.

	\subsection{\label{sec:levelEEF} Effects of External Electric Field}
	
	\begin{figure*}[!htb]
		\centering
		\begin{tabular}{lc}
			\includegraphics[scale=.6]{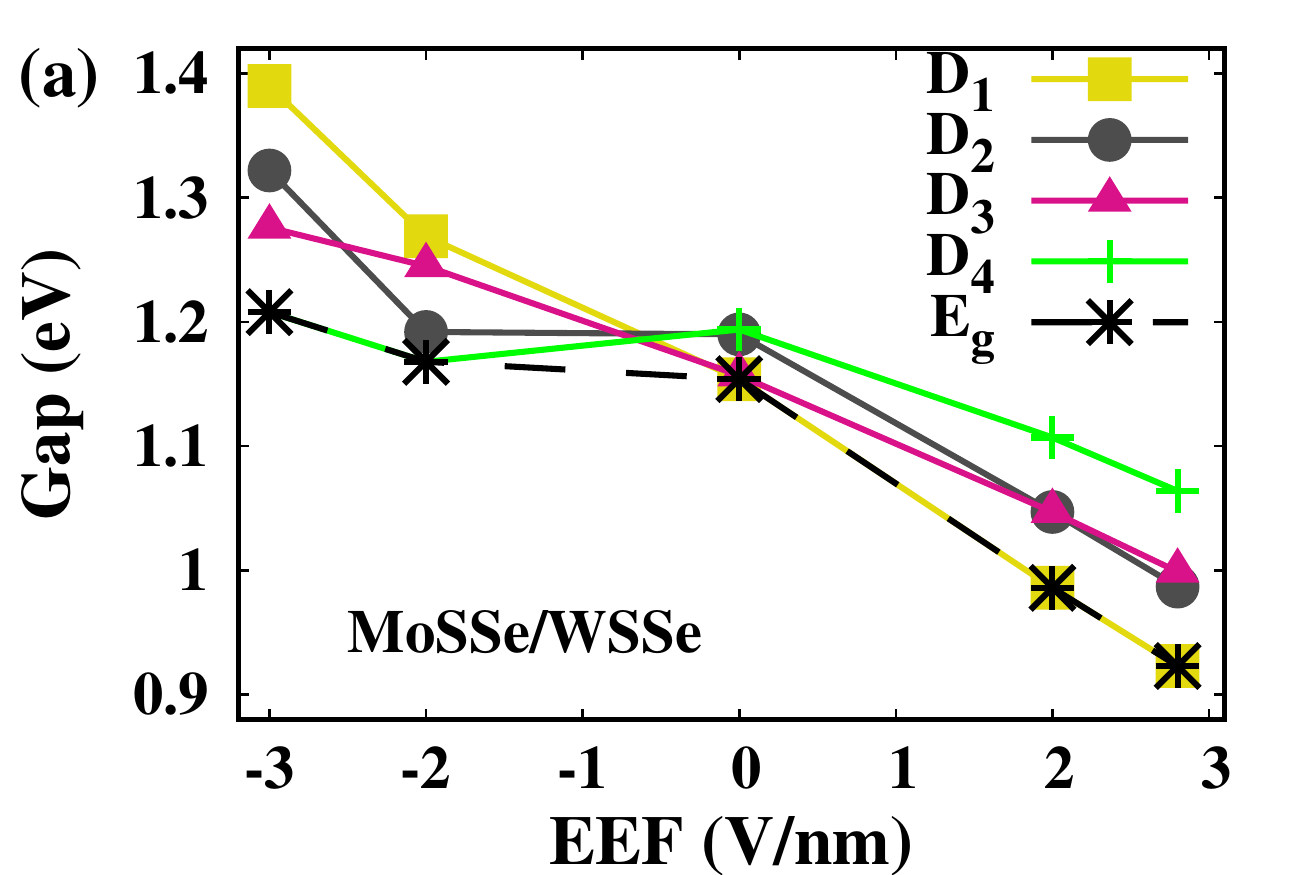}&
			\includegraphics[scale=.6]{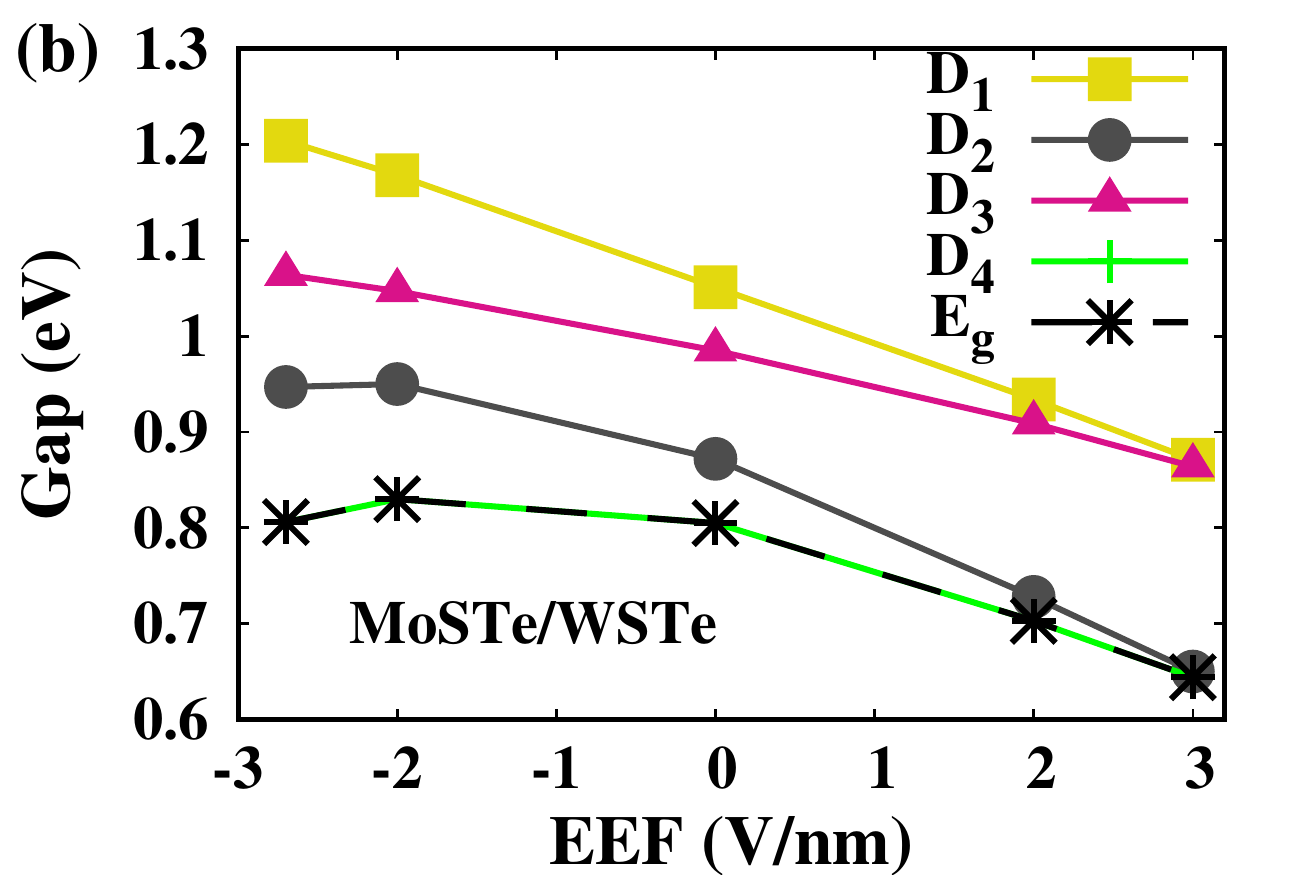}\\
			
			\includegraphics[scale=.6]{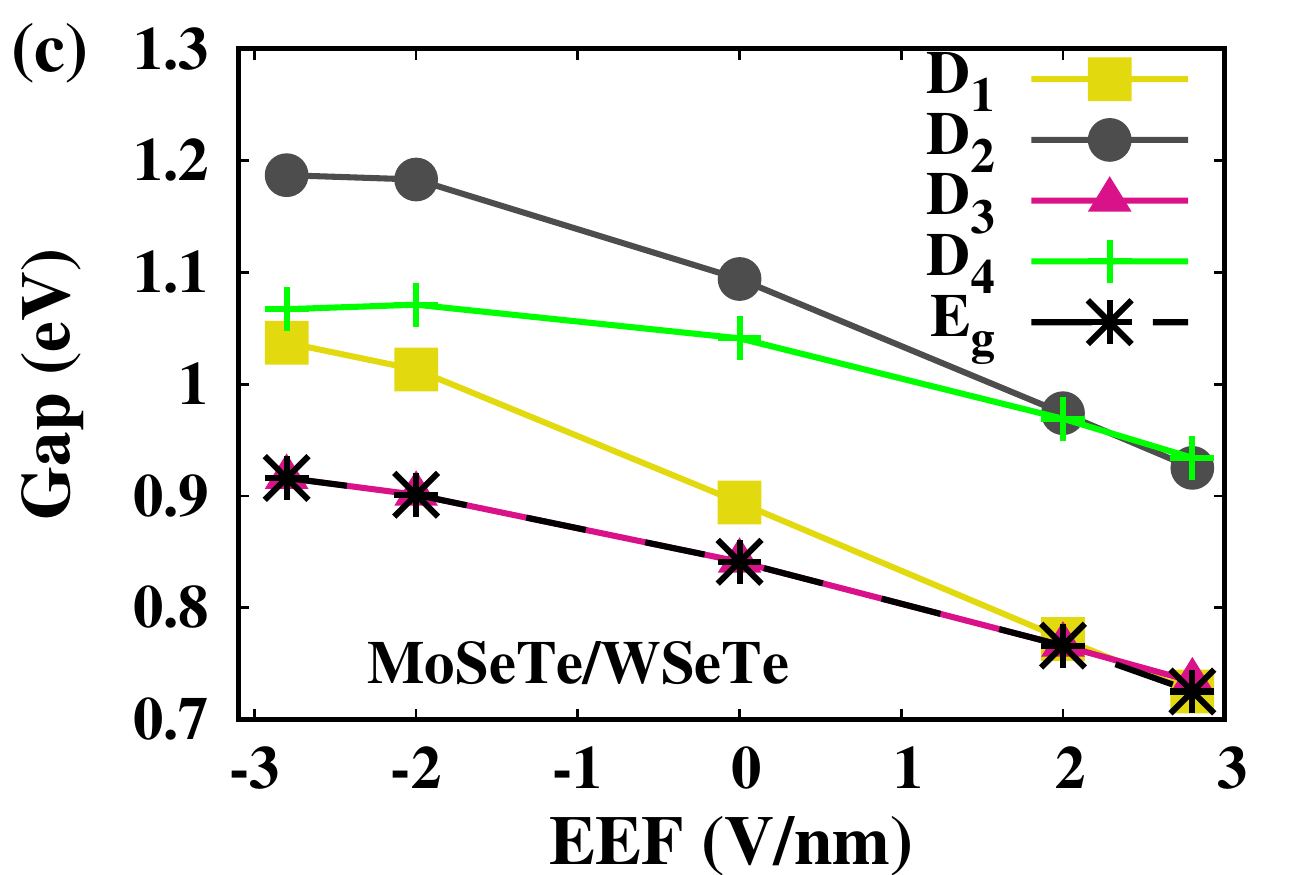}&
			\includegraphics[scale=.6]{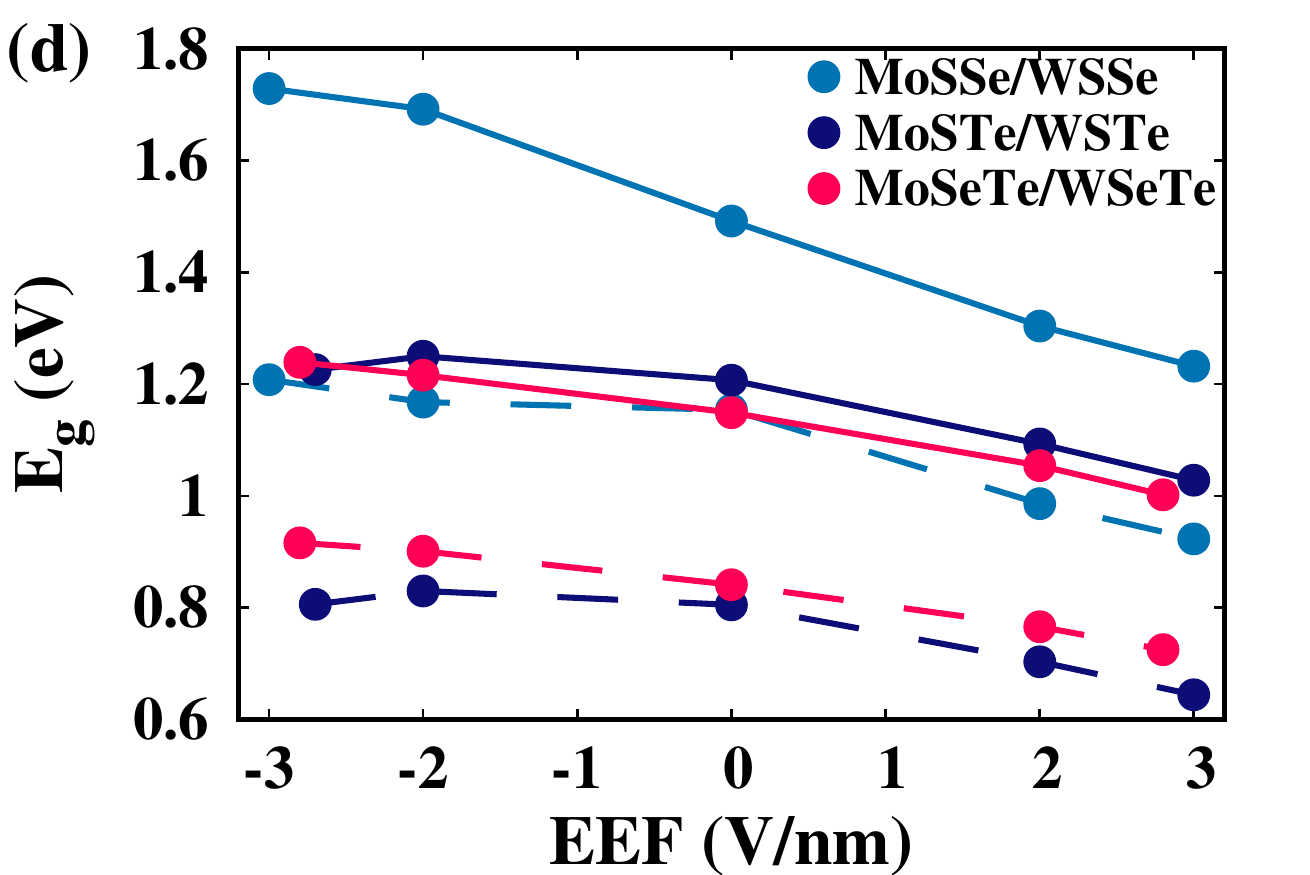}\\
		\end{tabular}
		\caption{(a-c) Variation of gaps with EEF for MoSSe/WSSe, MoSTe/WSTe and MoSeTe/WSeTe HSs, respectively. D$_1$ is the direct gap at the K point, D$_2$ is the gap between VBM at $\Gamma$ and CBM at the K point, and D$_3$(D$_4$) is the gap between VBM at K ($\Gamma$) and CBM in the $\Gamma-$K direction, as shown in Fig.~\ref{fig:efield_direction}(d). E$_g$ is the true energy band-gap indicated by black dashed lines. (d) System band gaps ($E_g$) with HSE06 (solid lines) and GGA- PBE (dashed lines) functionals.}  \label{fig:gaps_variation}
	\end{figure*}	
	
	As mentioned in the previous section, with the effect of mirror asymmetry in J-HSs, a vertical dipole moment is formed, which causes an intrinsic out-of-plane electric field, leading to the novel properties that are absent in conventional \textit{M}$X_2$ ML TMDs. It is possible to enhance or reduce the strength of the intrinsic electric field by applying an external electric field (EEF), either parallel or anti-parallel to it. The net electric field can then modify the Rashba splitting as well as the orbital overlaps resulting in the variation of band gaps. There are many studies dedicated in this field, varying from bulk \cite{shanavas2014electric} to 2D materials \cite{hu2018intrinsic,liu2021tuning, yao2017manipulation, liu2020manipulation} and to the HSs \cite{sharma2014strain}. Variation of the band-gaps and the Rashba parameter, $\alpha_R$, with external electric field in the case of J-MLs has been studied by Tao Hu et al. \cite{hu2018intrinsic}. 
	
	\begin{figure*}
		\centering
		\textbf{\Large Effects of External Electric Field}\\
		\vspace{0.4cm}
		\begin{tabular}{lcc}
			\includegraphics[scale=.23]{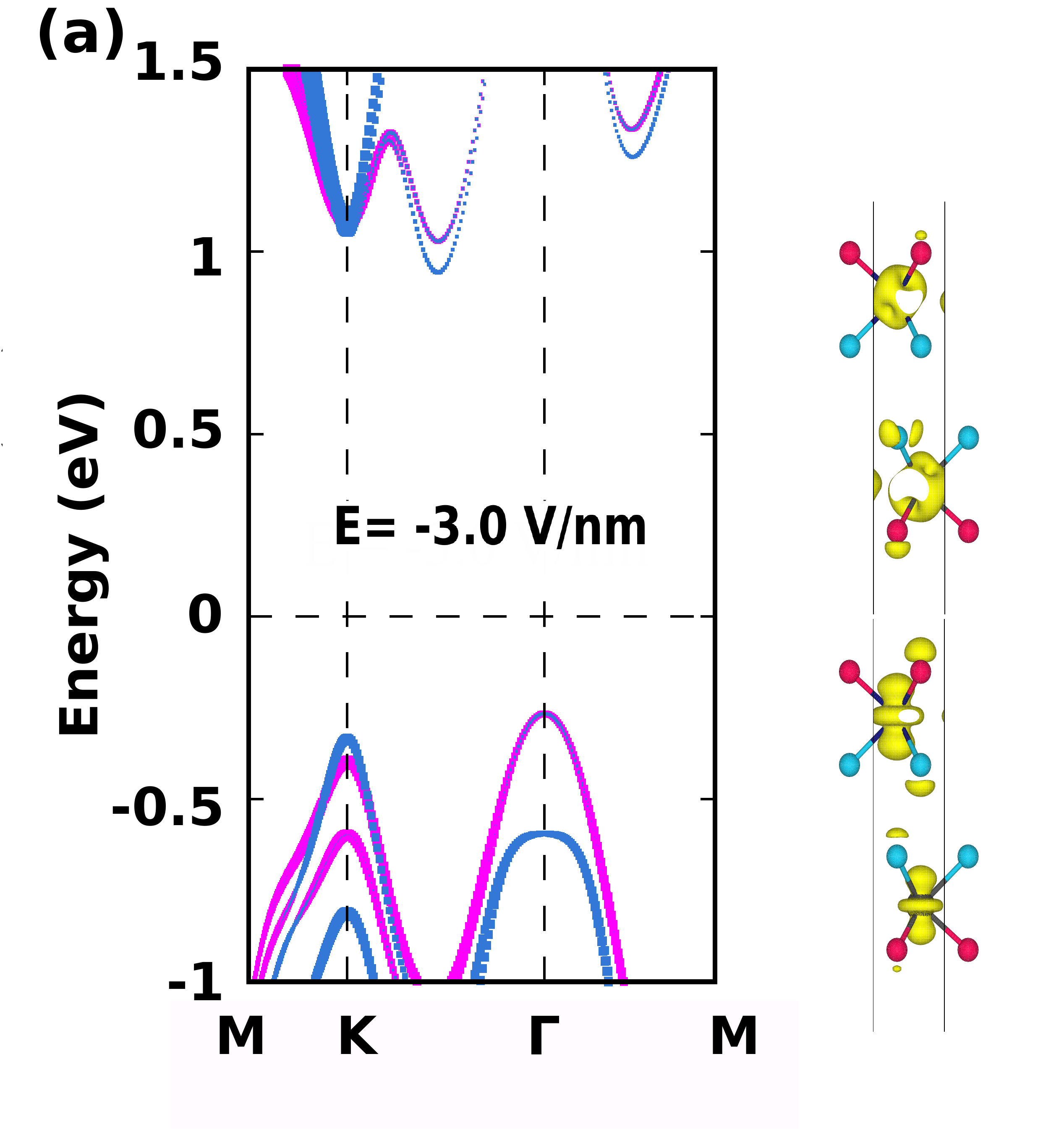}&
			\includegraphics[scale=.23]{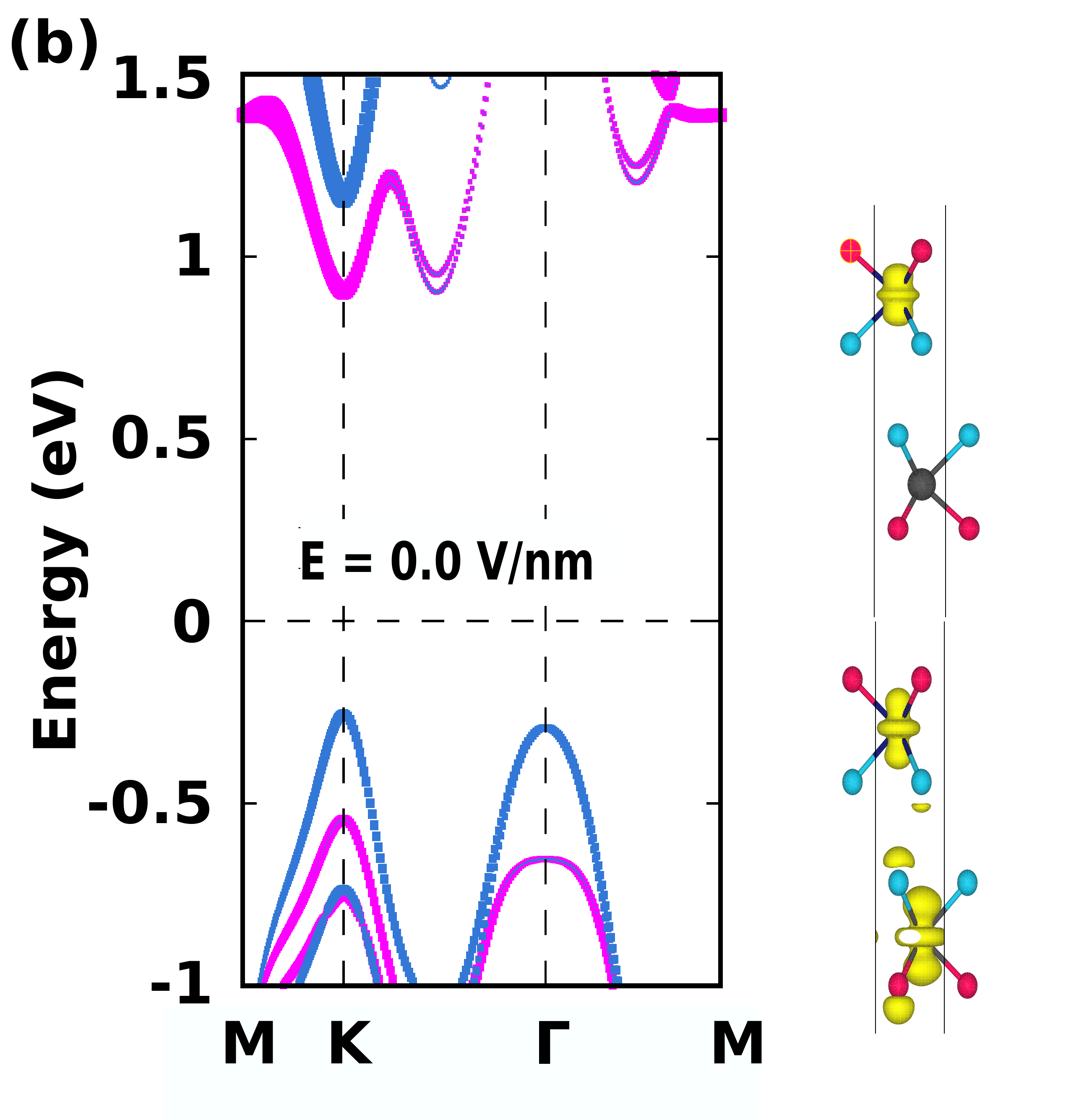}&
			\includegraphics[scale=.23]{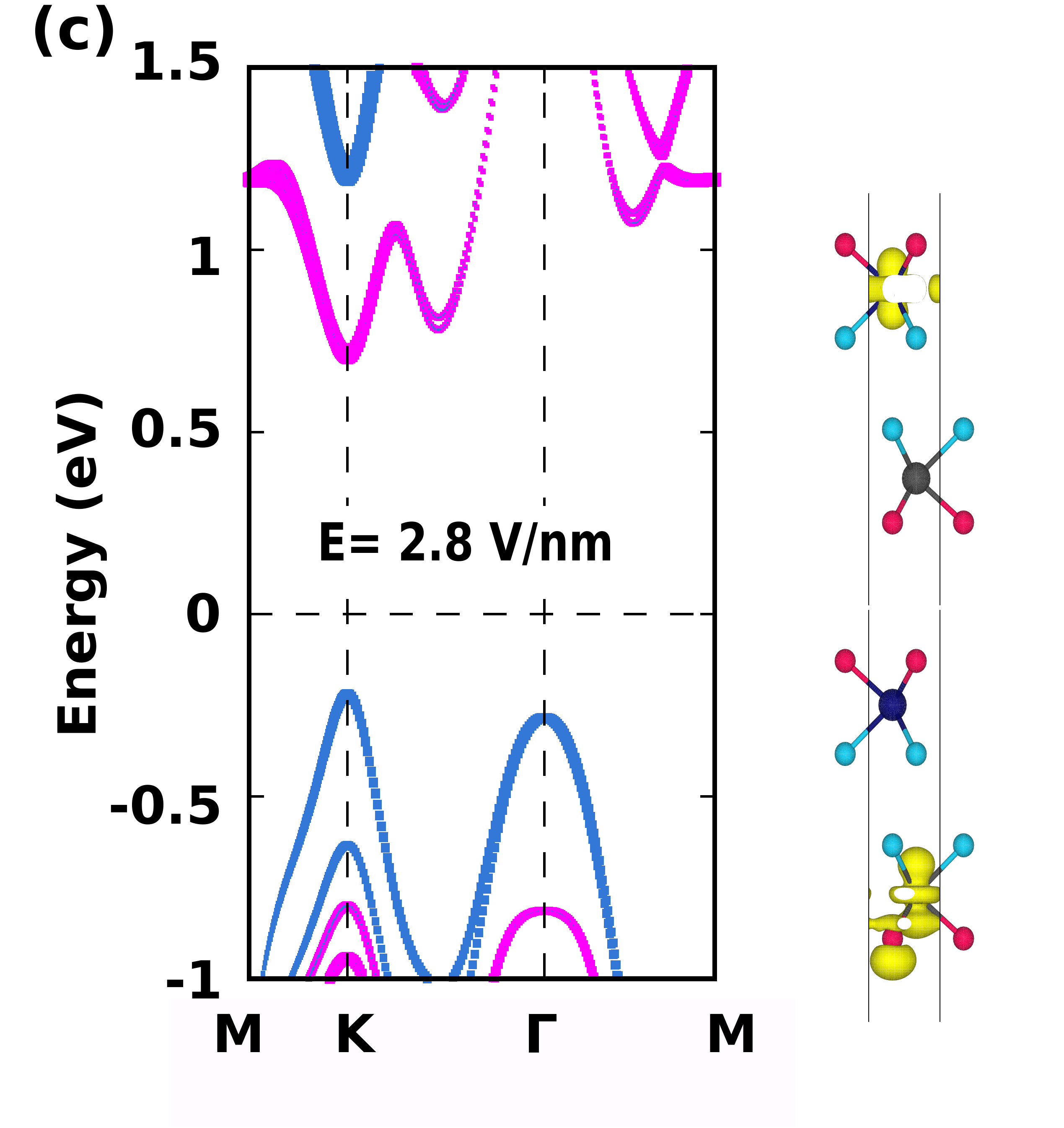}\\
			
			\includegraphics[scale=.23]{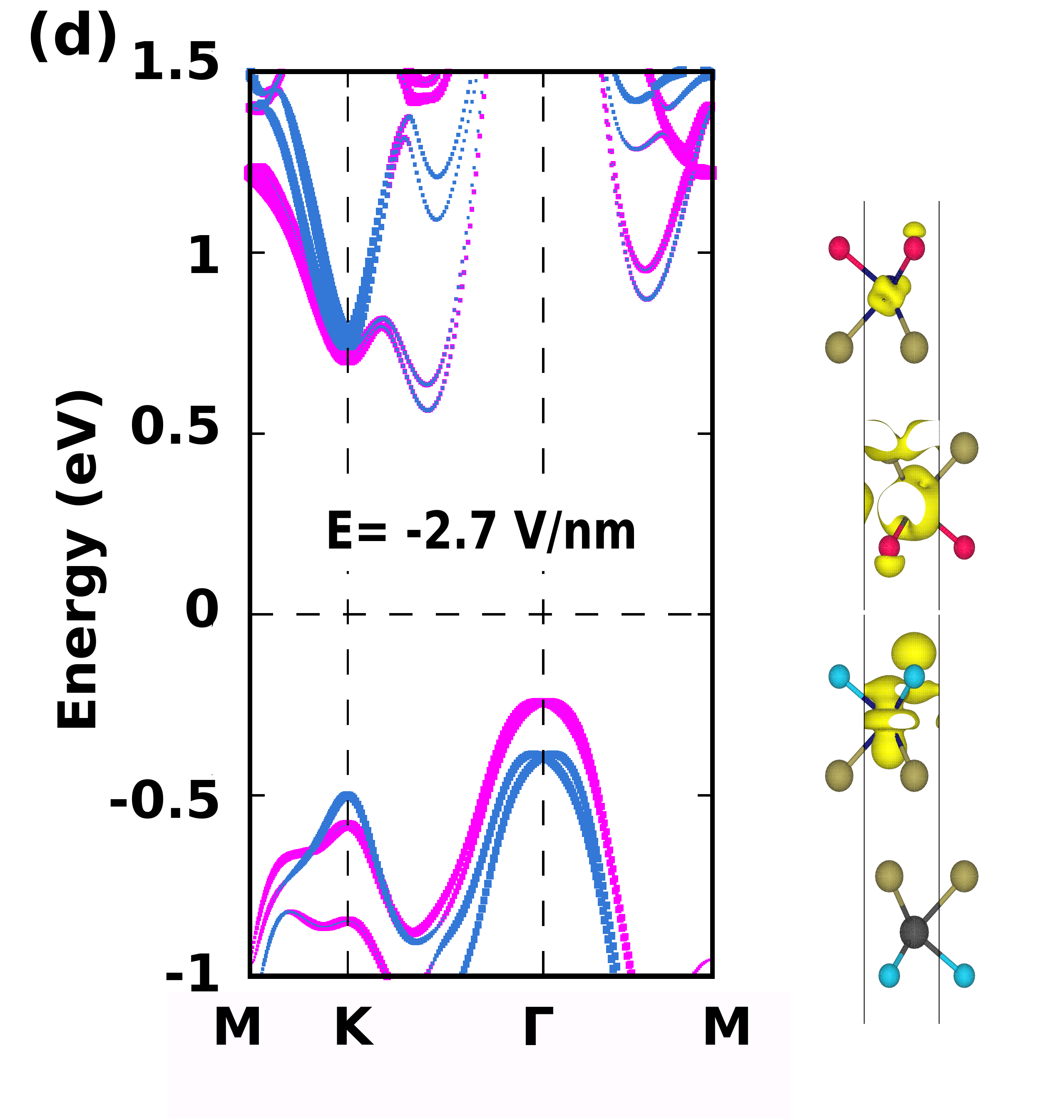}&
			\includegraphics[scale=.23]{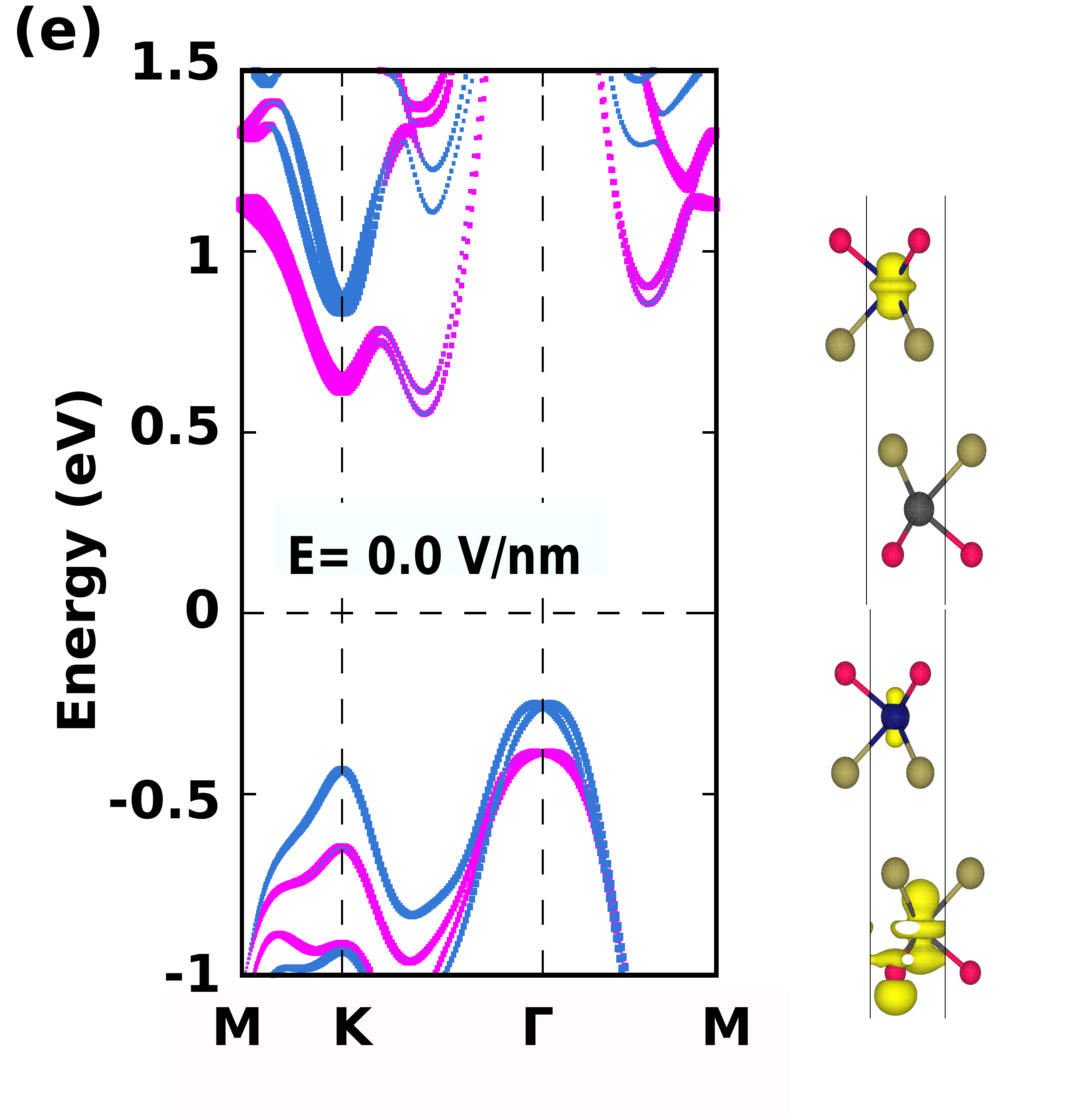}&
			\includegraphics[scale=.23]{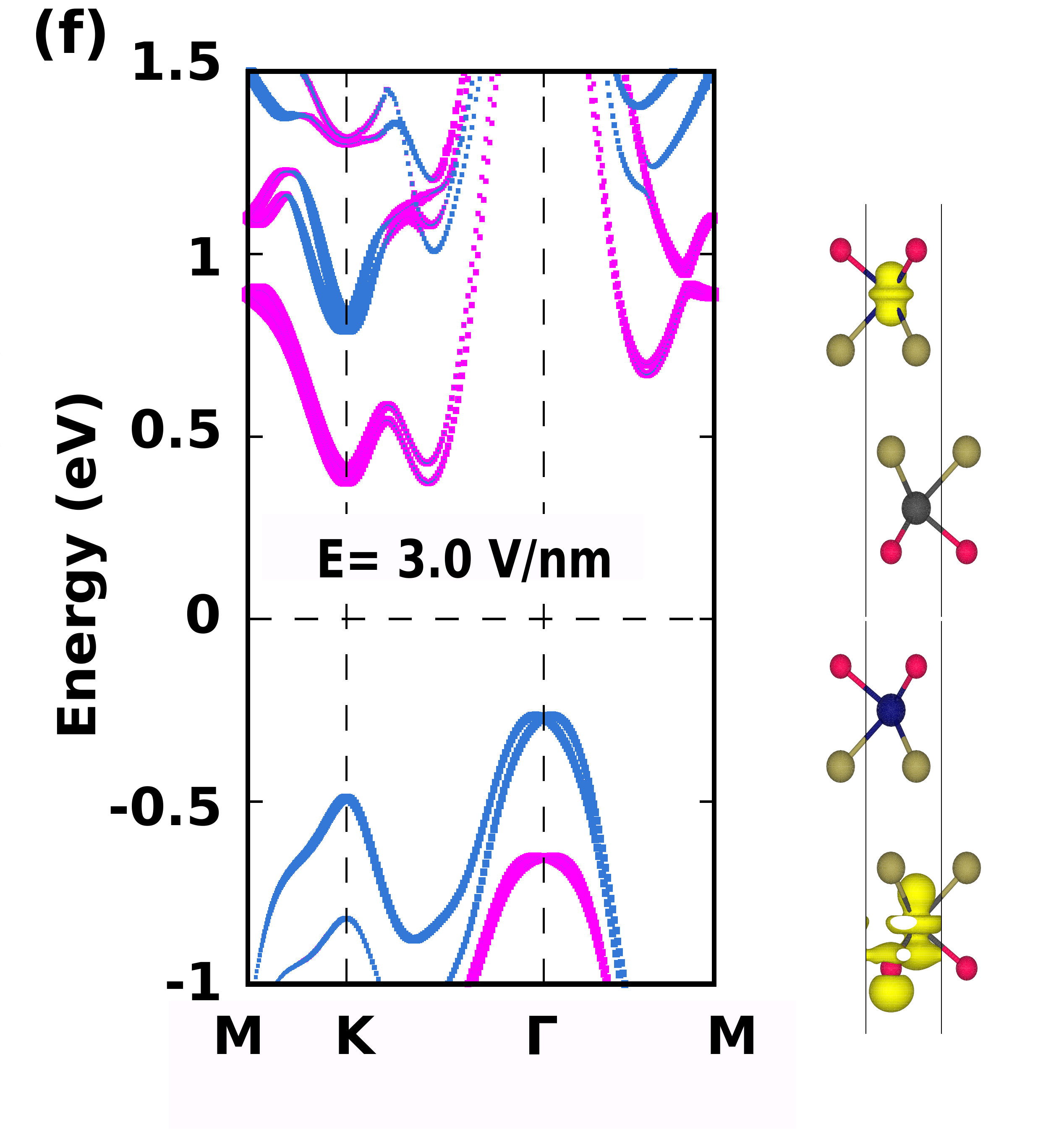}\\
			
			\includegraphics[scale=.23]{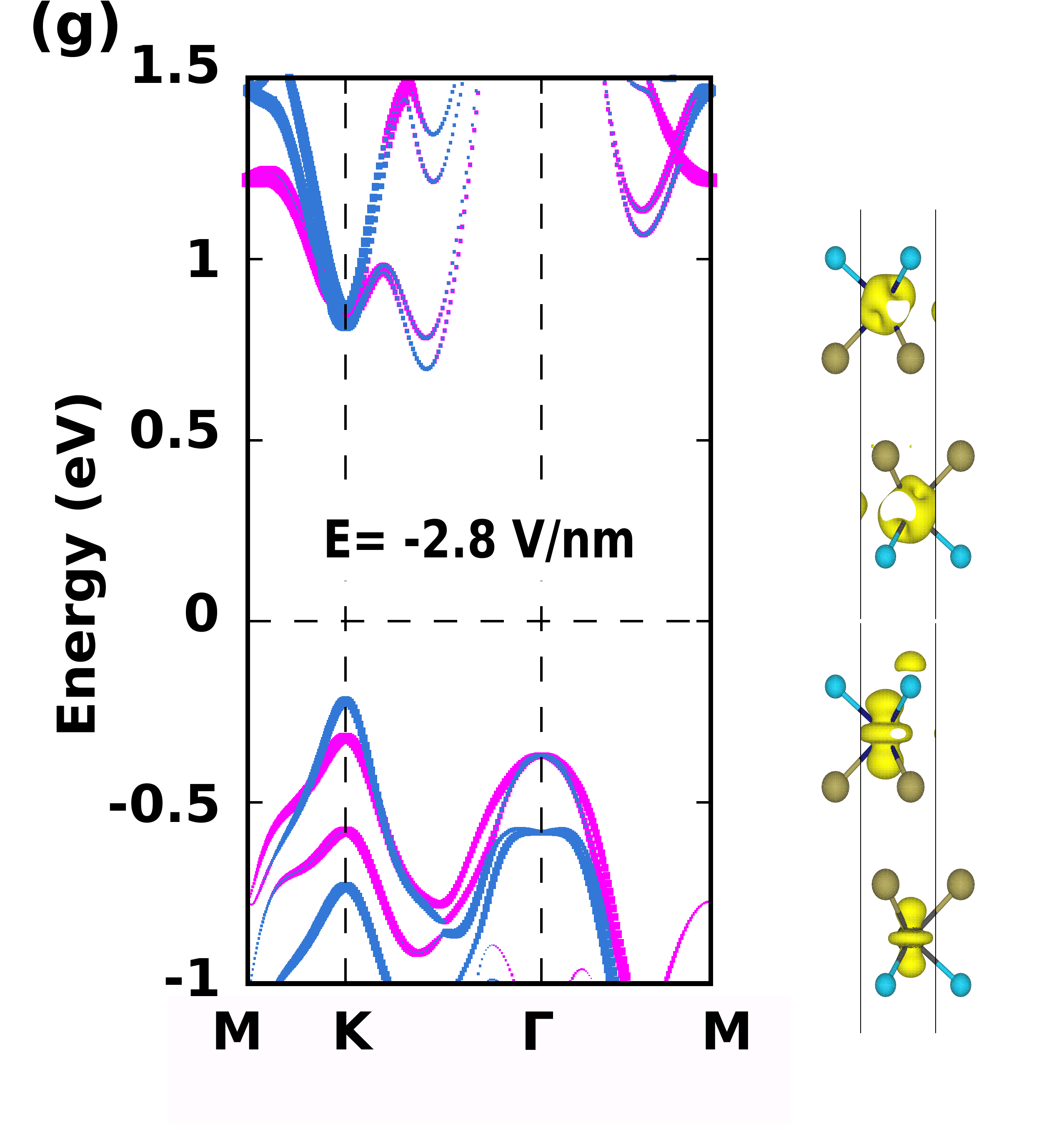}&
			\includegraphics[scale=.23]{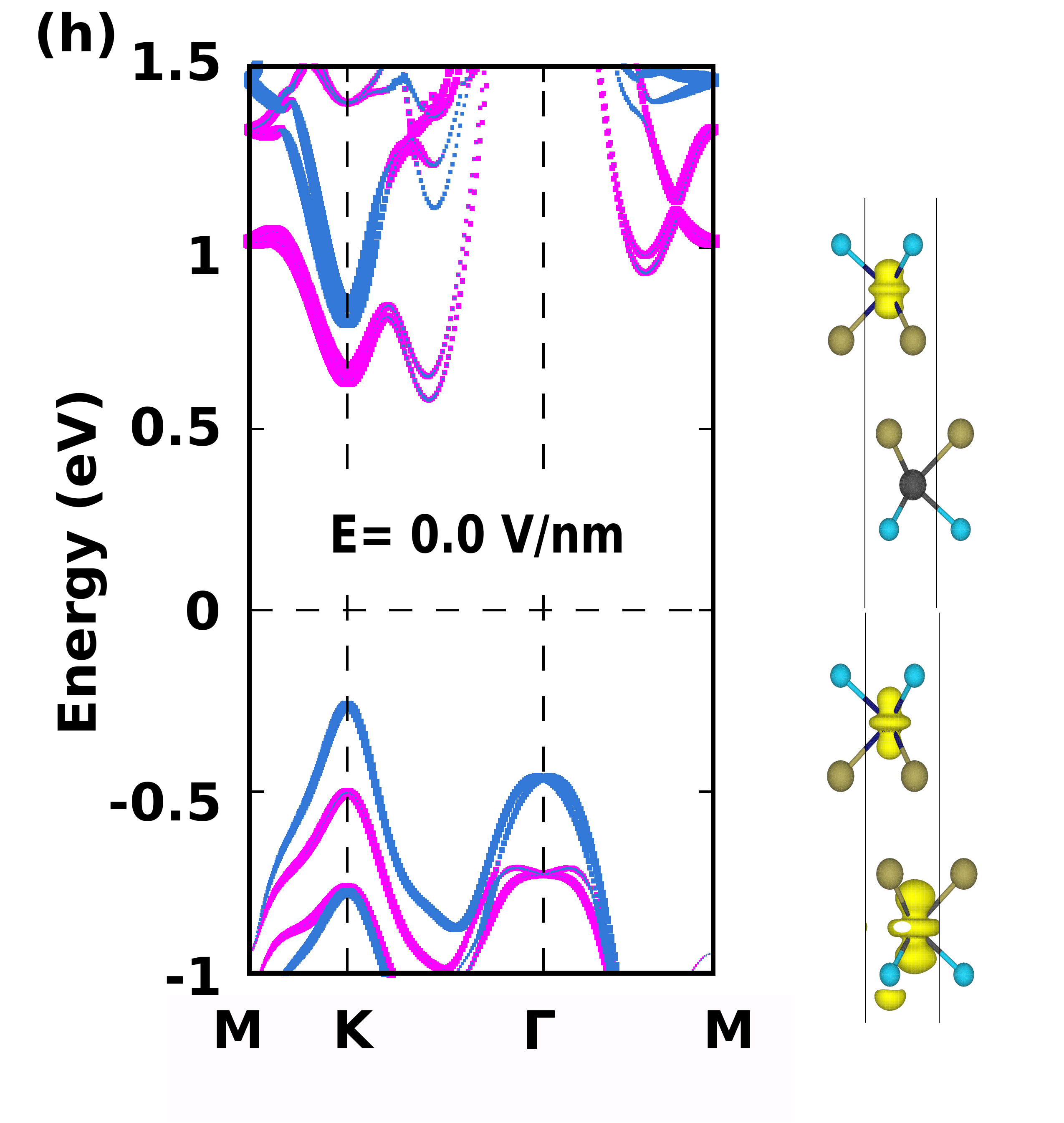}&
			\includegraphics[scale=.23]{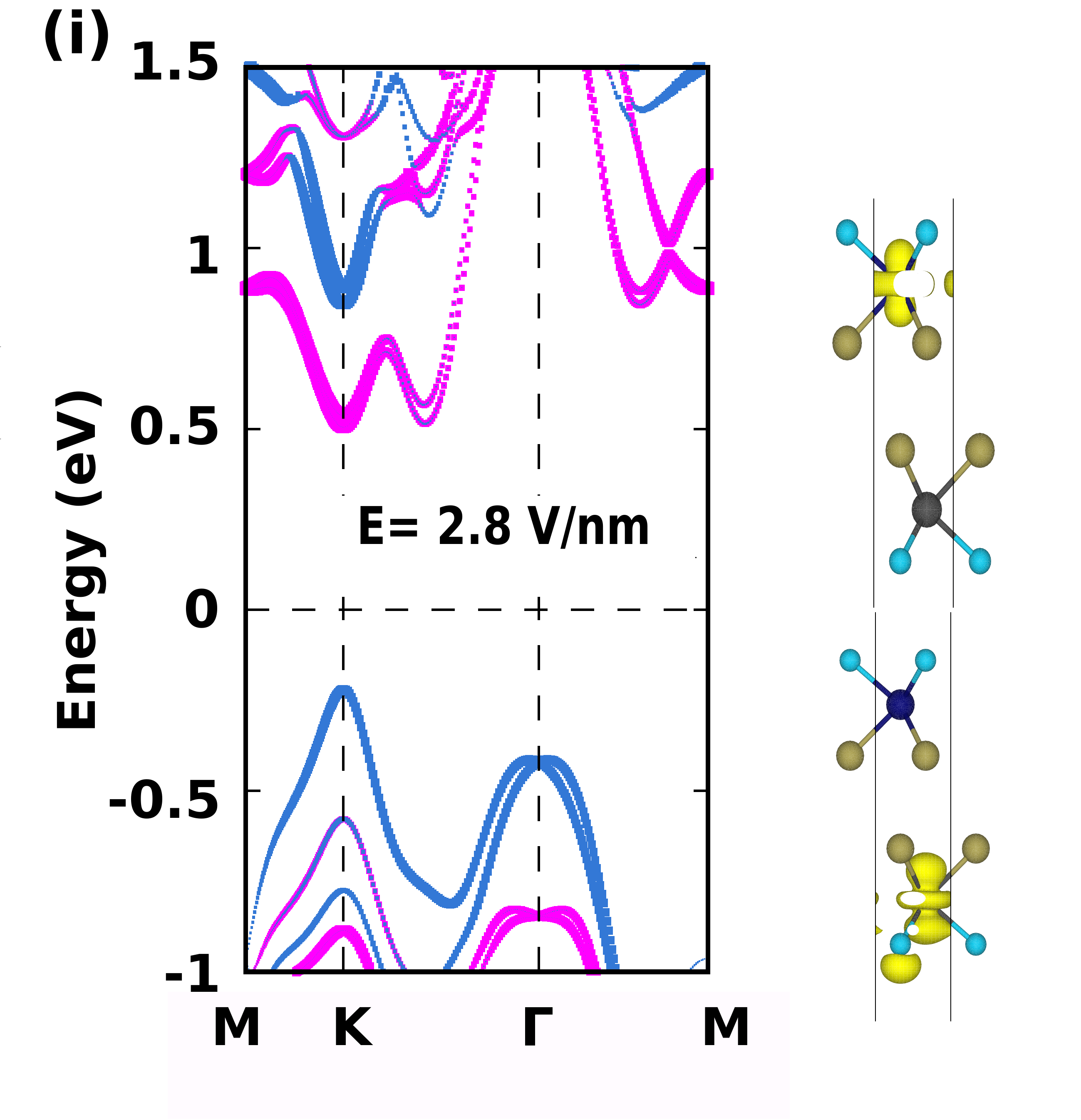}\\
		\end{tabular}
		\caption{Layer projected band structures and partial charge density plots (band-decomposed charge density plots) under different external electric field strengths for MoSSe/WSSe (upper row), MoSTe/WSTe (middle row) and MoSeTe/WSeTe (bottom row) HSs. The magenta and blue bands denote Mo$XY$ and W$XY$ layers, respectively. For all the cases, we plot the charge densities at the $\Gamma$-point for the valence band (bottom unit-cell), while for the conduction band (upper unit-cell), we plot charge densities at the K point under zero and positive EEF, and at CBM, which is located in between $\Gamma$ and K points, under reverse EEF. Isosurface value is set at 0.002 e/\AA$^3$ for charge density plots. Depending on the relative orientations of EEF and E$_\text{int}$, charges (yellow) accumulate or deplete from the W$XY$ bottom layer. Moreover, the layer-projected band structures indicate that the type-II band alignment in the pristine HSs remains robust under a positive electric field, whereas the nature of the band alignment changes by the application of a reverse electric field. }  
		\label{fig:partial_CD_efield}
	\end{figure*}	
	
	Here we first study the evolution of different types of band-gaps and their tuning with EEF. Since we have AB-stacked HSs with a $Y-Y$ interface, where the electronegativity of the $Y$ atom is less than that of $X$, the intrinsic electric field is directed from $Y$ to $X$. It is the opposite in both layers, which cancels the net polarization induced by the chalcogen atoms. However, the transition metal atoms in both layers have different electronegativities. As a result, there is a small intrinsic electric field induced from Mo to W atom (see Fig.~\ref{fig:efield_direction}(b)). This direction of the induced electric field can be verified from the local electrostatic potential profile in Fig.~\ref{fig:efield_direction}(c), in which a potential drop is clearly visible. 
	
	The troughs in the local potential plot for W and Mo are not at the same height, and it indicates that there is an intrinsic electric field in the direction from Mo to W atom. Now, if EEF is applied in the direction of the intrinsic field, it superimposes and shows us the modified effects in band-gaps and $\alpha_R$ as well. We apply the positive EEF in the direction from Mo to W as shown in Fig.~\ref{fig:efield_direction}(b). As discussed above, among all three HSs, MoSSe/WSSe has a direct band-gap at the K point, while the other two HSs have indirect band-gaps. The nature of the gap varies with EEF, and in most cases, it is indirect. The different types of gaps with PBE XC functional are illustrated in Fig.~\ref{fig:efield_direction}(d). The MoSSe/WSSe HS shows a direct band-gap of D$_1$ type in the positive EEF case while it is of D$_4$ type indirect band-gap in the reverse EEF. On the other hand, the nature of the band-gap in the case of  MoSTe/WSTe does not change with EEF, and it remains the D$_4$ type throughout. At 0.28 V/\AA, in the MoSeTe/WSeTe HS, there is a transition from D$_3$ type band-gap into a direct band-gap (D$_1$) semiconductor. The modulation of different kinds of gaps with EEF, along with the total system band-gap, is demonstrated in Fig.~\ref{fig:gaps_variation}. We can clearly see the band-gaps reduce with the strength of EEF, and the reason one can think of is the charge transfer from the upper (Mo$XY$) layer to the bottom (W$XY$) layer. Charges try to align them in the direction of an electric field. Thus as we increase EEF in the direction of the intrinsic electric field, there is more charge accumulation on the W$XY$ layer of the HSs. This charge transfer increases the interaction between both layers leading to the band-gap reduction. The HSE06 XC functional does not change the qualitative nature of the bands for MoSTe/WSTe and MoSeTe/WSeTe HSs except for an increment in the band-gap values, as shown in Fig. 2. They remain indirect band-gap semiconductors of D$_4$ (MoSTe/WSTe) and D$_3$ (MoSeTe/WSeTe) type for all values of EEF. On the other hand, MoSSe/WSSe HS remains a direct band-gap semiconductor of D$_1$ type in the reverse EEF up to -0.2 V/\AA\, and then converts to D$_3$ type (see Fig. S2(a) \cite{patel2022rashba}).
	
	We can also verify from the charge density plots that without the application of any EEF, there is more charge localization on the W$XY$ layer; this further clarifies the direction of the intrinsic electric field, which is downward. The layer projected band structures in Fig.~\ref{fig:partial_CD_efield} are in accordance with the charge density plots, and it can be confirmed that in the reverse EEF, the VBM at the $\Gamma$-point is occupied by the Mo$XY$ layer, unlike the case of zero and positive EEF where it is occupied by W$XY$ layer. The CBM at K point is occupied by the Mo$XY$ layer in case of positive EEF, while there is a mix contribution from both the layers at CBM, which is now located in between the $\Gamma$ and K points, in the reverse EEF strength. This indicates that the type-II band alignment in the pristine HSs remains robust under a positive electric field, whereas the nature of the band alignment changes by the application of a reverse electric field.
	
\begin{figure}
		\centering
		\includegraphics[scale=0.7]{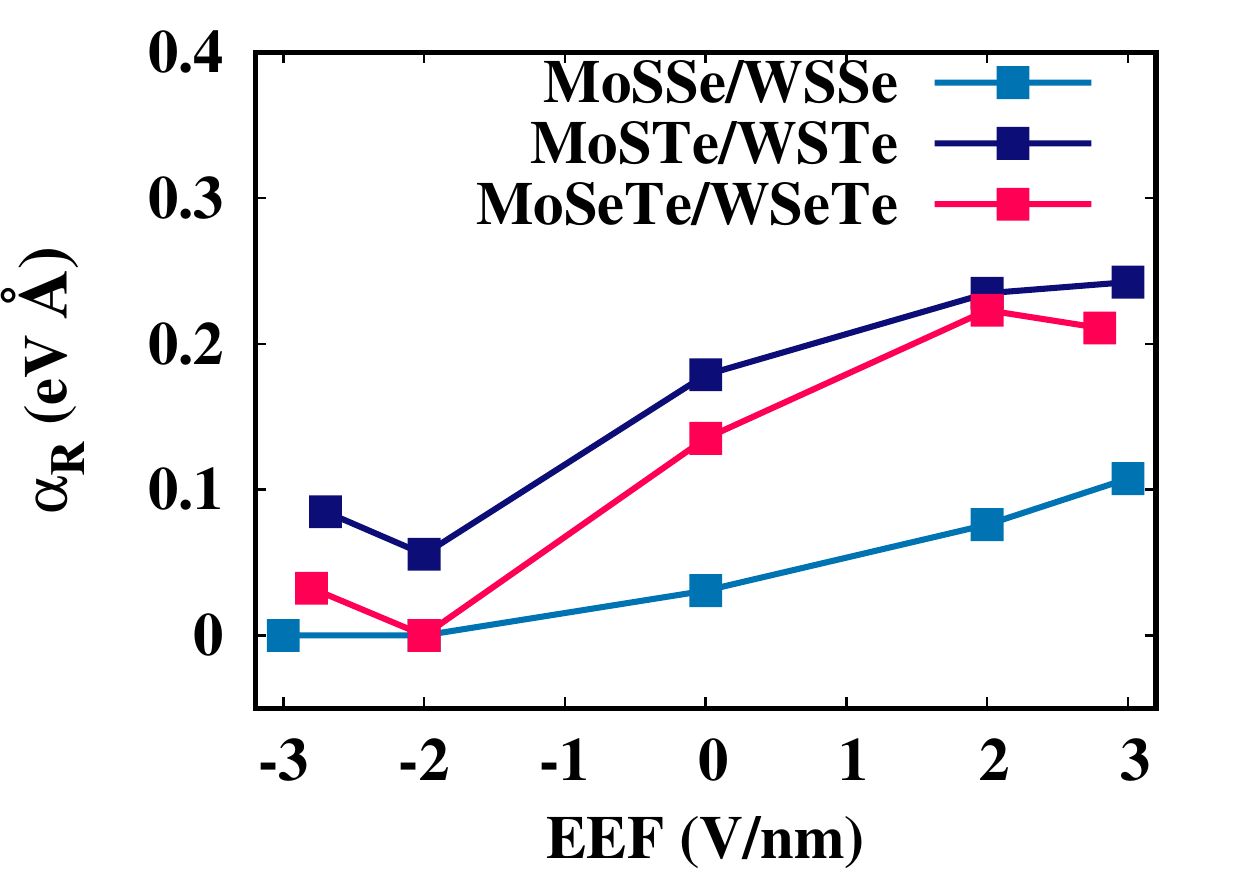}
		\caption{Variation of the Rashba parameter $\alpha_R$ with EEF. As seen, $\alpha_R$ increases with increasing EEF. In the AB-stacked HSs with $Y-Y$ interface, the small intrinsic electric field superimposes with positive EEF and amplifies $\alpha_R$.}
		\label{fig:rashba_HS_efield}
	\end{figure}

	The effect of EEF on $\alpha_R$ is summarized in Fig.~\ref{fig:rashba_HS_efield}. It can be noted that $\alpha_R$ is zero under reverse EEF for the chosen stacking of MoSSe/WSSe HS, and it increases with increasing EEF. Naturally, we also expect that $\alpha_R$ should increase with the strength of the electric field. In AB-stacking with a $Y-Y$ interface, the intrinsic electric field is small, which superimposes with EEF and amplifies $\alpha_R$. 
	In the case of J-MLs, the Rashba parameter increases with the atomic number of $Y$ atom \cite{hu2018intrinsic}. The value of $\alpha_R$ for \textit{M}SeTe (\textit{M} = Mo or W) is the highest, while for \textit{M}SSe, it is the lowest since the SOC strength is higher for atoms with larger atomic numbers \cite{cheng2013spin}. Therefore, it is expected that the MoSeTe/WSeTe HS would possess the highest $\alpha_R$.
	Now, let us compare the difference of electronegativities ($\Delta_{en}$) between different combinations of $X$ and $Y$. We find that Te-S has the largest $\Delta_{en}$, which results in the highest dipole moments due to chalcogen atoms in the opposite directions from the interface. This mechanism further results in the increment in the distance between the transition metals of both layers, which gives rise to an enhanced net dipole moment (net intrinsic electric field), and consequently highest $\alpha_R$ in AB-stacked MoSTe/WSTe HS.

	\subsection{\label{sec:levelstrain}Effects of strain}
	
	\begin{figure}[!htb]
		\centering
		\includegraphics[scale=.2]{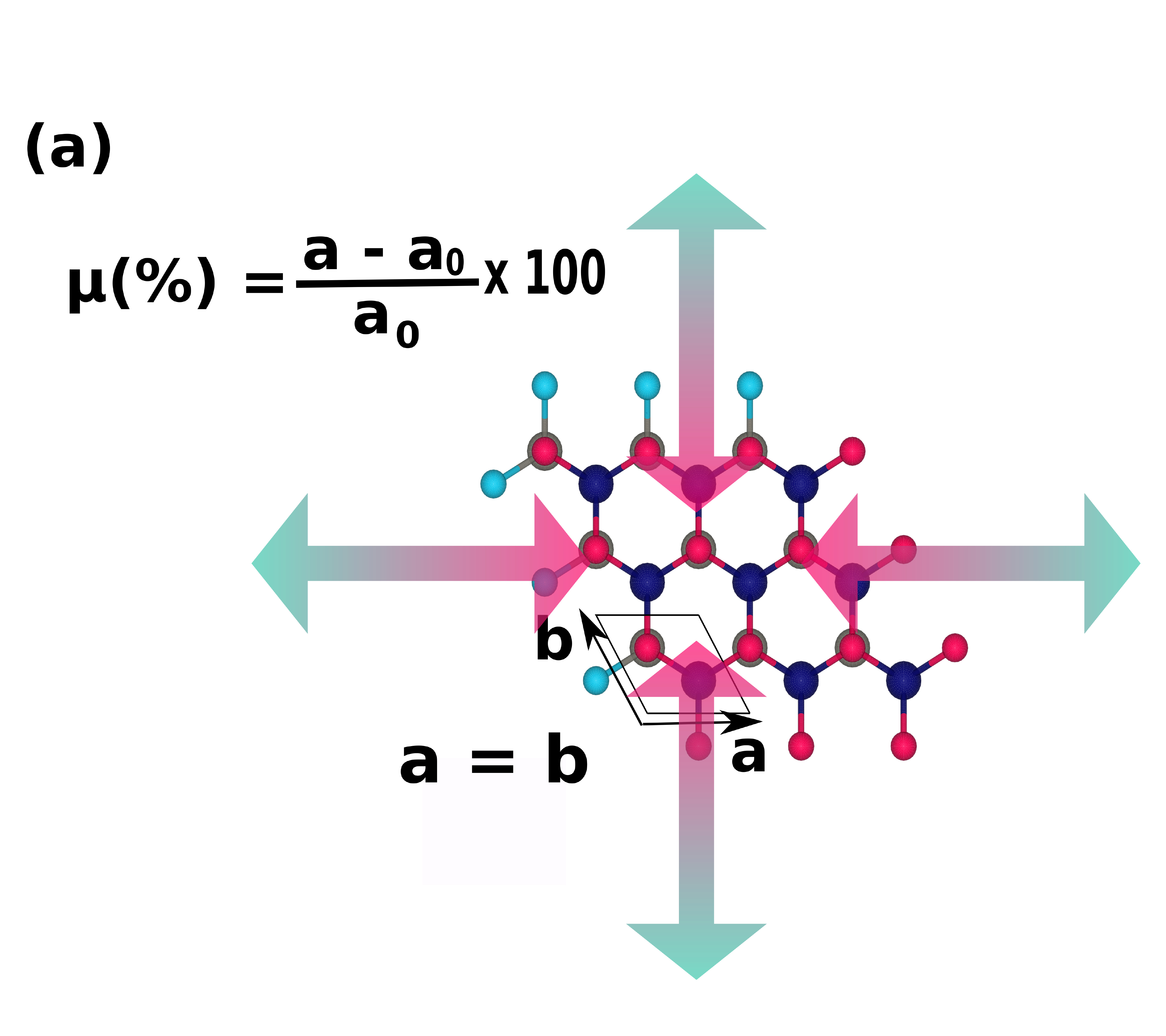}
		\includegraphics[scale=.68]{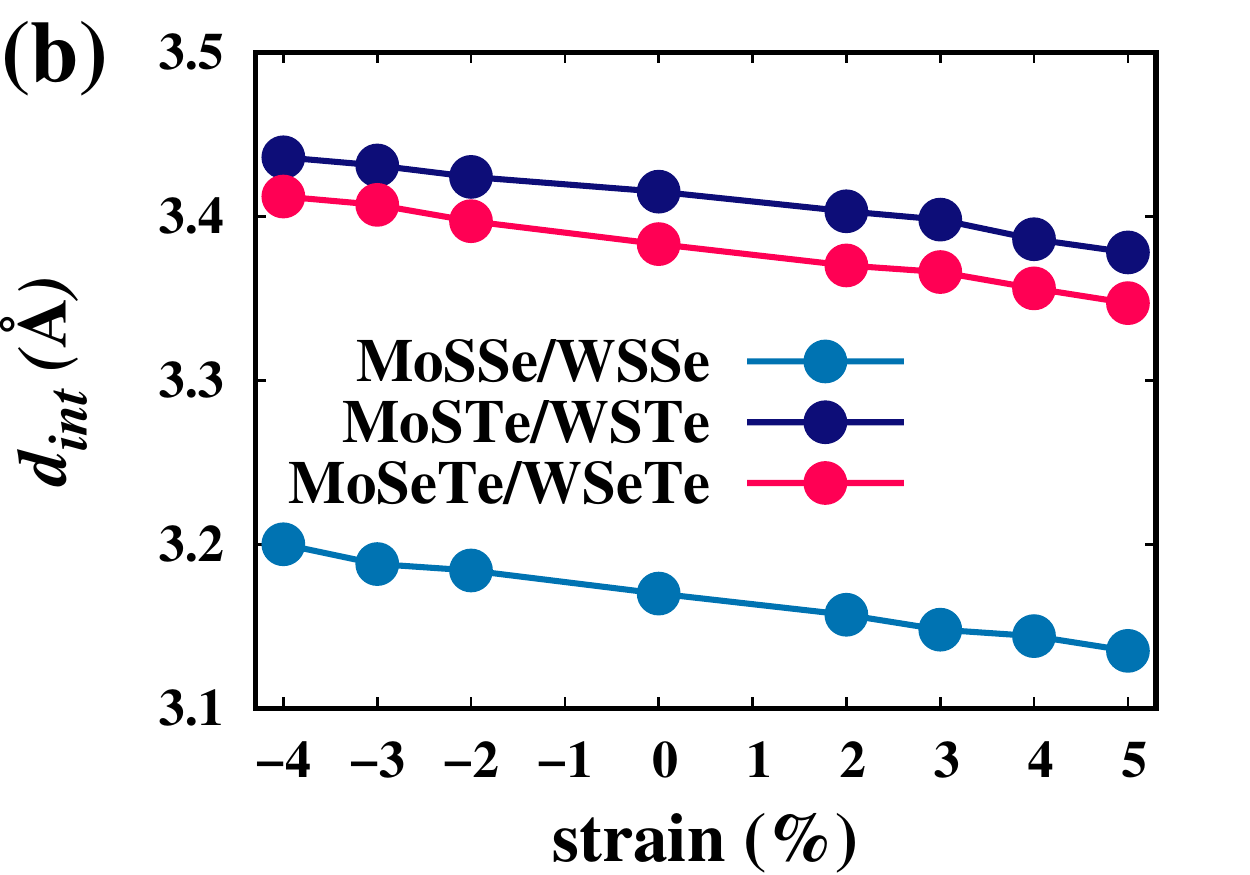}
		\caption{(a) Schematics for applying an in-plane biaxial strain in our AB-stacked HSs. The pink inward (green outward) arrow depicts the compressive (tensile) strain. a and b are the lattice parameters in the $x$ and $y$-directions, respectively, which are equally compressed and stretched when applying a biaxial strain. $\mu$ is the strength of the biaxial strain, a$_0$ and a are the in-plane lattice parameters of the strained and unstrained structures, respectively. (b) Variation of the distance between the chalcogen atoms at the interface ($d_{int}$) as a function of biaxial strain. Negative and positive values of the strain represent the compressive and the tensile strains, respectively. $d_{int}$ decreases (increases) linearly with tensile (compressive) strain.}
		\label{Fig:biaxial_scheme}
	\end{figure}
	
	\begin{figure}[!htb]
		\centering	
		\begin{tabular}{lc}
			\includegraphics[scale=0.33]{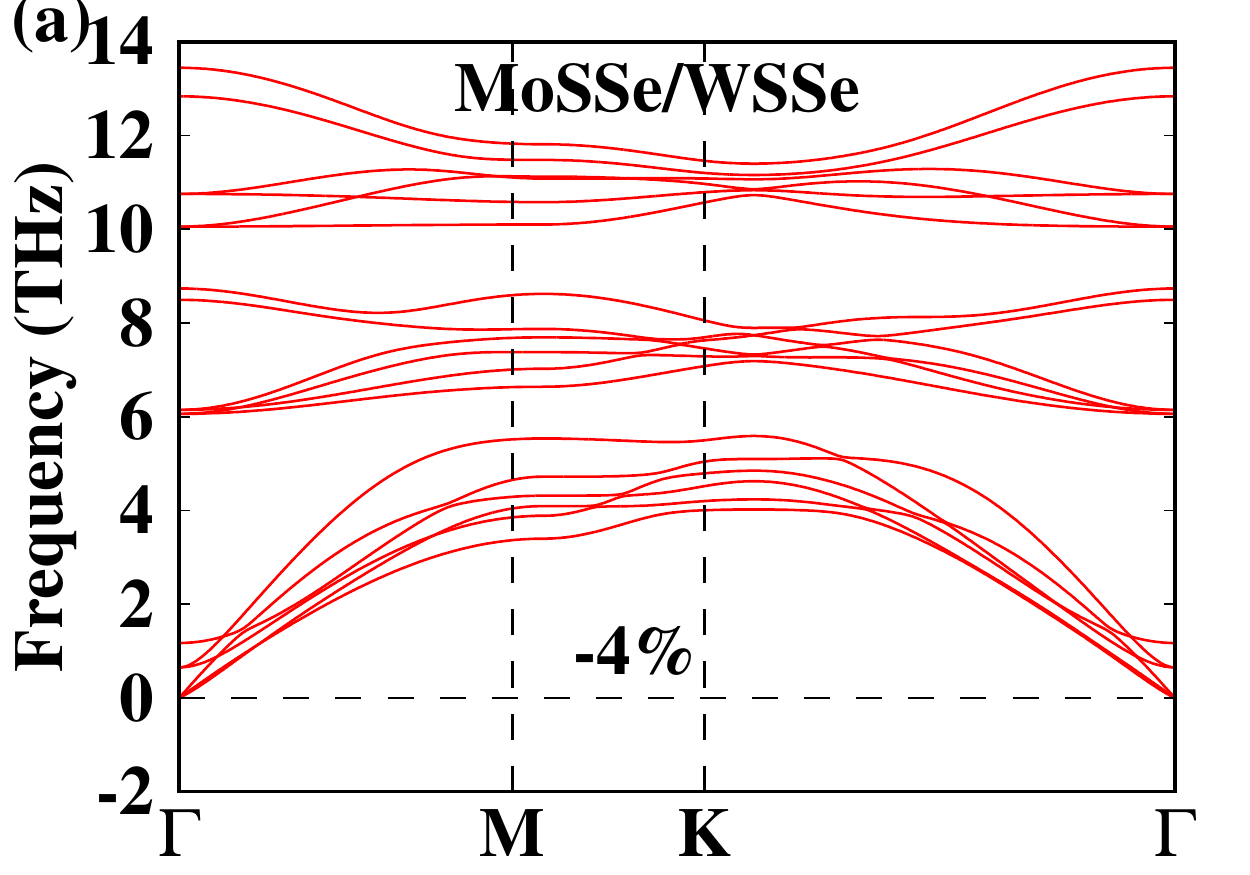}&
			\includegraphics[scale=0.33]{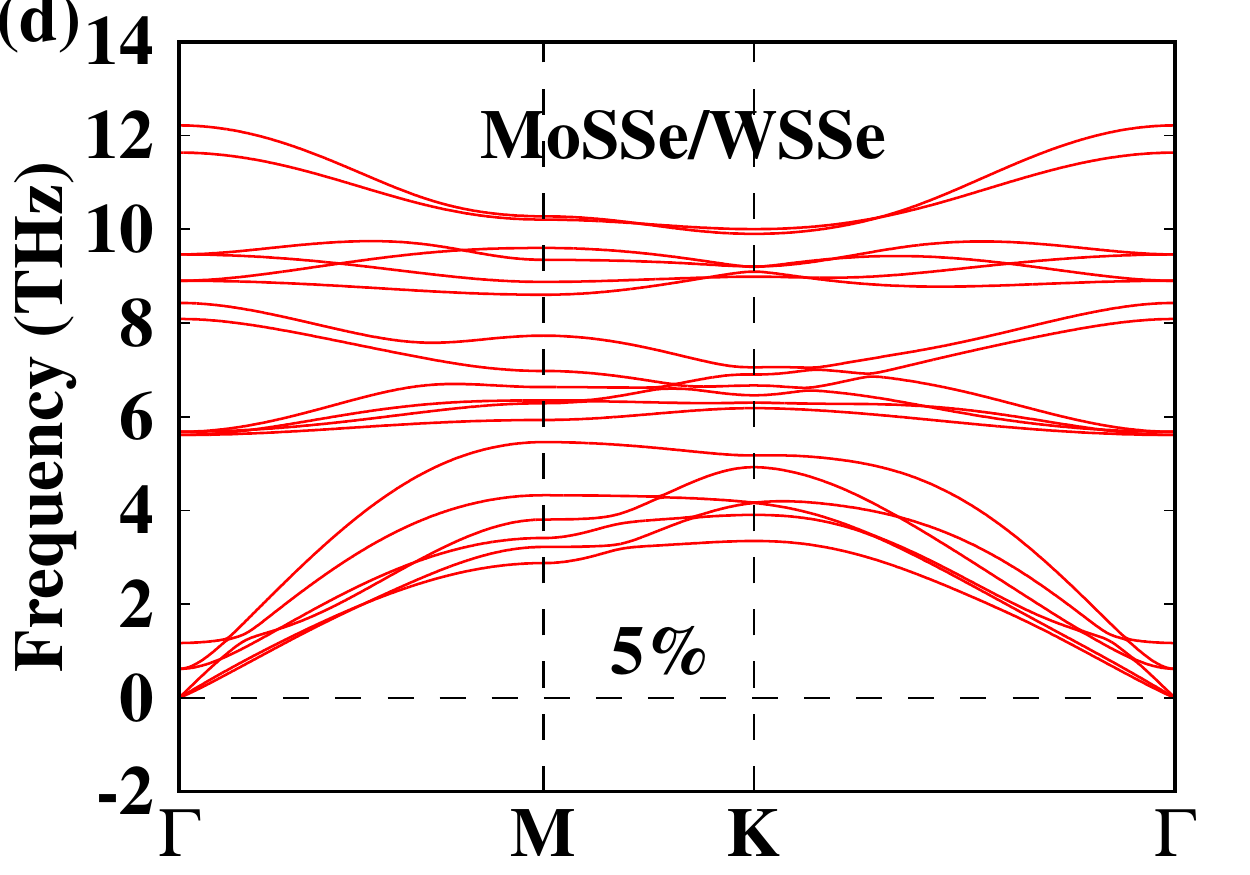}\\
			\includegraphics[scale=0.33]{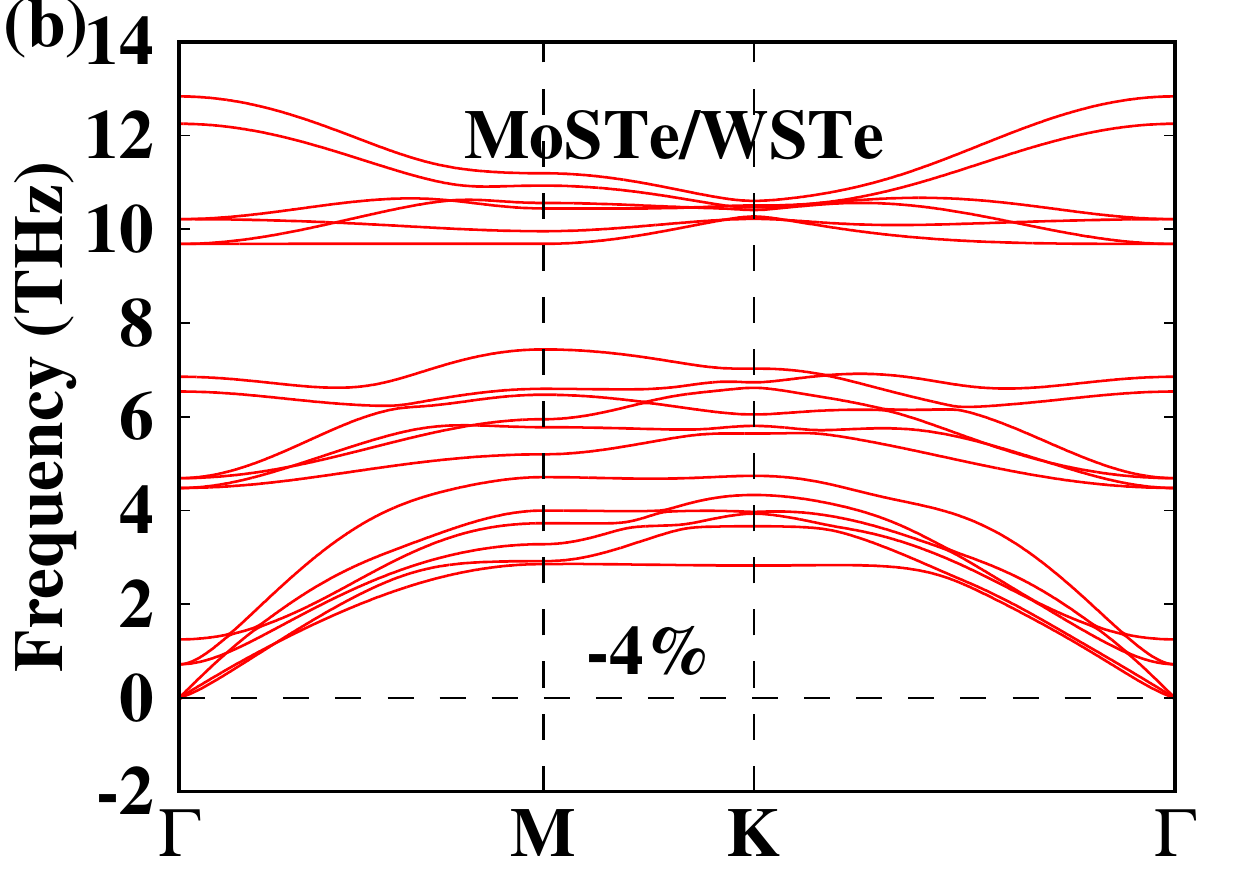}&
			\includegraphics[scale=0.33]{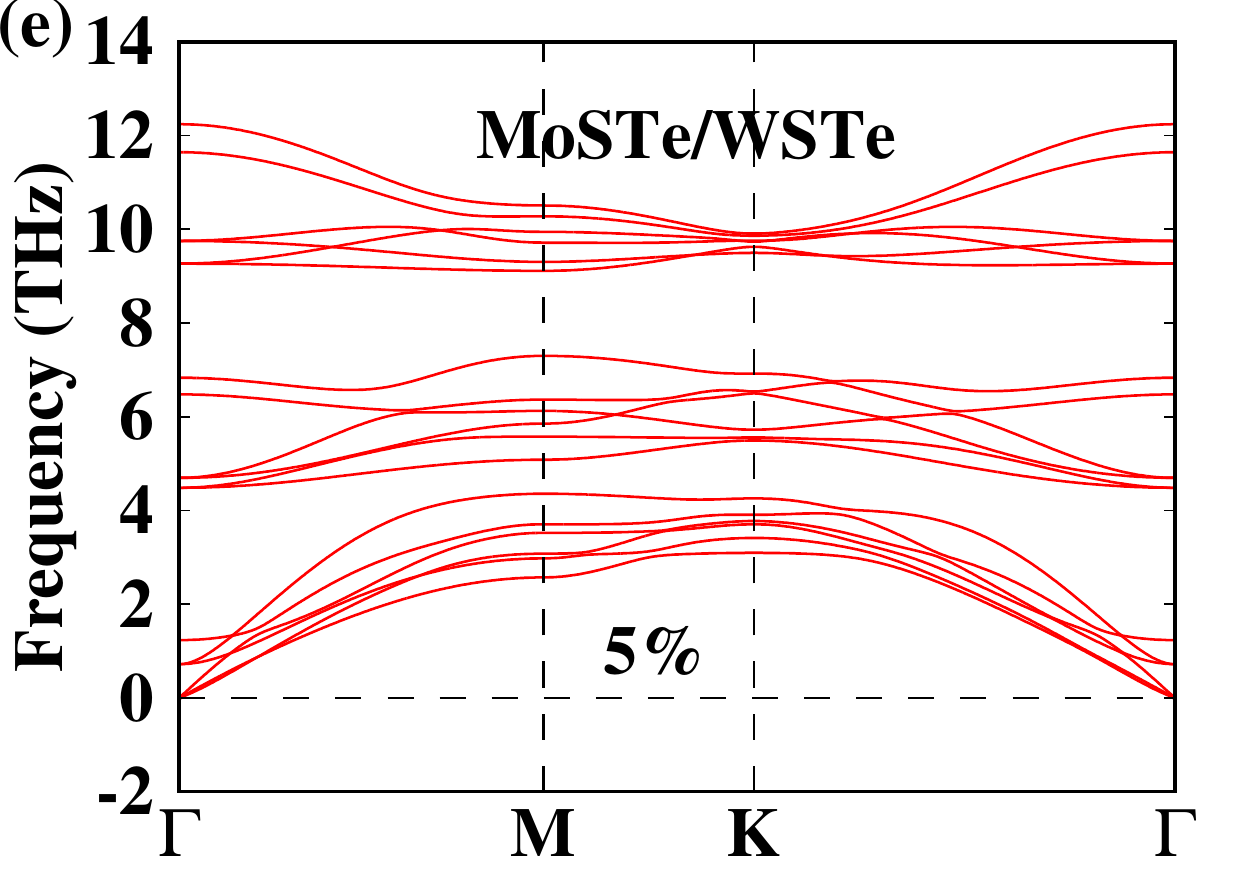}\\
			\includegraphics[scale=0.33]{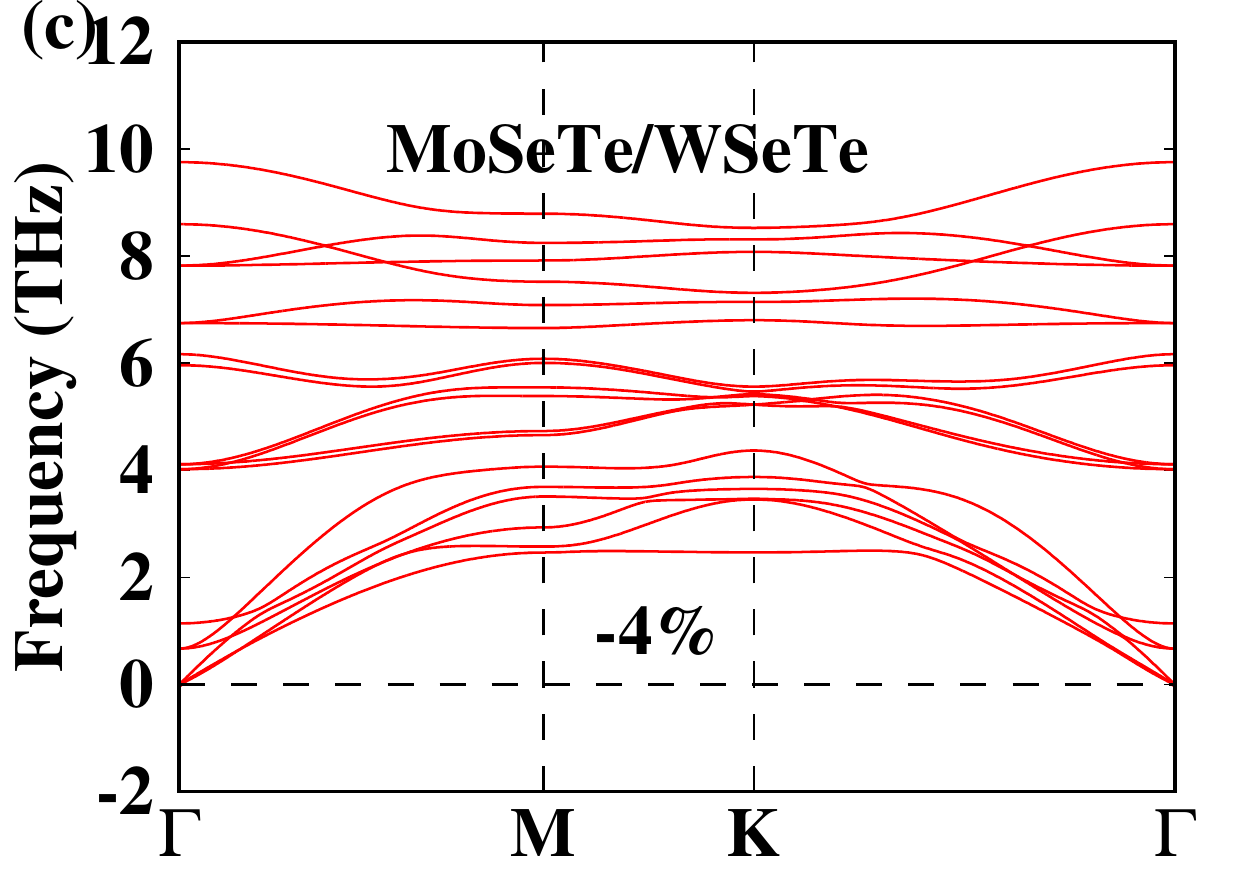}&
			\includegraphics[scale=0.33]{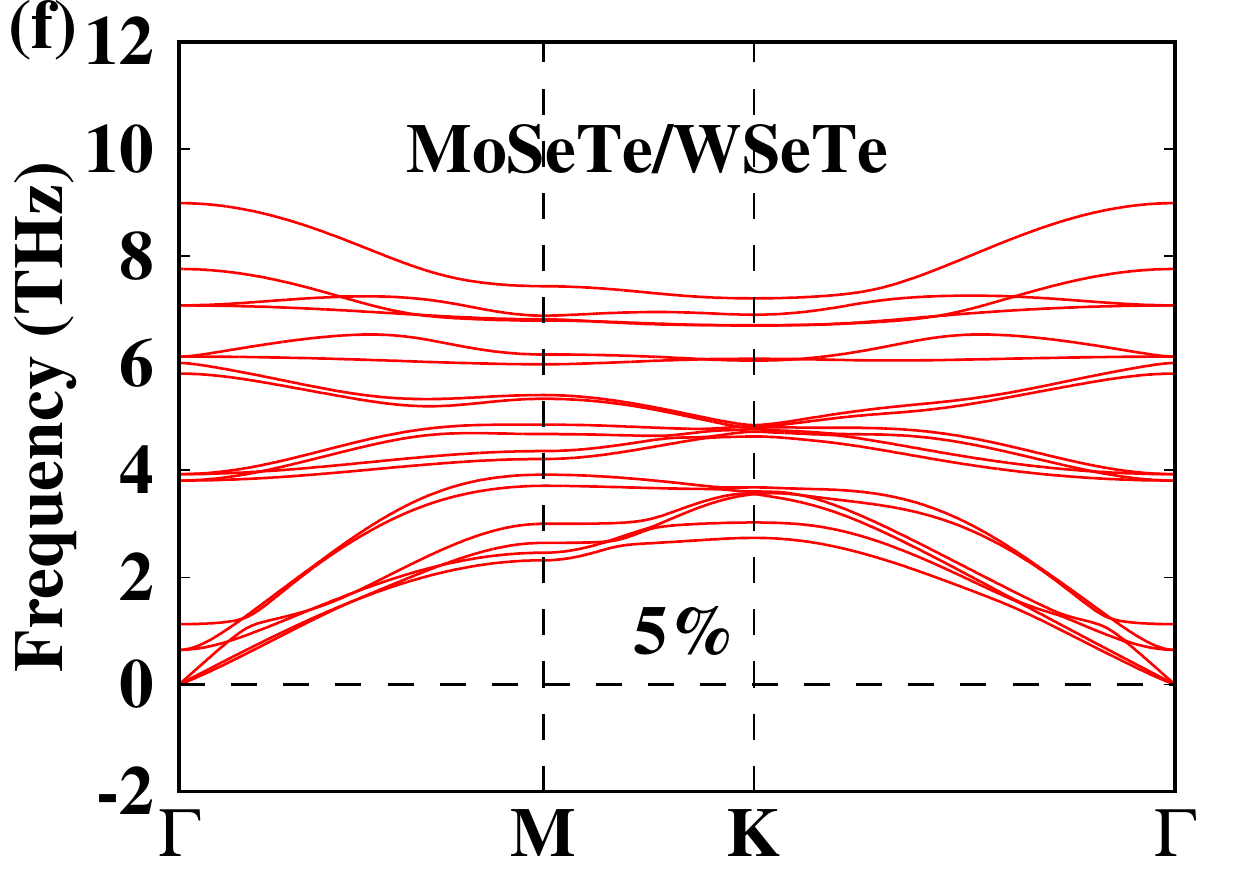}\\
		\end{tabular}	
		\caption{Phonon dispersions for all three HSs at the largest (a-c) compressive and  (d-f) tensile strains. The phonon frequencies remain positive for the HSs at the maximum strains showing their dynamical stability.}
		\label{fig:phonons}
	\end{figure}
	
	\begin{figure*}[!htb]
		\centering
		\begin{tabular}{lc}
			\includegraphics[scale=.74]{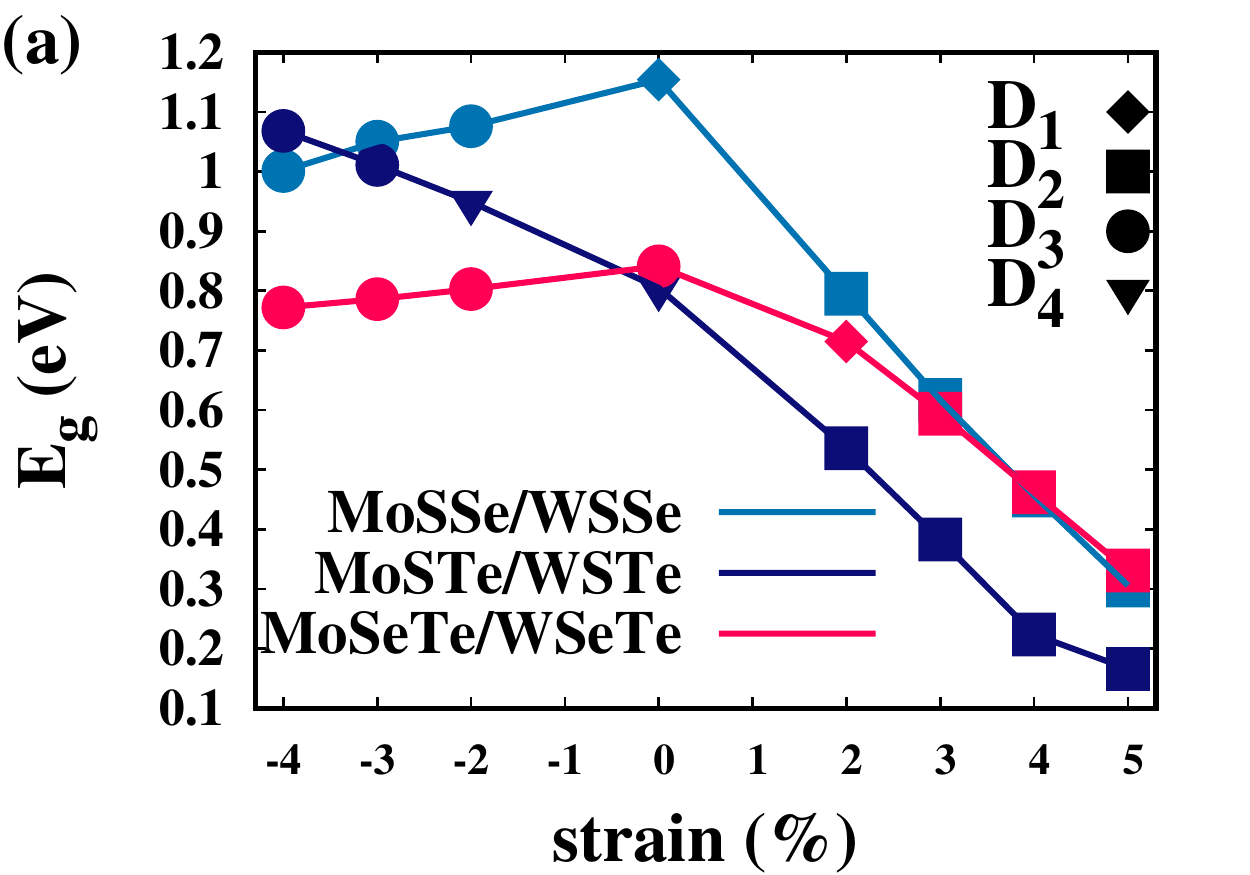}&
			\includegraphics[scale=.74]{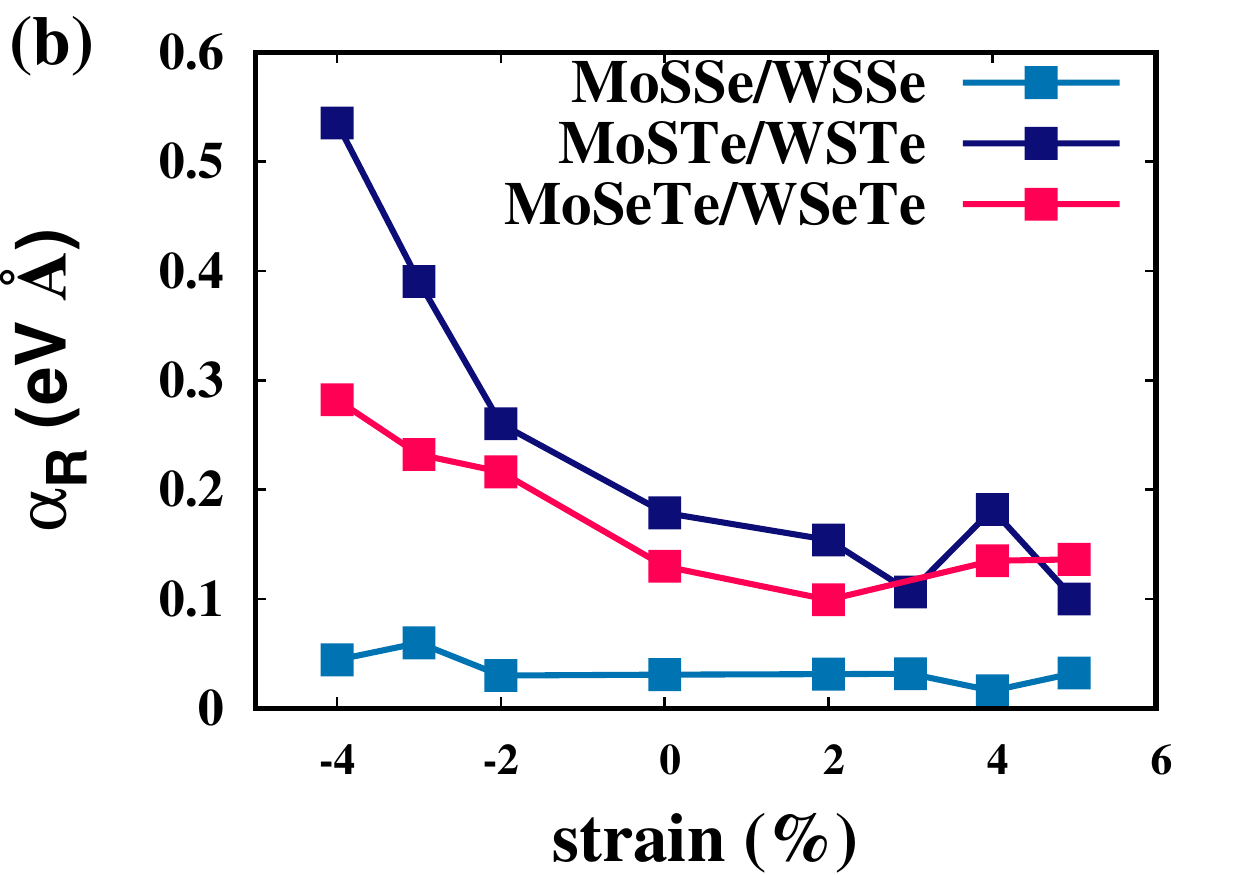}\\
		\end{tabular}
		
		\caption{Evolution of (a) band-gap and (b) Rashba parameter as a function of strain. Different symbols represent different types of gaps shown in Fig.~\ref{fig:efield_direction}(d) and different colors correspond to different vdW HSs. MoSTe/WSTe HS shows the highest $\alpha_R$ at 4\% compressive strain.}	\label{fig:strain}
	\end{figure*}
	\begin{figure*}[!htb]
		\centering
		\textbf{\Large Effects of in-plane Biaxial Strains}\\
		\vspace{0.35cm}
		\includegraphics[scale=0.77]{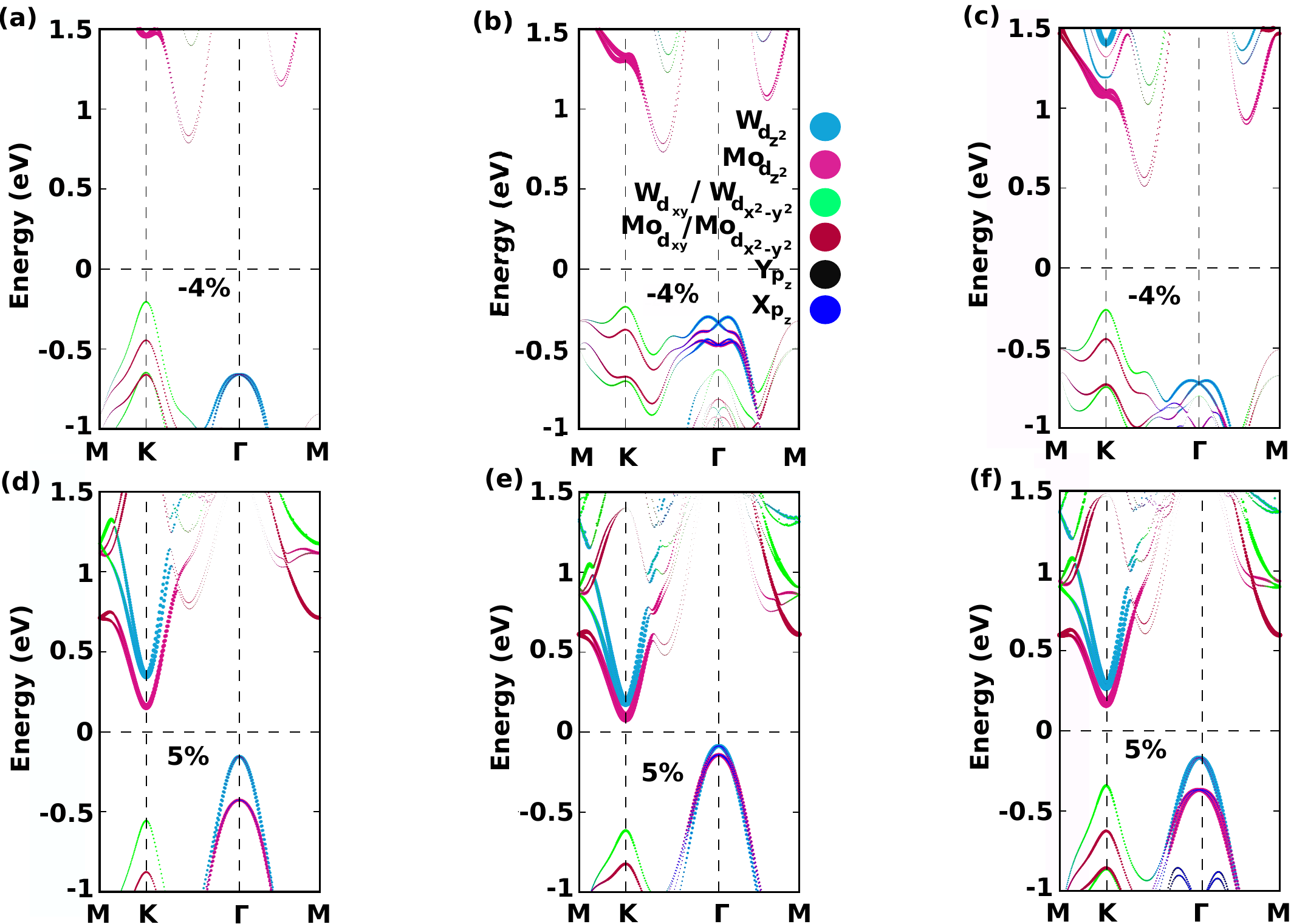}
		\caption{Projected band structures under -4\%\ (upper row) and 5\%\ (lower row) biaxial strains for MoSSe/WSSe, MoSTe/WSTe and MoSeTe/WSeTe, respectively from left to right. The weight of the orbitals is defined by the circles of different colors shown in the upper panel. In all cases, the CBM and VBM are located at the Mo$XY$ and W$XY$ layers, respectively, confirming that the type-II band alignment remains robust under biaxial strains.}
		\label{fig:projected_bands_strain}
	\end{figure*}
	
	An in-plane biaxial strain is applied to further investigate the manipulation of the band-gap and the Rashba spin splitting in J-HSs. The Schematic for this is shown in Fig.~\ref{Fig:biaxial_scheme}(a). The strain is applied from -4\% to +5\%, where negative (positive) means compressive (tensile) strain. We also present the variation of interfacial distances $d_{int}$ with strain in Fig.~\ref{Fig:biaxial_scheme}(b), which shows a monotonic decrease of the interlayer distances. We find that d$_{int}$ for $\pm 4$\% strains are not much different ($\sim\pm 0.03$ \AA) from that of the equilibrium structures. In Fig. S3(a) \cite{patel2022rashba} we show the ground state energies of the systems under different strains. The system energies remain negative throughout the range of the strains considered.  In order to check the dynamical stability of the systems, we also perform phonon calculations under the largest compressive (Fig.~\ref{fig:phonons}(a-c)) and tensile strains (Fig.~\ref{fig:phonons}(d-f)) applied in our study, and we find no imaginary frequency in the phonon spectra. This suggests that we are safe in applying the specified ranges of compressive and tensile strains on these heterostructures.
	In a study by W. Guo \textit{et al.} \cite{guo2020strain}, electronic properties of Janus MoSSe/WSSe are investigated under 10\% compressive strain to 10\% tensile biaxial strain. Their calculated values of Young's modulus and Poisson's ratio also show the mechanical stability of the systems.
	The band-gap and the evolution of Rashba splitting are studied with the band structures under the biaxial strain. The band-gap evolution as a function of strain is plotted in Fig.~\ref{fig:strain}(a). It should be noted that for MoSSe/WSSe and MoSeTe/WSeTe HSs, both tensile and compressive strains reduce the band gaps. Except for the zero strain condition, all the band-gaps in the case of MoSSe/WSSe HS are indirect. Under compressive strain, the band-gap is D$_3$ type, and for tensile strain, it is D$_2$ type for the MoSSe/WSSe HS. For the MoSTe/WSTe, the VBM moves from the K point at 4\%\ compressive strain to the $\Gamma$-point at -3 to 5\%\ strains, while the CBM transfers to the K point from the place in between K and $\Gamma$-points. This is the same case in MoSeTe/WSeTe HS, except the VBM, takes over the $\Gamma$ point at 4\%\ tensile strain. Therefore, we conclude that in all the HSs, there is a transition in band-gaps from the D$_3$ type to the D$_2$ type.
	
	Furthermore, it can be realized from Fig.~\ref{fig:strain}(a) that the band gaps will reduce all the way down to zero if the tensile strain is increased enough. A semiconductor to metal transition can be easily seen under higher tensile strain conditions in all three HSs. The band structures at maximum compressive and tensile strains are presented in Fig.~\ref{fig:projected_bands_strain} along with their orbital weights, which shows that under both tensile and compressive strains, the type-II band alignment of the HSs remains robust. We show the variation of band-gap with HSE06 XC functional in Fig. S3(b). The change in the band-gap for MoSSe/WSSe under compressive strain is not monotonic. It increases first and then decreases, as shown in Fig. S3(b), while it is clear from Fig.~\ref{fig:strain}(a) that the band-gap drops down under the compressive and the tensile strains with the PBE functional. On the other hand, the MoSeTe/WSeTe HS remains a direct band-gap semiconductor until 4\% of tensile strain with HSE06 and then changes its nature to an indirect D$_2$ type gap, in which the VBM at the $\Gamma$ point goes up in energy than the K point.

	Modulation of the Rashba parameter $\alpha_R$ as a function of biaxial strain is demonstrated in Fig.~\ref{fig:strain}(b), which shows that tensile strain suppresses $\alpha_R$, while a compressive strain enhances it significantly. The interlayer (or interfacial) distance and the net dipole moment are the lowest for the MoSSe/WSSe HS and the highest for the MoSTe/WSTe HS. As seen in Fig.~\ref{Fig:biaxial_scheme}(b), the interlayer distance reduces under tensile strain, and it clarifies the reduction in $\alpha_R$ under tensile strain. The smaller interlayer distance reduces the effective dipole moment between the layers and, consequently, the effective intrinsic electric field. On the other hand, a compressive strain increases the interlayer distance, which gives rise to the enhancement in $\alpha_R$. In the case of MoSTe/WSTe HS, the value of $\alpha_R$ is 178.8 meV\AA\, which increases significantly and reaches 535.3 meV\AA\, with 4\% of compressive strain. On the other hand, the MoSSe/WSSe HS is almost insensitive to the in-plane biaxial strain. This large Rashba splitting under 4\%\ compressive strain will have useful applications in experimental devices.

	\section{\label{sec:levelconclusion}Conclusions}
	
	To summarize, the structural, electronic, and spintronic properties of Janus AB-stacked Mo$XY$/W$XY$ heterostructures with $Y-Y$ interface have been investigated under external electric field and in-plane biaxial strain using first-principles calculations. We have studied the evolution of the band-gap and Rashba parameter, $\alpha_R$, for three different vdW HSs of this particular stacking and found the varying natures of band-gap and $\alpha_R$. The MoSSe/WSSe HS shows a direct band-gap at the K point under equilibrium state and positive EEF, while it is indirect under reverse EEF and both compressive and tensile strain conditions. The MoSTe/WSTe HS remains an indirect band-gap semiconductor of D$_4$ type throughout the range of EEF. The nature of the indirect band-gap in this HS is different under compressive and tensile strains. The MoSeTe/WSeTe HS shows a direct band-gap at 0.28 V/\AA\ and 2\% of tensile strain, and it possesses different natures of band-gap above and below this 2\% strain. All the HSs tend to a semiconductor to metal transition under large tensile strains. This kind of band-gap engineering can be beneficial in nanoelectronics, excitonic physics and valleytronics. Introduction of HSE06 XC functional is found to improve (as expected) the band-gaps significantly for all the HSs without a noticeable change in the qualitative nature of the bands. Calculations of the Rashba parameter show that $\alpha_R$ is the lowest for the MoSSe/WSSe HS, zero under reverse EEF, and remains almost insensitive to the strain. We note a considerable increment in $\alpha_R$ for the MoSTe/WSTe HS under 4\% compressive strain. Our results predict that a small $\alpha_R$ in AB-stacked Janus Mo$XY$/W$XY$ with a $Y-Y$ interface can be enhanced with the effect of an in-plane biaxial strain and can be utilized in the spintronic devices. The type-II band alignment, which results in the long lifetime of carriers, remains robust under biaxial strains and positive electric field, while its nature changes under reverse electric field. Thus, our work describes a way to control the electronic, spintronic, and optoelectronic properties of Janus vdW HSs, which have a strong bearing on potential technological applications. Manipulations of valley degree of freedom and optical properties of these kinds of Janus HSs are still under investigation, leading to new possibilities in the field of valleytronics and optoelectronics. Our DFT calculations for the pristine HSs are supplemented with $\mathbf{k\cdot p}$ model analyses. We define the minimal $\mathbf{k\cdot p}$ Hamiltonians around the $\Gamma$ and K points in the Brillouin zone and numerically fit the DFT bands to study the electronic and spin properties in detail.
	
	\section*{\label{sec:acknowledgement}Acknowledgements}
	
	We acknowledge National Supercomputing Mission (NSM) for providing computing resources of 'PARAM Shakti' at IIT Kharagpur, which is implemented by C-DAC and supported by the Ministry of Electronics and Information Technology (MeitY) and Department of Science and Technology (DST), Government of India. NPA acknowledges ICTP (Italy) OEA network program NT-14.
	
	\nocite{*}
	
	\bibliography{references}

\begin{thebibliography}{83}%
\makeatletter
\providecommand \@ifxundefined [1]{%
 \@ifx{#1\undefined}
}%
\providecommand \@ifnum [1]{%
 \ifnum #1\expandafter \@firstoftwo
 \else \expandafter \@secondoftwo
 \fi
}%
\providecommand \@ifx [1]{%
 \ifx #1\expandafter \@firstoftwo
 \else \expandafter \@secondoftwo
 \fi
}%
\providecommand \natexlab [1]{#1}%
\providecommand \enquote  [1]{``#1''}%
\providecommand \bibnamefont  [1]{#1}%
\providecommand \bibfnamefont [1]{#1}%
\providecommand \citenamefont [1]{#1}%
\providecommand \href@noop [0]{\@secondoftwo}%
\providecommand \href [0]{\begingroup \@sanitize@url \@href}%
\providecommand \@href[1]{\@@startlink{#1}\@@href}%
\providecommand \@@href[1]{\endgroup#1\@@endlink}%
\providecommand \@sanitize@url [0]{\catcode `\\12\catcode `\$12\catcode
  `\&12\catcode `\#12\catcode `\^12\catcode `\_12\catcode `\%12\relax}%
\providecommand \@@startlink[1]{}%
\providecommand \@@endlink[0]{}%
\providecommand \url  [0]{\begingroup\@sanitize@url \@url }%
\providecommand \@url [1]{\endgroup\@href {#1}{\urlprefix }}%
\providecommand \urlprefix  [0]{URL }%
\providecommand \Eprint [0]{\href }%
\providecommand \doibase [0]{https://doi.org/}%
\providecommand \selectlanguage [0]{\@gobble}%
\providecommand \bibinfo  [0]{\@secondoftwo}%
\providecommand \bibfield  [0]{\@secondoftwo}%
\providecommand \translation [1]{[#1]}%
\providecommand \BibitemOpen [0]{}%
\providecommand \bibitemStop [0]{}%
\providecommand \bibitemNoStop [0]{.\EOS\space}%
\providecommand \EOS [0]{\spacefactor3000\relax}%
\providecommand \BibitemShut  [1]{\csname bibitem#1\endcsname}%
\let\auto@bib@innerbib\@empty
\bibitem [{\citenamefont {Castro~Neto}\ \emph {et~al.}(2009)\citenamefont
  {Castro~Neto}, \citenamefont {Guinea}, \citenamefont {Peres}, \citenamefont
  {Novoselov},\ and\ \citenamefont {Geim}}]{castro2009electronic}%
  \BibitemOpen
  \bibfield  {author} {\bibinfo {author} {\bibfnamefont {A.~H.}\ \bibnamefont
  {Castro~Neto}}, \bibinfo {author} {\bibfnamefont {F.}~\bibnamefont {Guinea}},
  \bibinfo {author} {\bibfnamefont {N.~M.~R.}\ \bibnamefont {Peres}}, \bibinfo
  {author} {\bibfnamefont {K.~S.}\ \bibnamefont {Novoselov}},\ and\ \bibinfo
  {author} {\bibfnamefont {A.~K.}\ \bibnamefont {Geim}},\ }\bibfield  {title}
  {\bibinfo {title} {The electronic properties of graphene},\ }\href
  {https://doi.org/10.1103/RevModPhys.81.109} {\bibfield  {journal} {\bibinfo
  {journal} {Rev. Mod. Phys.}\ }\textbf {\bibinfo {volume} {81}},\ \bibinfo
  {pages} {109} (\bibinfo {year} {2009})}\BibitemShut {NoStop}%
\bibitem [{\citenamefont {Lee}\ \emph {et~al.}(2012)\citenamefont {Lee},
  \citenamefont {Zhang}, \citenamefont {Zhang}, \citenamefont {Chang},
  \citenamefont {Lin}, \citenamefont {Chang}, \citenamefont {Yu}, \citenamefont
  {Wang}, \citenamefont {Chang}, \citenamefont {Li} \emph
  {et~al.}}]{lee2012synthesis}%
  \BibitemOpen
  \bibfield  {author} {\bibinfo {author} {\bibfnamefont {Y.-H.}\ \bibnamefont
  {Lee}}, \bibinfo {author} {\bibfnamefont {X.-Q.}\ \bibnamefont {Zhang}},
  \bibinfo {author} {\bibfnamefont {W.}~\bibnamefont {Zhang}}, \bibinfo
  {author} {\bibfnamefont {M.-T.}\ \bibnamefont {Chang}}, \bibinfo {author}
  {\bibfnamefont {C.-T.}\ \bibnamefont {Lin}}, \bibinfo {author} {\bibfnamefont
  {K.-D.}\ \bibnamefont {Chang}}, \bibinfo {author} {\bibfnamefont {Y.-C.}\
  \bibnamefont {Yu}}, \bibinfo {author} {\bibfnamefont {J.~T.-W.}\ \bibnamefont
  {Wang}}, \bibinfo {author} {\bibfnamefont {C.-S.}\ \bibnamefont {Chang}},
  \bibinfo {author} {\bibfnamefont {L.-J.}\ \bibnamefont {Li}}, \emph
  {et~al.},\ }\bibfield  {title} {\bibinfo {title} {{Synthesis of large-area
  {MoS$_2$} atomic layers with chemical vapor deposition}},\ }\href
  {https://doi.org/https://doi.org/10.1002/adma.201104798} {\bibfield
  {journal} {\bibinfo  {journal} {Advanced materials}\ }\textbf {\bibinfo
  {volume} {24}},\ \bibinfo {pages} {2320} (\bibinfo {year}
  {2012})}\BibitemShut {NoStop}%
\bibitem [{\citenamefont {Shi}\ \emph {et~al.}(2015)\citenamefont {Shi},
  \citenamefont {Li},\ and\ \citenamefont {Li}}]{shi2015recent}%
  \BibitemOpen
  \bibfield  {author} {\bibinfo {author} {\bibfnamefont {Y.}~\bibnamefont
  {Shi}}, \bibinfo {author} {\bibfnamefont {H.}~\bibnamefont {Li}},\ and\
  \bibinfo {author} {\bibfnamefont {L.-J.}\ \bibnamefont {Li}},\ }\bibfield
  {title} {\bibinfo {title} {Recent advances in controlled synthesis of
  two-dimensional transition metal dichalcogenides via vapour deposition
  techniques},\ }\href {https://doi.org/https://doi.org/10.1039/C4CS00256C}
  {\bibfield  {journal} {\bibinfo  {journal} {Chemical Society Reviews}\
  }\textbf {\bibinfo {volume} {44}},\ \bibinfo {pages} {2744} (\bibinfo {year}
  {2015})}\BibitemShut {NoStop}%
\bibitem [{\citenamefont {Li}\ \emph {et~al.}(2013)\citenamefont {Li},
  \citenamefont {Lu}, \citenamefont {Wang}, \citenamefont {Yin}, \citenamefont
  {Cong}, \citenamefont {He}, \citenamefont {Wang}, \citenamefont {Ding},
  \citenamefont {Yu},\ and\ \citenamefont {Zhang}}]{li2013mechanical}%
  \BibitemOpen
  \bibfield  {author} {\bibinfo {author} {\bibfnamefont {H.}~\bibnamefont
  {Li}}, \bibinfo {author} {\bibfnamefont {G.}~\bibnamefont {Lu}}, \bibinfo
  {author} {\bibfnamefont {Y.}~\bibnamefont {Wang}}, \bibinfo {author}
  {\bibfnamefont {Z.}~\bibnamefont {Yin}}, \bibinfo {author} {\bibfnamefont
  {C.}~\bibnamefont {Cong}}, \bibinfo {author} {\bibfnamefont {Q.}~\bibnamefont
  {He}}, \bibinfo {author} {\bibfnamefont {L.}~\bibnamefont {Wang}}, \bibinfo
  {author} {\bibfnamefont {F.}~\bibnamefont {Ding}}, \bibinfo {author}
  {\bibfnamefont {T.}~\bibnamefont {Yu}},\ and\ \bibinfo {author}
  {\bibfnamefont {H.}~\bibnamefont {Zhang}},\ }\bibfield  {title} {\bibinfo
  {title} {{Mechanical exfoliation and characterization of single-and few-layer
  nanosheets of {WSe$_2$, TaS$_2$, and TaSe$_2$}}},\ }\href
  {https://doi.org/https://doi.org/10.1002/smll.201202919} {\bibfield
  {journal} {\bibinfo  {journal} {small}\ }\textbf {\bibinfo {volume} {9}},\
  \bibinfo {pages} {1974} (\bibinfo {year} {2013})}\BibitemShut {NoStop}%
\bibitem [{\citenamefont {Zeng}\ \emph {et~al.}(2012)\citenamefont {Zeng},
  \citenamefont {Sun}, \citenamefont {Zhu}, \citenamefont {Huang},
  \citenamefont {Yin}, \citenamefont {Lu}, \citenamefont {Fan}, \citenamefont
  {Yan}, \citenamefont {Hng},\ and\ \citenamefont {Zhang}}]{zeng2012effective}%
  \BibitemOpen
  \bibfield  {author} {\bibinfo {author} {\bibfnamefont {Z.}~\bibnamefont
  {Zeng}}, \bibinfo {author} {\bibfnamefont {T.}~\bibnamefont {Sun}}, \bibinfo
  {author} {\bibfnamefont {J.}~\bibnamefont {Zhu}}, \bibinfo {author}
  {\bibfnamefont {X.}~\bibnamefont {Huang}}, \bibinfo {author} {\bibfnamefont
  {Z.}~\bibnamefont {Yin}}, \bibinfo {author} {\bibfnamefont {G.}~\bibnamefont
  {Lu}}, \bibinfo {author} {\bibfnamefont {Z.}~\bibnamefont {Fan}}, \bibinfo
  {author} {\bibfnamefont {Q.}~\bibnamefont {Yan}}, \bibinfo {author}
  {\bibfnamefont {H.~H.}\ \bibnamefont {Hng}},\ and\ \bibinfo {author}
  {\bibfnamefont {H.}~\bibnamefont {Zhang}},\ }\bibfield  {title} {\bibinfo
  {title} {An effective method for the fabrication of few-layer-thick inorganic
  nanosheets},\ }\href {https://doi.org/https://doi.org/10.1002/anie.201204208}
  {\bibfield  {journal} {\bibinfo  {journal} {Angewandte Chemie International
  Edition}\ }\textbf {\bibinfo {volume} {51}},\ \bibinfo {pages} {9052}
  (\bibinfo {year} {2012})}\BibitemShut {NoStop}%
\bibitem [{\citenamefont {Feng}\ \emph {et~al.}(2015)\citenamefont {Feng},
  \citenamefont {Mao}, \citenamefont {Wu}, \citenamefont {Xu}, \citenamefont
  {Wang}, \citenamefont {Zhang},\ and\ \citenamefont {Xie}}]{feng2015growth}%
  \BibitemOpen
  \bibfield  {author} {\bibinfo {author} {\bibfnamefont {Q.}~\bibnamefont
  {Feng}}, \bibinfo {author} {\bibfnamefont {N.}~\bibnamefont {Mao}}, \bibinfo
  {author} {\bibfnamefont {J.}~\bibnamefont {Wu}}, \bibinfo {author}
  {\bibfnamefont {H.}~\bibnamefont {Xu}}, \bibinfo {author} {\bibfnamefont
  {C.}~\bibnamefont {Wang}}, \bibinfo {author} {\bibfnamefont {J.}~\bibnamefont
  {Zhang}},\ and\ \bibinfo {author} {\bibfnamefont {L.}~\bibnamefont {Xie}},\
  }\bibfield  {title} {\bibinfo {title} {{Growth of {MoS$_{2 (1-x)}$ Se$_{2x}$}
  (x= 0.41-1.00) monolayer alloys with controlled morphology by physical vapor
  deposition}},\ }\href
  {https://doi.org/https://doi.org/10.1021/acsnano.5b02506} {\bibfield
  {journal} {\bibinfo  {journal} {ACS nano}\ }\textbf {\bibinfo {volume} {9}},\
  \bibinfo {pages} {7450} (\bibinfo {year} {2015})}\BibitemShut {NoStop}%
\bibitem [{\citenamefont {Coleman}\ \emph {et~al.}(2011)\citenamefont
  {Coleman}, \citenamefont {Lotya}, \citenamefont {O’Neill}, \citenamefont
  {Bergin}, \citenamefont {King}, \citenamefont {Khan}, \citenamefont {Young},
  \citenamefont {Gaucher}, \citenamefont {De}, \citenamefont {Smith} \emph
  {et~al.}}]{coleman2011two}%
  \BibitemOpen
  \bibfield  {author} {\bibinfo {author} {\bibfnamefont {J.~N.}\ \bibnamefont
  {Coleman}}, \bibinfo {author} {\bibfnamefont {M.}~\bibnamefont {Lotya}},
  \bibinfo {author} {\bibfnamefont {A.}~\bibnamefont {O’Neill}}, \bibinfo
  {author} {\bibfnamefont {S.~D.}\ \bibnamefont {Bergin}}, \bibinfo {author}
  {\bibfnamefont {P.~J.}\ \bibnamefont {King}}, \bibinfo {author}
  {\bibfnamefont {U.}~\bibnamefont {Khan}}, \bibinfo {author} {\bibfnamefont
  {K.}~\bibnamefont {Young}}, \bibinfo {author} {\bibfnamefont
  {A.}~\bibnamefont {Gaucher}}, \bibinfo {author} {\bibfnamefont
  {S.}~\bibnamefont {De}}, \bibinfo {author} {\bibfnamefont {R.~J.}\
  \bibnamefont {Smith}}, \emph {et~al.},\ }\bibfield  {title} {\bibinfo {title}
  {Two-dimensional nanosheets produced by liquid exfoliation of layered
  materials},\ }\href {https://doi.org/10.1126/science.1194975} {\bibfield
  {journal} {\bibinfo  {journal} {Science}\ }\textbf {\bibinfo {volume}
  {331}},\ \bibinfo {pages} {568} (\bibinfo {year} {2011})}\BibitemShut
  {NoStop}%
\bibitem [{\citenamefont {Li}\ \emph {et~al.}(2018)\citenamefont {Li},
  \citenamefont {Cheng},\ and\ \citenamefont {Huang}}]{li2018recent}%
  \BibitemOpen
  \bibfield  {author} {\bibinfo {author} {\bibfnamefont {R.}~\bibnamefont
  {Li}}, \bibinfo {author} {\bibfnamefont {Y.}~\bibnamefont {Cheng}},\ and\
  \bibinfo {author} {\bibfnamefont {W.}~\bibnamefont {Huang}},\ }\bibfield
  {title} {\bibinfo {title} {Recent progress of janus 2d transition metal
  chalcogenides: from theory to experiments},\ }\href
  {https://doi.org/https://doi.org/10.1002/smll.201802091} {\bibfield
  {journal} {\bibinfo  {journal} {Small}\ }\textbf {\bibinfo {volume} {14}},\
  \bibinfo {pages} {1802091} (\bibinfo {year} {2018})}\BibitemShut {NoStop}%
\bibitem [{\citenamefont {Cheng}\ \emph {et~al.}(2013)\citenamefont {Cheng},
  \citenamefont {Zhu}, \citenamefont {Tahir},\ and\ \citenamefont
  {Schwingenschl{\"o}gl}}]{cheng2013spin}%
  \BibitemOpen
  \bibfield  {author} {\bibinfo {author} {\bibfnamefont {Y.}~\bibnamefont
  {Cheng}}, \bibinfo {author} {\bibfnamefont {Z.}~\bibnamefont {Zhu}}, \bibinfo
  {author} {\bibfnamefont {M.}~\bibnamefont {Tahir}},\ and\ \bibinfo {author}
  {\bibfnamefont {U.}~\bibnamefont {Schwingenschl{\"o}gl}},\ }\bibfield
  {title} {\bibinfo {title} {Spin-orbit--induced spin splittings in polar
  transition metal dichalcogenide monolayers},\ }\href
  {https://doi.org/10.1209/0295-5075/102/57001} {\bibfield  {journal} {\bibinfo
   {journal} {EPL (Europhysics Letters)}\ }\textbf {\bibinfo {volume} {102}},\
  \bibinfo {pages} {57001} (\bibinfo {year} {2013})}\BibitemShut {NoStop}%
\bibitem [{\citenamefont {Lu}\ \emph {et~al.}(2017)\citenamefont {Lu},
  \citenamefont {Zhu}, \citenamefont {Xiao}, \citenamefont {Chuu},
  \citenamefont {Han}, \citenamefont {Chiu}, \citenamefont {Cheng},
  \citenamefont {Yang}, \citenamefont {Wei}, \citenamefont {Yang} \emph
  {et~al.}}]{lu2017janus}%
  \BibitemOpen
  \bibfield  {author} {\bibinfo {author} {\bibfnamefont {A.-Y.}\ \bibnamefont
  {Lu}}, \bibinfo {author} {\bibfnamefont {H.}~\bibnamefont {Zhu}}, \bibinfo
  {author} {\bibfnamefont {J.}~\bibnamefont {Xiao}}, \bibinfo {author}
  {\bibfnamefont {C.-P.}\ \bibnamefont {Chuu}}, \bibinfo {author}
  {\bibfnamefont {Y.}~\bibnamefont {Han}}, \bibinfo {author} {\bibfnamefont
  {M.-H.}\ \bibnamefont {Chiu}}, \bibinfo {author} {\bibfnamefont {C.-C.}\
  \bibnamefont {Cheng}}, \bibinfo {author} {\bibfnamefont {C.-W.}\ \bibnamefont
  {Yang}}, \bibinfo {author} {\bibfnamefont {K.-H.}\ \bibnamefont {Wei}},
  \bibinfo {author} {\bibfnamefont {Y.}~\bibnamefont {Yang}}, \emph {et~al.},\
  }\bibfield  {title} {\bibinfo {title} {Janus monolayers of transition metal
  dichalcogenides},\ }\href
  {https://doi.org/https://doi.org/10.1038/nnano.2017.100} {\bibfield
  {journal} {\bibinfo  {journal} {Nature nanotechnology}\ }\textbf {\bibinfo
  {volume} {12}},\ \bibinfo {pages} {744} (\bibinfo {year} {2017})}\BibitemShut
  {NoStop}%
\bibitem [{\citenamefont {Zhang}\ \emph {et~al.}(2017)\citenamefont {Zhang},
  \citenamefont {Jia}, \citenamefont {Kholmanov}, \citenamefont {Dong},
  \citenamefont {Er}, \citenamefont {Chen}, \citenamefont {Guo}, \citenamefont
  {Jin}, \citenamefont {Shenoy}, \citenamefont {Shi} \emph
  {et~al.}}]{zhang2017janus}%
  \BibitemOpen
  \bibfield  {author} {\bibinfo {author} {\bibfnamefont {J.}~\bibnamefont
  {Zhang}}, \bibinfo {author} {\bibfnamefont {S.}~\bibnamefont {Jia}}, \bibinfo
  {author} {\bibfnamefont {I.}~\bibnamefont {Kholmanov}}, \bibinfo {author}
  {\bibfnamefont {L.}~\bibnamefont {Dong}}, \bibinfo {author} {\bibfnamefont
  {D.}~\bibnamefont {Er}}, \bibinfo {author} {\bibfnamefont {W.}~\bibnamefont
  {Chen}}, \bibinfo {author} {\bibfnamefont {H.}~\bibnamefont {Guo}}, \bibinfo
  {author} {\bibfnamefont {Z.}~\bibnamefont {Jin}}, \bibinfo {author}
  {\bibfnamefont {V.~B.}\ \bibnamefont {Shenoy}}, \bibinfo {author}
  {\bibfnamefont {L.}~\bibnamefont {Shi}}, \emph {et~al.},\ }\bibfield  {title}
  {\bibinfo {title} {Janus monolayer transition-metal dichalcogenides},\ }\href
  {https://doi.org/https://doi.org/10.1021/acsnano.7b03186} {\bibfield
  {journal} {\bibinfo  {journal} {ACS nano}\ }\textbf {\bibinfo {volume}
  {11}},\ \bibinfo {pages} {8192} (\bibinfo {year} {2017})}\BibitemShut
  {NoStop}%
\bibitem [{\citenamefont {Lin}\ \emph {et~al.}(2020)\citenamefont {Lin},
  \citenamefont {Liu}, \citenamefont {Yu}, \citenamefont {Zarkadoula},
  \citenamefont {Yoon}, \citenamefont {Puretzky}, \citenamefont {Liang},
  \citenamefont {Kong}, \citenamefont {Gu}, \citenamefont {Strasser} \emph
  {et~al.}}]{lin2020low}%
  \BibitemOpen
  \bibfield  {author} {\bibinfo {author} {\bibfnamefont {Y.-C.}\ \bibnamefont
  {Lin}}, \bibinfo {author} {\bibfnamefont {C.}~\bibnamefont {Liu}}, \bibinfo
  {author} {\bibfnamefont {Y.}~\bibnamefont {Yu}}, \bibinfo {author}
  {\bibfnamefont {E.}~\bibnamefont {Zarkadoula}}, \bibinfo {author}
  {\bibfnamefont {M.}~\bibnamefont {Yoon}}, \bibinfo {author} {\bibfnamefont
  {A.~A.}\ \bibnamefont {Puretzky}}, \bibinfo {author} {\bibfnamefont
  {L.}~\bibnamefont {Liang}}, \bibinfo {author} {\bibfnamefont
  {X.}~\bibnamefont {Kong}}, \bibinfo {author} {\bibfnamefont {Y.}~\bibnamefont
  {Gu}}, \bibinfo {author} {\bibfnamefont {A.}~\bibnamefont {Strasser}}, \emph
  {et~al.},\ }\bibfield  {title} {\bibinfo {title} {Low energy implantation
  into transition-metal dichalcogenide monolayers to form janus structures},\
  }\href {https://doi.org/https://doi.org/10.1021/acsnano.9b10196} {\bibfield
  {journal} {\bibinfo  {journal} {ACS nano}\ }\textbf {\bibinfo {volume}
  {14}},\ \bibinfo {pages} {3896} (\bibinfo {year} {2020})}\BibitemShut
  {NoStop}%
\bibitem [{\citenamefont {Dresselhaus}(1955)}]{dresselhaus1955spin}%
  \BibitemOpen
  \bibfield  {author} {\bibinfo {author} {\bibfnamefont {G.}~\bibnamefont
  {Dresselhaus}},\ }\bibfield  {title} {\bibinfo {title} {Spin-orbit coupling
  effects in zinc blende structures},\ }\href
  {https://doi.org/10.1103/PhysRev.100.580} {\bibfield  {journal} {\bibinfo
  {journal} {Phys. Rev.}\ }\textbf {\bibinfo {volume} {100}},\ \bibinfo {pages}
  {580} (\bibinfo {year} {1955})}\BibitemShut {NoStop}%
\bibitem [{\citenamefont {{Bychkov}}\ and\ \citenamefont
  {{Rashba}}(1984)}]{bychkov1984properties}%
  \BibitemOpen
  \bibfield  {author} {\bibinfo {author} {\bibfnamefont {Y.~A.}\ \bibnamefont
  {{Bychkov}}}\ and\ \bibinfo {author} {\bibfnamefont {{\'E}.~I.}\ \bibnamefont
  {{Rashba}}},\ }\bibfield  {title} {\bibinfo {title} {{Properties of a 2D
  electron gas with lifted spectral degeneracy}},\ }\href
  {http://jetpletters.ru/ps/1264/article_19121.shtml} {\bibfield  {journal}
  {\bibinfo  {journal} {Soviet Journal of Experimental and Theoretical Physics
  Letters}\ }\textbf {\bibinfo {volume} {39}},\ \bibinfo {pages} {78} (\bibinfo
  {year} {1984})}\BibitemShut {NoStop}%
\bibitem [{\citenamefont {Manchon}\ \emph {et~al.}(2015)\citenamefont
  {Manchon}, \citenamefont {Koo}, \citenamefont {Nitta}, \citenamefont
  {Frolov},\ and\ \citenamefont {Duine}}]{manchon2015new}%
  \BibitemOpen
  \bibfield  {author} {\bibinfo {author} {\bibfnamefont {A.}~\bibnamefont
  {Manchon}}, \bibinfo {author} {\bibfnamefont {H.~C.}\ \bibnamefont {Koo}},
  \bibinfo {author} {\bibfnamefont {J.}~\bibnamefont {Nitta}}, \bibinfo
  {author} {\bibfnamefont {S.}~\bibnamefont {Frolov}},\ and\ \bibinfo {author}
  {\bibfnamefont {R.}~\bibnamefont {Duine}},\ }\bibfield  {title} {\bibinfo
  {title} {New perspectives for rashba spin--orbit coupling},\ }\href
  {https://doi.org/https://doi.org/10.1038/nmat4360} {\bibfield  {journal}
  {\bibinfo  {journal} {Nature materials}\ }\textbf {\bibinfo {volume} {14}},\
  \bibinfo {pages} {871} (\bibinfo {year} {2015})}\BibitemShut {NoStop}%
\bibitem [{\citenamefont {Bihlmayer}\ \emph {et~al.}(2015)\citenamefont
  {Bihlmayer}, \citenamefont {Rader},\ and\ \citenamefont
  {Winkler}}]{bihlmayer2015focus}%
  \BibitemOpen
  \bibfield  {author} {\bibinfo {author} {\bibfnamefont {G.}~\bibnamefont
  {Bihlmayer}}, \bibinfo {author} {\bibfnamefont {O.}~\bibnamefont {Rader}},\
  and\ \bibinfo {author} {\bibfnamefont {R.}~\bibnamefont {Winkler}},\
  }\bibfield  {title} {\bibinfo {title} {Focus on the rashba effect},\ }\href
  {https://doi.org/10.1088/1367-2630/17/5/050202} {\bibfield  {journal}
  {\bibinfo  {journal} {New journal of physics}\ }\textbf {\bibinfo {volume}
  {17}},\ \bibinfo {pages} {050202} (\bibinfo {year} {2015})}\BibitemShut
  {NoStop}%
\bibitem [{\citenamefont {Riis-Jensen}\ \emph {et~al.}(2019)\citenamefont
  {Riis-Jensen}, \citenamefont {Deilmann}, \citenamefont {Olsen},\ and\
  \citenamefont {Thygesen}}]{riis2019classifying}%
  \BibitemOpen
  \bibfield  {author} {\bibinfo {author} {\bibfnamefont {A.~C.}\ \bibnamefont
  {Riis-Jensen}}, \bibinfo {author} {\bibfnamefont {T.}~\bibnamefont
  {Deilmann}}, \bibinfo {author} {\bibfnamefont {T.}~\bibnamefont {Olsen}},\
  and\ \bibinfo {author} {\bibfnamefont {K.~S.}\ \bibnamefont {Thygesen}},\
  }\bibfield  {title} {\bibinfo {title} {{Classifying the electronic and
  optical properties of Janus monolayers}},\ }\href
  {https://doi.org/https://doi.org/10.1021/acsnano.9b06698} {\bibfield
  {journal} {\bibinfo  {journal} {ACS nano}\ }\textbf {\bibinfo {volume}
  {13}},\ \bibinfo {pages} {13354} (\bibinfo {year} {2019})}\BibitemShut
  {NoStop}%
\bibitem [{\citenamefont {Li}\ and\ \citenamefont
  {Li}(2015)}]{li2015piezoelectricity}%
  \BibitemOpen
  \bibfield  {author} {\bibinfo {author} {\bibfnamefont {W.}~\bibnamefont
  {Li}}\ and\ \bibinfo {author} {\bibfnamefont {J.}~\bibnamefont {Li}},\
  }\bibfield  {title} {\bibinfo {title} {{Piezoelectricity in two-dimensional
  group-III monochalcogenides}},\ }\href
  {https://doi.org/https://doi.org/10.1007/s12274-015-0878-8} {\bibfield
  {journal} {\bibinfo  {journal} {Nano Research}\ }\textbf {\bibinfo {volume}
  {8}},\ \bibinfo {pages} {3796} (\bibinfo {year} {2015})}\BibitemShut
  {NoStop}%
\bibitem [{\citenamefont {Dong}\ \emph {et~al.}(2017)\citenamefont {Dong},
  \citenamefont {Lou},\ and\ \citenamefont {Shenoy}}]{dong2017large}%
  \BibitemOpen
  \bibfield  {author} {\bibinfo {author} {\bibfnamefont {L.}~\bibnamefont
  {Dong}}, \bibinfo {author} {\bibfnamefont {J.}~\bibnamefont {Lou}},\ and\
  \bibinfo {author} {\bibfnamefont {V.~B.}\ \bibnamefont {Shenoy}},\ }\bibfield
   {title} {\bibinfo {title} {{Large in-plane and vertical piezoelectricity in
  Janus transition metal dichalchogenides}},\ }\href
  {https://doi.org/https://doi.org/10.1021/acsnano.7b03313} {\bibfield
  {journal} {\bibinfo  {journal} {ACS nano}\ }\textbf {\bibinfo {volume}
  {11}},\ \bibinfo {pages} {8242} (\bibinfo {year} {2017})}\BibitemShut
  {NoStop}%
\bibitem [{\citenamefont {Yagmurcukardes}\ \emph {et~al.}(2019)\citenamefont
  {Yagmurcukardes}, \citenamefont {Sevik},\ and\ \citenamefont
  {Peeters}}]{yagmurcukardes2019electronic}%
  \BibitemOpen
  \bibfield  {author} {\bibinfo {author} {\bibfnamefont {M.}~\bibnamefont
  {Yagmurcukardes}}, \bibinfo {author} {\bibfnamefont {C.}~\bibnamefont
  {Sevik}},\ and\ \bibinfo {author} {\bibfnamefont {F.~M.}\ \bibnamefont
  {Peeters}},\ }\bibfield  {title} {\bibinfo {title} {{Electronic, vibrational,
  elastic, and piezoelectric properties of monolayer Janus MoSTe phases: A
  first-principles study}},\ }\href
  {https://doi.org/10.1103/PhysRevB.100.045415} {\bibfield  {journal} {\bibinfo
   {journal} {Phys. Rev. B}\ }\textbf {\bibinfo {volume} {100}},\ \bibinfo
  {pages} {045415} (\bibinfo {year} {2019})}\BibitemShut {NoStop}%
\bibitem [{\citenamefont {Sato}\ and\ \citenamefont
  {Fujimoto}(2009)}]{sato2009topological}%
  \BibitemOpen
  \bibfield  {author} {\bibinfo {author} {\bibfnamefont {M.}~\bibnamefont
  {Sato}}\ and\ \bibinfo {author} {\bibfnamefont {S.}~\bibnamefont
  {Fujimoto}},\ }\bibfield  {title} {\bibinfo {title} {{Topological phases of
  noncentrosymmetric superconductors: Edge states, Majorana fermions, and
  non-Abelian statistics}},\ }\href
  {https://doi.org/https://doi.org/10.1103/PhysRevB.79.094504} {\bibfield
  {journal} {\bibinfo  {journal} {Phys. Rev. B}\ }\textbf {\bibinfo {volume}
  {79}},\ \bibinfo {pages} {094504} (\bibinfo {year} {2009})}\BibitemShut
  {NoStop}%
\bibitem [{\citenamefont {Maghirang}\ \emph {et~al.}(2019)\citenamefont
  {Maghirang}, \citenamefont {Huang}, \citenamefont {Villaos}, \citenamefont
  {Hsu}, \citenamefont {Feng}, \citenamefont {Florido}, \citenamefont {Lin},
  \citenamefont {Bansil},\ and\ \citenamefont
  {Chuang}}]{maghirang2019predicting}%
  \BibitemOpen
  \bibfield  {author} {\bibinfo {author} {\bibfnamefont {A.~B.}\ \bibnamefont
  {Maghirang}}, \bibinfo {author} {\bibfnamefont {Z.-Q.}\ \bibnamefont
  {Huang}}, \bibinfo {author} {\bibfnamefont {R.~A.~B.}\ \bibnamefont
  {Villaos}}, \bibinfo {author} {\bibfnamefont {C.-H.}\ \bibnamefont {Hsu}},
  \bibinfo {author} {\bibfnamefont {L.-Y.}\ \bibnamefont {Feng}}, \bibinfo
  {author} {\bibfnamefont {E.}~\bibnamefont {Florido}}, \bibinfo {author}
  {\bibfnamefont {H.}~\bibnamefont {Lin}}, \bibinfo {author} {\bibfnamefont
  {A.}~\bibnamefont {Bansil}},\ and\ \bibinfo {author} {\bibfnamefont {F.-C.}\
  \bibnamefont {Chuang}},\ }\bibfield  {title} {\bibinfo {title} {{Predicting
  two-dimensional topological phases in Janus materials by substitutional
  doping in transition metal dichalcogenide monolayers}},\ }\href
  {https://doi.org/https://doi.org/10.1038/s41699-019-0118-2} {\bibfield
  {journal} {\bibinfo  {journal} {npj 2D Materials and Applications}\ }\textbf
  {\bibinfo {volume} {3}},\ \bibinfo {pages} {1} (\bibinfo {year}
  {2019})}\BibitemShut {NoStop}%
\bibitem [{\citenamefont {Dey}\ and\ \citenamefont
  {Botana}(2020)}]{dey2020structural}%
  \BibitemOpen
  \bibfield  {author} {\bibinfo {author} {\bibfnamefont {D.}~\bibnamefont
  {Dey}}\ and\ \bibinfo {author} {\bibfnamefont {A.~S.}\ \bibnamefont
  {Botana}},\ }\bibfield  {title} {\bibinfo {title} {{Structural, electronic,
  and magnetic properties of vanadium-based Janus dichalcogenide monolayers: A
  first-principles study}},\ }\href
  {https://doi.org/10.1103/PhysRevMaterials.4.074002} {\bibfield  {journal}
  {\bibinfo  {journal} {Phys. Rev. Materials}\ }\textbf {\bibinfo {volume}
  {4}},\ \bibinfo {pages} {074002} (\bibinfo {year} {2020})}\BibitemShut
  {NoStop}%
\bibitem [{\citenamefont {Georgiou}\ \emph {et~al.}(2013)\citenamefont
  {Georgiou}, \citenamefont {Jalil}, \citenamefont {Belle}, \citenamefont
  {Britnell}, \citenamefont {Gorbachev}, \citenamefont {Morozov}, \citenamefont
  {Kim}, \citenamefont {Gholinia}, \citenamefont {Haigh}, \citenamefont
  {Makarovsky} \emph {et~al.}}]{georgiou2013vertical}%
  \BibitemOpen
  \bibfield  {author} {\bibinfo {author} {\bibfnamefont {T.}~\bibnamefont
  {Georgiou}}, \bibinfo {author} {\bibfnamefont {R.}~\bibnamefont {Jalil}},
  \bibinfo {author} {\bibfnamefont {B.~D.}\ \bibnamefont {Belle}}, \bibinfo
  {author} {\bibfnamefont {L.}~\bibnamefont {Britnell}}, \bibinfo {author}
  {\bibfnamefont {R.~V.}\ \bibnamefont {Gorbachev}}, \bibinfo {author}
  {\bibfnamefont {S.~V.}\ \bibnamefont {Morozov}}, \bibinfo {author}
  {\bibfnamefont {Y.-J.}\ \bibnamefont {Kim}}, \bibinfo {author} {\bibfnamefont
  {A.}~\bibnamefont {Gholinia}}, \bibinfo {author} {\bibfnamefont {S.~J.}\
  \bibnamefont {Haigh}}, \bibinfo {author} {\bibfnamefont {O.}~\bibnamefont
  {Makarovsky}}, \emph {et~al.},\ }\bibfield  {title} {\bibinfo {title}
  {{Vertical field-effect transistor based on graphene--{WS$_2$}
  heterostructures for flexible and transparent electronics}},\ }\href
  {https://doi.org/https://doi.org/10.1038/nnano.2012.224} {\bibfield
  {journal} {\bibinfo  {journal} {Nature nanotechnology}\ }\textbf {\bibinfo
  {volume} {8}},\ \bibinfo {pages} {100} (\bibinfo {year} {2013})}\BibitemShut
  {NoStop}%
\bibitem [{\citenamefont {Anto~Jeffery}\ \emph {et~al.}(2014)\citenamefont
  {Anto~Jeffery}, \citenamefont {Nethravathi},\ and\ \citenamefont
  {Rajamathi}}]{anto2014two}%
  \BibitemOpen
  \bibfield  {author} {\bibinfo {author} {\bibfnamefont {A.}~\bibnamefont
  {Anto~Jeffery}}, \bibinfo {author} {\bibfnamefont {C.}~\bibnamefont
  {Nethravathi}},\ and\ \bibinfo {author} {\bibfnamefont {M.}~\bibnamefont
  {Rajamathi}},\ }\bibfield  {title} {\bibinfo {title} {{Two-dimensional
  nanosheets and layered hybrids of MoS{$_2$} and WS{$_2$} through exfoliation
  of ammoniated {MS$_2$ (M= Mo, W)}}},\ }\href
  {https://doi.org/https://doi.org/10.1021/jp410918c} {\bibfield  {journal}
  {\bibinfo  {journal} {The Journal of Physical Chemistry C}\ }\textbf
  {\bibinfo {volume} {118}},\ \bibinfo {pages} {1386} (\bibinfo {year}
  {2014})}\BibitemShut {NoStop}%
\bibitem [{\citenamefont {Zheng}\ \emph {et~al.}(2018)\citenamefont {Zheng},
  \citenamefont {Ma}, \citenamefont {Li}, \citenamefont {Lan}, \citenamefont
  {Zhang}, \citenamefont {Sun}, \citenamefont {Zheng}, \citenamefont {Yang},
  \citenamefont {Zhu}, \citenamefont {Ouyang} \emph {et~al.}}]{zheng2018band}%
  \BibitemOpen
  \bibfield  {author} {\bibinfo {author} {\bibfnamefont {B.}~\bibnamefont
  {Zheng}}, \bibinfo {author} {\bibfnamefont {C.}~\bibnamefont {Ma}}, \bibinfo
  {author} {\bibfnamefont {D.}~\bibnamefont {Li}}, \bibinfo {author}
  {\bibfnamefont {J.}~\bibnamefont {Lan}}, \bibinfo {author} {\bibfnamefont
  {Z.}~\bibnamefont {Zhang}}, \bibinfo {author} {\bibfnamefont
  {X.}~\bibnamefont {Sun}}, \bibinfo {author} {\bibfnamefont {W.}~\bibnamefont
  {Zheng}}, \bibinfo {author} {\bibfnamefont {T.}~\bibnamefont {Yang}},
  \bibinfo {author} {\bibfnamefont {C.}~\bibnamefont {Zhu}}, \bibinfo {author}
  {\bibfnamefont {G.}~\bibnamefont {Ouyang}}, \emph {et~al.},\ }\bibfield
  {title} {\bibinfo {title} {{Band alignment engineering in two-dimensional
  lateral heterostructures}},\ }\href
  {https://doi.org/https://doi.org/10.1021/jacs.8b07401} {\bibfield  {journal}
  {\bibinfo  {journal} {Journal of the American Chemical Society}\ }\textbf
  {\bibinfo {volume} {140}},\ \bibinfo {pages} {11193} (\bibinfo {year}
  {2018})}\BibitemShut {NoStop}%
\bibitem [{\citenamefont {Kou}\ \emph {et~al.}(2013)\citenamefont {Kou},
  \citenamefont {Frauenheim},\ and\ \citenamefont {Chen}}]{kou2013nanoscale}%
  \BibitemOpen
  \bibfield  {author} {\bibinfo {author} {\bibfnamefont {L.}~\bibnamefont
  {Kou}}, \bibinfo {author} {\bibfnamefont {T.}~\bibnamefont {Frauenheim}},\
  and\ \bibinfo {author} {\bibfnamefont {C.}~\bibnamefont {Chen}},\ }\bibfield
  {title} {\bibinfo {title} {{Nanoscale multilayer transition-metal
  dichalcogenide heterostructures: band gap modulation by interfacial strain
  and spontaneous polarization}},\ }\href
  {https://doi.org/https://doi.org/10.1021/jz400668d} {\bibfield  {journal}
  {\bibinfo  {journal} {{The journal of physical chemistry letters}}\ }\textbf
  {\bibinfo {volume} {4}},\ \bibinfo {pages} {1730} (\bibinfo {year}
  {2013})}\BibitemShut {NoStop}%
\bibitem [{\citenamefont {Komsa}\ and\ \citenamefont
  {Krasheninnikov}(2013)}]{komsa2013electronic}%
  \BibitemOpen
  \bibfield  {author} {\bibinfo {author} {\bibfnamefont {H.-P.}\ \bibnamefont
  {Komsa}}\ and\ \bibinfo {author} {\bibfnamefont {A.~V.}\ \bibnamefont
  {Krasheninnikov}},\ }\bibfield  {title} {\bibinfo {title} {{Electronic
  structures and optical properties of realistic transition metal
  dichalcogenide heterostructures from first principles}},\ }\href
  {https://doi.org/10.1103/PhysRevB.88.085318} {\bibfield  {journal} {\bibinfo
  {journal} {Phys. Rev. B}\ }\textbf {\bibinfo {volume} {88}},\ \bibinfo
  {pages} {085318} (\bibinfo {year} {2013})}\BibitemShut {NoStop}%
\bibitem [{\citenamefont {Terrones}\ \emph {et~al.}(2013)\citenamefont
  {Terrones}, \citenamefont {L{\'o}pez-Ur{\'\i}as},\ and\ \citenamefont
  {Terrones}}]{terrones2013novel}%
  \BibitemOpen
  \bibfield  {author} {\bibinfo {author} {\bibfnamefont {H.}~\bibnamefont
  {Terrones}}, \bibinfo {author} {\bibfnamefont {F.}~\bibnamefont
  {L{\'o}pez-Ur{\'\i}as}},\ and\ \bibinfo {author} {\bibfnamefont
  {M.}~\bibnamefont {Terrones}},\ }\bibfield  {title} {\bibinfo {title} {{Novel
  hetero-layered materials with tunable direct band gaps by sandwiching
  different metal disulfides and diselenides}},\ }\href
  {https://doi.org/https://doi.org/10.1038/srep01549} {\bibfield  {journal}
  {\bibinfo  {journal} {Scientific reports}\ }\textbf {\bibinfo {volume} {3}},\
  \bibinfo {pages} {1} (\bibinfo {year} {2013})}\BibitemShut {NoStop}%
\bibitem [{\citenamefont {Long}\ and\ \citenamefont
  {Prezhdo}(2016)}]{long2016quantum}%
  \BibitemOpen
  \bibfield  {author} {\bibinfo {author} {\bibfnamefont {R.}~\bibnamefont
  {Long}}\ and\ \bibinfo {author} {\bibfnamefont {O.~V.}\ \bibnamefont
  {Prezhdo}},\ }\bibfield  {title} {\bibinfo {title} {{Quantum coherence
  facilitates efficient charge separation at a {MoS$_2$/MoSe$_2$} van der Waals
  junction}},\ }\href
  {https://doi.org/https://doi.org/10.1021/acs.nanolett.5b05264} {\bibfield
  {journal} {\bibinfo  {journal} {Nano letters}\ }\textbf {\bibinfo {volume}
  {16}},\ \bibinfo {pages} {1996} (\bibinfo {year} {2016})}\BibitemShut
  {NoStop}%
\bibitem [{\citenamefont {Chhowalla}\ \emph {et~al.}(2013)\citenamefont
  {Chhowalla}, \citenamefont {Shin}, \citenamefont {Eda}, \citenamefont {Li},
  \citenamefont {Loh},\ and\ \citenamefont {Zhang}}]{chhowalla2013chemistry}%
  \BibitemOpen
  \bibfield  {author} {\bibinfo {author} {\bibfnamefont {M.}~\bibnamefont
  {Chhowalla}}, \bibinfo {author} {\bibfnamefont {H.~S.}\ \bibnamefont {Shin}},
  \bibinfo {author} {\bibfnamefont {G.}~\bibnamefont {Eda}}, \bibinfo {author}
  {\bibfnamefont {L.-J.}\ \bibnamefont {Li}}, \bibinfo {author} {\bibfnamefont
  {K.~P.}\ \bibnamefont {Loh}},\ and\ \bibinfo {author} {\bibfnamefont
  {H.}~\bibnamefont {Zhang}},\ }\bibfield  {title} {\bibinfo {title} {The
  chemistry of two-dimensional layered transition metal dichalcogenide
  nanosheets},\ }\href {https://doi.org/https://doi.org/10.1038/nchem.1589}
  {\bibfield  {journal} {\bibinfo  {journal} {{Nature chemistry}}\ }\textbf
  {\bibinfo {volume} {5}},\ \bibinfo {pages} {263} (\bibinfo {year}
  {2013})}\BibitemShut {NoStop}%
\bibitem [{\citenamefont {Kou}\ \emph {et~al.}(2014)\citenamefont {Kou},
  \citenamefont {Du}, \citenamefont {Chen},\ and\ \citenamefont
  {Frauenheim}}]{kou2014strain}%
  \BibitemOpen
  \bibfield  {author} {\bibinfo {author} {\bibfnamefont {L.}~\bibnamefont
  {Kou}}, \bibinfo {author} {\bibfnamefont {A.}~\bibnamefont {Du}}, \bibinfo
  {author} {\bibfnamefont {C.}~\bibnamefont {Chen}},\ and\ \bibinfo {author}
  {\bibfnamefont {T.}~\bibnamefont {Frauenheim}},\ }\bibfield  {title}
  {\bibinfo {title} {{Strain engineering of selective chemical adsorption on
  monolayer {MoS$_2$}}},\ }\href
  {https://doi.org/https://doi.org/10.1039/C3NR06670C} {\bibfield  {journal}
  {\bibinfo  {journal} {{Nanoscale}}\ }\textbf {\bibinfo {volume} {6}},\
  \bibinfo {pages} {5156} (\bibinfo {year} {2014})}\BibitemShut {NoStop}%
\bibitem [{\citenamefont {Arora}\ \emph {et~al.}(2017)\citenamefont {Arora},
  \citenamefont {Dr{\"u}ppel}, \citenamefont {Schmidt}, \citenamefont
  {Deilmann}, \citenamefont {Schneider}, \citenamefont {Molas}, \citenamefont
  {Marauhn}, \citenamefont {de~Vasconcellos}, \citenamefont {Potemski},
  \citenamefont {Rohlfing} \emph {et~al.}}]{arora2017interlayer}%
  \BibitemOpen
  \bibfield  {author} {\bibinfo {author} {\bibfnamefont {A.}~\bibnamefont
  {Arora}}, \bibinfo {author} {\bibfnamefont {M.}~\bibnamefont {Dr{\"u}ppel}},
  \bibinfo {author} {\bibfnamefont {R.}~\bibnamefont {Schmidt}}, \bibinfo
  {author} {\bibfnamefont {T.}~\bibnamefont {Deilmann}}, \bibinfo {author}
  {\bibfnamefont {R.}~\bibnamefont {Schneider}}, \bibinfo {author}
  {\bibfnamefont {M.~R.}\ \bibnamefont {Molas}}, \bibinfo {author}
  {\bibfnamefont {P.}~\bibnamefont {Marauhn}}, \bibinfo {author} {\bibfnamefont
  {S.~M.}\ \bibnamefont {de~Vasconcellos}}, \bibinfo {author} {\bibfnamefont
  {M.}~\bibnamefont {Potemski}}, \bibinfo {author} {\bibfnamefont
  {M.}~\bibnamefont {Rohlfing}}, \emph {et~al.},\ }\bibfield  {title} {\bibinfo
  {title} {{Interlayer excitons in a bulk van der Waals semiconductor}},\
  }\href {https://doi.org/https://doi.org/10.1038/s41467-017-00691-5}
  {\bibfield  {journal} {\bibinfo  {journal} {Nature communications}\ }\textbf
  {\bibinfo {volume} {8}},\ \bibinfo {pages} {1} (\bibinfo {year}
  {2017})}\BibitemShut {NoStop}%
\bibitem [{\citenamefont {Liu}\ \emph {et~al.}(2020{\natexlab{a}})\citenamefont
  {Liu}, \citenamefont {Zong}, \citenamefont {Wang}, \citenamefont {Wen},
  \citenamefont {Wu}, \citenamefont {Xia},\ and\ \citenamefont
  {Wei}}]{liu2020excitons}%
  \BibitemOpen
  \bibfield  {author} {\bibinfo {author} {\bibfnamefont {H.}~\bibnamefont
  {Liu}}, \bibinfo {author} {\bibfnamefont {Y.-X.}\ \bibnamefont {Zong}},
  \bibinfo {author} {\bibfnamefont {P.}~\bibnamefont {Wang}}, \bibinfo {author}
  {\bibfnamefont {H.}~\bibnamefont {Wen}}, \bibinfo {author} {\bibfnamefont
  {H.-B.}\ \bibnamefont {Wu}}, \bibinfo {author} {\bibfnamefont {J.-B.}\
  \bibnamefont {Xia}},\ and\ \bibinfo {author} {\bibfnamefont {Z.}~\bibnamefont
  {Wei}},\ }\bibfield  {title} {\bibinfo {title} {{Excitons in two-dimensional
  van der Waals heterostructures}},\ }\href
  {https://iopscience.iop.org/article/10.1088/1361-6463/abbf75} {\bibfield
  {journal} {\bibinfo  {journal} {Journal of Physics D: Applied Physics}\
  }\textbf {\bibinfo {volume} {54}},\ \bibinfo {pages} {053001} (\bibinfo
  {year} {2020}{\natexlab{a}})}\BibitemShut {NoStop}%
\bibitem [{\citenamefont {Suzuki}\ \emph {et~al.}(2014)\citenamefont {Suzuki},
  \citenamefont {Sakano}, \citenamefont {Zhang}, \citenamefont {Akashi},
  \citenamefont {Morikawa}, \citenamefont {Harasawa}, \citenamefont {Yaji},
  \citenamefont {Kuroda}, \citenamefont {Miyamoto}, \citenamefont {Okuda} \emph
  {et~al.}}]{suzuki2014valley}%
  \BibitemOpen
  \bibfield  {author} {\bibinfo {author} {\bibfnamefont {R.}~\bibnamefont
  {Suzuki}}, \bibinfo {author} {\bibfnamefont {M.}~\bibnamefont {Sakano}},
  \bibinfo {author} {\bibfnamefont {Y.}~\bibnamefont {Zhang}}, \bibinfo
  {author} {\bibfnamefont {R.}~\bibnamefont {Akashi}}, \bibinfo {author}
  {\bibfnamefont {D.}~\bibnamefont {Morikawa}}, \bibinfo {author}
  {\bibfnamefont {A.}~\bibnamefont {Harasawa}}, \bibinfo {author}
  {\bibfnamefont {K.}~\bibnamefont {Yaji}}, \bibinfo {author} {\bibfnamefont
  {K.}~\bibnamefont {Kuroda}}, \bibinfo {author} {\bibfnamefont
  {K.}~\bibnamefont {Miyamoto}}, \bibinfo {author} {\bibfnamefont
  {T.}~\bibnamefont {Okuda}}, \emph {et~al.},\ }\bibfield  {title} {\bibinfo
  {title} {{Valley-dependent spin polarization in bulk {MoS$_2$} with broken
  inversion symmetry}},\ }\href
  {https://doi.org/https://doi.org/10.1038/nnano.2014.148} {\bibfield
  {journal} {\bibinfo  {journal} {Nature nanotechnology}\ }\textbf {\bibinfo
  {volume} {9}},\ \bibinfo {pages} {611} (\bibinfo {year} {2014})}\BibitemShut
  {NoStop}%
\bibitem [{\citenamefont {Li}\ \emph {et~al.}(2017)\citenamefont {Li},
  \citenamefont {Wei}, \citenamefont {Zhao}, \citenamefont {Huang},\ and\
  \citenamefont {Dai}}]{li2017electronic}%
  \BibitemOpen
  \bibfield  {author} {\bibinfo {author} {\bibfnamefont {F.}~\bibnamefont
  {Li}}, \bibinfo {author} {\bibfnamefont {W.}~\bibnamefont {Wei}}, \bibinfo
  {author} {\bibfnamefont {P.}~\bibnamefont {Zhao}}, \bibinfo {author}
  {\bibfnamefont {B.}~\bibnamefont {Huang}},\ and\ \bibinfo {author}
  {\bibfnamefont {Y.}~\bibnamefont {Dai}},\ }\bibfield  {title} {\bibinfo
  {title} {{Electronic and optical properties of pristine and vertical and
  lateral heterostructures of Janus MoSSe and WSSe}},\ }\href
  {https://doi.org/https://doi.org/10.1021/acs.jpclett.7b02841} {\bibfield
  {journal} {\bibinfo  {journal} {The journal of physical chemistry letters}\
  }\textbf {\bibinfo {volume} {8}},\ \bibinfo {pages} {5959} (\bibinfo {year}
  {2017})}\BibitemShut {NoStop}%
\bibitem [{\citenamefont {Zhou}\ \emph {et~al.}(2019)\citenamefont {Zhou},
  \citenamefont {Chen}, \citenamefont {Yang}, \citenamefont {Liu},\ and\
  \citenamefont {Ouyang}}]{zhou2019geometry}%
  \BibitemOpen
  \bibfield  {author} {\bibinfo {author} {\bibfnamefont {W.}~\bibnamefont
  {Zhou}}, \bibinfo {author} {\bibfnamefont {J.}~\bibnamefont {Chen}}, \bibinfo
  {author} {\bibfnamefont {Z.}~\bibnamefont {Yang}}, \bibinfo {author}
  {\bibfnamefont {J.}~\bibnamefont {Liu}},\ and\ \bibinfo {author}
  {\bibfnamefont {F.}~\bibnamefont {Ouyang}},\ }\bibfield  {title} {\bibinfo
  {title} {{Geometry and electronic structure of monolayer, bilayer, and
  multilayer Janus WSSe}},\ }\href {https://doi.org/10.1103/PhysRevB.99.075160}
  {\bibfield  {journal} {\bibinfo  {journal} {Phys. Rev. B}\ }\textbf {\bibinfo
  {volume} {99}},\ \bibinfo {pages} {075160} (\bibinfo {year}
  {2019})}\BibitemShut {NoStop}%
\bibitem [{\citenamefont {Rezavand}\ and\ \citenamefont
  {Ghobadi}(2021)}]{rezavand2021stacking}%
  \BibitemOpen
  \bibfield  {author} {\bibinfo {author} {\bibfnamefont {A.}~\bibnamefont
  {Rezavand}}\ and\ \bibinfo {author} {\bibfnamefont {N.}~\bibnamefont
  {Ghobadi}},\ }\bibfield  {title} {\bibinfo {title} {{Stacking-dependent
  Rashba spin-splitting in Janus bilayer transition metal dichalcogenides: The
  role of in-plane strain and out-of-plane electric field}},\ }\href
  {https://doi.org/https://doi.org/10.1016/j.physe.2021.114768} {\bibfield
  {journal} {\bibinfo  {journal} {Physica E: Low-dimensional Systems and
  Nanostructures}\ }\textbf {\bibinfo {volume} {132}},\ \bibinfo {pages}
  {114768} (\bibinfo {year} {2021})}\BibitemShut {NoStop}%
\bibitem [{\citenamefont {Guo}\ \emph {et~al.}(2020)\citenamefont {Guo},
  \citenamefont {Ge}, \citenamefont {Sun}, \citenamefont {Xie},\ and\
  \citenamefont {Ye}}]{guo2020strain}%
  \BibitemOpen
  \bibfield  {author} {\bibinfo {author} {\bibfnamefont {W.}~\bibnamefont
  {Guo}}, \bibinfo {author} {\bibfnamefont {X.}~\bibnamefont {Ge}}, \bibinfo
  {author} {\bibfnamefont {S.}~\bibnamefont {Sun}}, \bibinfo {author}
  {\bibfnamefont {Y.}~\bibnamefont {Xie}},\ and\ \bibinfo {author}
  {\bibfnamefont {X.}~\bibnamefont {Ye}},\ }\bibfield  {title} {\bibinfo
  {title} {{The strain effect on the electronic properties of the MoSSe/WSSe
  van der Waals heterostructure: a first-principles study}},\ }\href
  {https://doi.org/https://doi.org/10.1039/D0CP00403K} {\bibfield  {journal}
  {\bibinfo  {journal} {Physical Chemistry Chemical Physics}\ }\textbf
  {\bibinfo {volume} {22}},\ \bibinfo {pages} {4946} (\bibinfo {year}
  {2020})}\BibitemShut {NoStop}%
\bibitem [{\citenamefont {Wang}\ \emph {et~al.}(2019)\citenamefont {Wang},
  \citenamefont {Wei}, \citenamefont {Huang},\ and\ \citenamefont
  {Dai}}]{wang2019mirror}%
  \BibitemOpen
  \bibfield  {author} {\bibinfo {author} {\bibfnamefont {Y.}~\bibnamefont
  {Wang}}, \bibinfo {author} {\bibfnamefont {W.}~\bibnamefont {Wei}}, \bibinfo
  {author} {\bibfnamefont {B.}~\bibnamefont {Huang}},\ and\ \bibinfo {author}
  {\bibfnamefont {Y.}~\bibnamefont {Dai}},\ }\bibfield  {title} {\bibinfo
  {title} {{The mirror asymmetry induced nontrivial properties of polar
  {WSSe/MoSSe} heterostructures}},\ }\href
  {https://doi.org/10.1088/1361-648x/aaffb7} {\bibfield  {journal} {\bibinfo
  {journal} {Journal of Physics: Condensed Matter}\ }\textbf {\bibinfo {volume}
  {31}},\ \bibinfo {pages} {125003} (\bibinfo {year} {2019})}\BibitemShut
  {NoStop}%
\bibitem [{\citenamefont {Hirsch}(1999)}]{hirsch1999spin}%
  \BibitemOpen
  \bibfield  {author} {\bibinfo {author} {\bibfnamefont {J.~E.}\ \bibnamefont
  {Hirsch}},\ }\bibfield  {title} {\bibinfo {title} {Spin hall effect},\ }\href
  {https://doi.org/10.1103/PhysRevLett.83.1834} {\bibfield  {journal} {\bibinfo
   {journal} {Phys. Rev. Lett.}\ }\textbf {\bibinfo {volume} {83}},\ \bibinfo
  {pages} {1834} (\bibinfo {year} {1999})}\BibitemShut {NoStop}%
\bibitem [{\citenamefont {Sinova}\ \emph {et~al.}(2004)\citenamefont {Sinova},
  \citenamefont {Culcer}, \citenamefont {Niu}, \citenamefont {Sinitsyn},
  \citenamefont {Jungwirth},\ and\ \citenamefont
  {MacDonald}}]{sinova2004universal}%
  \BibitemOpen
  \bibfield  {author} {\bibinfo {author} {\bibfnamefont {J.}~\bibnamefont
  {Sinova}}, \bibinfo {author} {\bibfnamefont {D.}~\bibnamefont {Culcer}},
  \bibinfo {author} {\bibfnamefont {Q.}~\bibnamefont {Niu}}, \bibinfo {author}
  {\bibfnamefont {N.~A.}\ \bibnamefont {Sinitsyn}}, \bibinfo {author}
  {\bibfnamefont {T.}~\bibnamefont {Jungwirth}},\ and\ \bibinfo {author}
  {\bibfnamefont {A.~H.}\ \bibnamefont {MacDonald}},\ }\bibfield  {title}
  {\bibinfo {title} {Universal intrinsic spin hall effect},\ }\href
  {https://doi.org/10.1103/PhysRevLett.92.126603} {\bibfield  {journal}
  {\bibinfo  {journal} {Phys. Rev. Lett.}\ }\textbf {\bibinfo {volume} {92}},\
  \bibinfo {pages} {126603} (\bibinfo {year} {2004})}\BibitemShut {NoStop}%
\bibitem [{\citenamefont {Wang}\ \emph {et~al.}(2020)\citenamefont {Wang},
  \citenamefont {Gopal}, \citenamefont {Picozzi}, \citenamefont {Curtarolo},
  \citenamefont {Nardelli},\ and\ \citenamefont
  {S{\l}awi{\'n}ska}}]{wang2020spin}%
  \BibitemOpen
  \bibfield  {author} {\bibinfo {author} {\bibfnamefont {H.}~\bibnamefont
  {Wang}}, \bibinfo {author} {\bibfnamefont {P.}~\bibnamefont {Gopal}},
  \bibinfo {author} {\bibfnamefont {S.}~\bibnamefont {Picozzi}}, \bibinfo
  {author} {\bibfnamefont {S.}~\bibnamefont {Curtarolo}}, \bibinfo {author}
  {\bibfnamefont {M.~B.}\ \bibnamefont {Nardelli}},\ and\ \bibinfo {author}
  {\bibfnamefont {J.}~\bibnamefont {S{\l}awi{\'n}ska}},\ }\bibfield  {title}
  {\bibinfo {title} {{Spin Hall effect in prototype Rashba ferroelectrics GeTe
  and SnTe}},\ }\href
  {https://doi.org/https://doi.org/10.1038/s41524-020-0274-0} {\bibfield
  {journal} {\bibinfo  {journal} {npj Computational Materials}\ }\textbf
  {\bibinfo {volume} {6}},\ \bibinfo {pages} {1} (\bibinfo {year}
  {2020})}\BibitemShut {NoStop}%
\bibitem [{\citenamefont {Yu}\ \emph {et~al.}(2021)\citenamefont {Yu},
  \citenamefont {Zhou}, \citenamefont {Zhang},\ and\ \citenamefont
  {Chang}}]{yu2021spin}%
  \BibitemOpen
  \bibfield  {author} {\bibinfo {author} {\bibfnamefont {S.-B.}\ \bibnamefont
  {Yu}}, \bibinfo {author} {\bibfnamefont {M.}~\bibnamefont {Zhou}}, \bibinfo
  {author} {\bibfnamefont {D.}~\bibnamefont {Zhang}},\ and\ \bibinfo {author}
  {\bibfnamefont {K.}~\bibnamefont {Chang}},\ }\bibfield  {title} {\bibinfo
  {title} {{Spin Hall effect in the monolayer Janus compound MoSSe enhanced by
  Rashba spin-orbit coupling}},\ }\href
  {https://doi.org/10.1103/PhysRevB.104.075435} {\bibfield  {journal} {\bibinfo
   {journal} {Phys. Rev. B}\ }\textbf {\bibinfo {volume} {104}},\ \bibinfo
  {pages} {075435} (\bibinfo {year} {2021})}\BibitemShut {NoStop}%
\bibitem [{\citenamefont {Datta}\ and\ \citenamefont
  {Das}(1990)}]{datta1990electronic}%
  \BibitemOpen
  \bibfield  {author} {\bibinfo {author} {\bibfnamefont {S.}~\bibnamefont
  {Datta}}\ and\ \bibinfo {author} {\bibfnamefont {B.}~\bibnamefont {Das}},\
  }\bibfield  {title} {\bibinfo {title} {Electronic analog of the electro-optic
  modulator},\ }\href {https://doi.org/https://doi.org/10.1063/1.102730}
  {\bibfield  {journal} {\bibinfo  {journal} {Applied Physics Letters}\
  }\textbf {\bibinfo {volume} {56}},\ \bibinfo {pages} {665} (\bibinfo {year}
  {1990})}\BibitemShut {NoStop}%
\bibitem [{\citenamefont {Shanavas}\ and\ \citenamefont
  {Satpathy}(2014)}]{shanavas2014electric}%
  \BibitemOpen
  \bibfield  {author} {\bibinfo {author} {\bibfnamefont {K.~V.}\ \bibnamefont
  {Shanavas}}\ and\ \bibinfo {author} {\bibfnamefont {S.}~\bibnamefont
  {Satpathy}},\ }\bibfield  {title} {\bibinfo {title} {Electric field tuning of
  the rashba effect in the polar perovskite structures},\ }\href
  {https://doi.org/10.1103/PhysRevLett.112.086802} {\bibfield  {journal}
  {\bibinfo  {journal} {Phys. Rev. Lett.}\ }\textbf {\bibinfo {volume} {112}},\
  \bibinfo {pages} {086802} (\bibinfo {year} {2014})}\BibitemShut {NoStop}%
\bibitem [{\citenamefont {Liu}\ \emph {et~al.}(2021)\citenamefont {Liu},
  \citenamefont {Gong}, \citenamefont {He},\ and\ \citenamefont
  {Cao}}]{liu2021tuning}%
  \BibitemOpen
  \bibfield  {author} {\bibinfo {author} {\bibfnamefont {M.-Y.}\ \bibnamefont
  {Liu}}, \bibinfo {author} {\bibfnamefont {L.}~\bibnamefont {Gong}}, \bibinfo
  {author} {\bibfnamefont {Y.}~\bibnamefont {He}},\ and\ \bibinfo {author}
  {\bibfnamefont {C.}~\bibnamefont {Cao}},\ }\bibfield  {title} {\bibinfo
  {title} {{Tuning Rashba effect, band inversion, and spin-charge conversion of
  Janus $X{\mathrm{Sn}}_{2}Y$ monolayers via an external field}},\ }\href
  {https://doi.org/10.1103/PhysRevB.103.075421} {\bibfield  {journal} {\bibinfo
   {journal} {Phys. Rev. B}\ }\textbf {\bibinfo {volume} {103}},\ \bibinfo
  {pages} {075421} (\bibinfo {year} {2021})}\BibitemShut {NoStop}%
\bibitem [{\citenamefont {Singh}\ and\ \citenamefont
  {Romero}(2017)}]{singh2017giant}%
  \BibitemOpen
  \bibfield  {author} {\bibinfo {author} {\bibfnamefont {S.}~\bibnamefont
  {Singh}}\ and\ \bibinfo {author} {\bibfnamefont {A.~H.}\ \bibnamefont
  {Romero}},\ }\bibfield  {title} {\bibinfo {title} {{Giant tunable Rashba spin
  splitting in a two-dimensional BiSb monolayer and in BiSb/AlN
  heterostructures}},\ }\href {https://doi.org/10.1103/PhysRevB.95.165444}
  {\bibfield  {journal} {\bibinfo  {journal} {Phys. Rev. B}\ }\textbf {\bibinfo
  {volume} {95}},\ \bibinfo {pages} {165444} (\bibinfo {year}
  {2017})}\BibitemShut {NoStop}%
\bibitem [{\citenamefont {Absor}\ \emph {et~al.}(2018)\citenamefont {Absor},
  \citenamefont {Santoso}, \citenamefont {Harsojo}, \citenamefont {Abraha},
  \citenamefont {Kotaka}, \citenamefont {Ishii},\ and\ \citenamefont
  {Saito}}]{absor2018strong}%
  \BibitemOpen
  \bibfield  {author} {\bibinfo {author} {\bibfnamefont {M.~A.~U.}\
  \bibnamefont {Absor}}, \bibinfo {author} {\bibfnamefont {I.}~\bibnamefont
  {Santoso}}, \bibinfo {author} {\bibnamefont {Harsojo}}, \bibinfo {author}
  {\bibfnamefont {K.}~\bibnamefont {Abraha}}, \bibinfo {author} {\bibfnamefont
  {H.}~\bibnamefont {Kotaka}}, \bibinfo {author} {\bibfnamefont
  {F.}~\bibnamefont {Ishii}},\ and\ \bibinfo {author} {\bibfnamefont
  {M.}~\bibnamefont {Saito}},\ }\bibfield  {title} {\bibinfo {title} {{Strong
  Rashba effect in the localized impurity states of halogen-doped monolayer
  ${\mathrm{PtSe}}_{2}$}},\ }\href {https://doi.org/10.1103/PhysRevB.97.205138}
  {\bibfield  {journal} {\bibinfo  {journal} {Phys. Rev. B}\ }\textbf {\bibinfo
  {volume} {97}},\ \bibinfo {pages} {205138} (\bibinfo {year}
  {2018})}\BibitemShut {NoStop}%
\bibitem [{\citenamefont {Chen}\ \emph {et~al.}(2020)\citenamefont {Chen},
  \citenamefont {Wu}, \citenamefont {Ma}, \citenamefont {Hu},\ and\
  \citenamefont {Yang}}]{chen2020tunable}%
  \BibitemOpen
  \bibfield  {author} {\bibinfo {author} {\bibfnamefont {J.}~\bibnamefont
  {Chen}}, \bibinfo {author} {\bibfnamefont {K.}~\bibnamefont {Wu}}, \bibinfo
  {author} {\bibfnamefont {H.}~\bibnamefont {Ma}}, \bibinfo {author}
  {\bibfnamefont {W.}~\bibnamefont {Hu}},\ and\ \bibinfo {author}
  {\bibfnamefont {J.}~\bibnamefont {Yang}},\ }\bibfield  {title} {\bibinfo
  {title} {{Tunable Rashba spin splitting in Janus transition-metal
  dichalcogenide monolayers via charge doping}},\ }\href
  {https://doi.org/10.1039/D0RA00674B} {\bibfield  {journal} {\bibinfo
  {journal} {RSC Advances}\ }\textbf {\bibinfo {volume} {10}},\ \bibinfo
  {pages} {6388} (\bibinfo {year} {2020})}\BibitemShut {NoStop}%
\bibitem [{\citenamefont {Kresse}\ and\ \citenamefont
  {Furthm{\"u}ller}(1996{\natexlab{a}})}]{kresse1996efficiency}%
  \BibitemOpen
  \bibfield  {author} {\bibinfo {author} {\bibfnamefont {G.}~\bibnamefont
  {Kresse}}\ and\ \bibinfo {author} {\bibfnamefont {J.}~\bibnamefont
  {Furthm{\"u}ller}},\ }\bibfield  {title} {\bibinfo {title} {Efficiency of
  ab-initio total energy calculations for metals and semiconductors using a
  plane-wave basis set},\ }\href
  {https://doi.org/https://doi.org/10.1016/0927-0256(96)00008-0} {\bibfield
  {journal} {\bibinfo  {journal} {Computational materials science}\ }\textbf
  {\bibinfo {volume} {6}},\ \bibinfo {pages} {15} (\bibinfo {year}
  {1996}{\natexlab{a}})}\BibitemShut {NoStop}%
\bibitem [{\citenamefont {Kresse}\ and\ \citenamefont
  {Furthm{\"u}ller}(1996{\natexlab{b}})}]{kresse1996efficient}%
  \BibitemOpen
  \bibfield  {author} {\bibinfo {author} {\bibfnamefont {G.}~\bibnamefont
  {Kresse}}\ and\ \bibinfo {author} {\bibfnamefont {J.}~\bibnamefont
  {Furthm{\"u}ller}},\ }\bibfield  {title} {\bibinfo {title} {Efficient
  iterative schemes for ab initio total-energy calculations using a plane-wave
  basis set},\ }\href {https://doi.org/10.1103/PhysRevB.54.11169} {\bibfield
  {journal} {\bibinfo  {journal} {Phys. Rev. B}\ }\textbf {\bibinfo {volume}
  {54}},\ \bibinfo {pages} {11169} (\bibinfo {year}
  {1996}{\natexlab{b}})}\BibitemShut {NoStop}%
\bibitem [{\citenamefont {Perdew}\ \emph {et~al.}(1996)\citenamefont {Perdew},
  \citenamefont {Burke},\ and\ \citenamefont
  {Ernzerhof}}]{perdew1996generalized}%
  \BibitemOpen
  \bibfield  {author} {\bibinfo {author} {\bibfnamefont {J.~P.}\ \bibnamefont
  {Perdew}}, \bibinfo {author} {\bibfnamefont {K.}~\bibnamefont {Burke}},\ and\
  \bibinfo {author} {\bibfnamefont {M.}~\bibnamefont {Ernzerhof}},\ }\bibfield
  {title} {\bibinfo {title} {Generalized gradient approximation made simple},\
  }\href {https://doi.org/10.1103/PhysRevLett.77.3865} {\bibfield  {journal}
  {\bibinfo  {journal} {Phys. Rev. Lett.}\ }\textbf {\bibinfo {volume} {77}},\
  \bibinfo {pages} {3865} (\bibinfo {year} {1996})}\BibitemShut {NoStop}%
\bibitem [{\citenamefont {Bl{\"o}chl}(1994)}]{blochl1994projector}%
  \BibitemOpen
  \bibfield  {author} {\bibinfo {author} {\bibfnamefont {P.~E.}\ \bibnamefont
  {Bl{\"o}chl}},\ }\bibfield  {title} {\bibinfo {title} {Projector
  augmented-wave method},\ }\href {https://doi.org/10.1103/PhysRevB.50.17953}
  {\bibfield  {journal} {\bibinfo  {journal} {Phys. Rev. B}\ }\textbf {\bibinfo
  {volume} {50}},\ \bibinfo {pages} {17953} (\bibinfo {year}
  {1994})}\BibitemShut {NoStop}%
\bibitem [{\citenamefont {Kresse}\ and\ \citenamefont
  {Joubert}(1999)}]{kresse1999ultrasoft}%
  \BibitemOpen
  \bibfield  {author} {\bibinfo {author} {\bibfnamefont {G.}~\bibnamefont
  {Kresse}}\ and\ \bibinfo {author} {\bibfnamefont {D.}~\bibnamefont
  {Joubert}},\ }\bibfield  {title} {\bibinfo {title} {From ultrasoft
  pseudopotentials to the projector augmented-wave method},\ }\href
  {https://doi.org/10.1103/PhysRevB.59.1758} {\bibfield  {journal} {\bibinfo
  {journal} {Phys. Rev. B}\ }\textbf {\bibinfo {volume} {59}},\ \bibinfo
  {pages} {1758} (\bibinfo {year} {1999})}\BibitemShut {NoStop}%
\bibitem [{\citenamefont {Heyd}\ \emph {et~al.}(2003)\citenamefont {Heyd},
  \citenamefont {Scuseria},\ and\ \citenamefont {Ernzerhof}}]{heyd2003hybrid}%
  \BibitemOpen
  \bibfield  {author} {\bibinfo {author} {\bibfnamefont {J.}~\bibnamefont
  {Heyd}}, \bibinfo {author} {\bibfnamefont {G.~E.}\ \bibnamefont {Scuseria}},\
  and\ \bibinfo {author} {\bibfnamefont {M.}~\bibnamefont {Ernzerhof}},\
  }\bibfield  {title} {\bibinfo {title} {{Hybrid functionals based on a
  screened Coulomb potential}},\ }\href
  {https://aip.scitation.org/doi/10.1063/1.1564060} {\bibfield  {journal}
  {\bibinfo  {journal} {The Journal of chemical physics}\ }\textbf {\bibinfo
  {volume} {118}},\ \bibinfo {pages} {8207} (\bibinfo {year}
  {2003})}\BibitemShut {NoStop}%
\bibitem [{\citenamefont {Heyd}\ and\ \citenamefont
  {Scuseria}(2004)}]{heyd2004efficient}%
  \BibitemOpen
  \bibfield  {author} {\bibinfo {author} {\bibfnamefont {J.}~\bibnamefont
  {Heyd}}\ and\ \bibinfo {author} {\bibfnamefont {G.~E.}\ \bibnamefont
  {Scuseria}},\ }\bibfield  {title} {\bibinfo {title} {{Efficient hybrid
  density functional calculations in solids: Assessment of the
  Heyd--Scuseria--Ernzerhof screened Coulomb hybrid functional}},\ }\href
  {https://aip.scitation.org/doi/10.1063/1.1760074} {\bibfield  {journal}
  {\bibinfo  {journal} {The Journal of chemical physics}\ }\textbf {\bibinfo
  {volume} {121}},\ \bibinfo {pages} {1187} (\bibinfo {year}
  {2004})}\BibitemShut {NoStop}%
\bibitem [{\citenamefont {Grimme}(2006)}]{grimme2006semiempirical}%
  \BibitemOpen
  \bibfield  {author} {\bibinfo {author} {\bibfnamefont {S.}~\bibnamefont
  {Grimme}},\ }\bibfield  {title} {\bibinfo {title} {Semiempirical gga-type
  density functional constructed with a long-range dispersion correction},\
  }\href {https://doi.org/https://doi.org/10.1002/jcc.20495} {\bibfield
  {journal} {\bibinfo  {journal} {Journal of computational chemistry}\ }\textbf
  {\bibinfo {volume} {27}},\ \bibinfo {pages} {1787} (\bibinfo {year}
  {2006})}\BibitemShut {NoStop}%
\bibitem [{\citenamefont {Togo}\ and\ \citenamefont
  {Tanaka}(2015)}]{togo2015first}%
  \BibitemOpen
  \bibfield  {author} {\bibinfo {author} {\bibfnamefont {A.}~\bibnamefont
  {Togo}}\ and\ \bibinfo {author} {\bibfnamefont {I.}~\bibnamefont {Tanaka}},\
  }\bibfield  {title} {\bibinfo {title} {First principles phonon calculations
  in materials science},\ }\href
  {https://www.sciencedirect.com/science/article/pii/S1359646215003127}
  {\bibfield  {journal} {\bibinfo  {journal} {Scripta Materialia}\ }\textbf
  {\bibinfo {volume} {108}},\ \bibinfo {pages} {1} (\bibinfo {year}
  {2015})}\BibitemShut {NoStop}%
\bibitem [{\citenamefont {Neugebauer}\ and\ \citenamefont
  {Scheffler}(1992)}]{neugebauer1992adsorbate}%
  \BibitemOpen
  \bibfield  {author} {\bibinfo {author} {\bibfnamefont {J.}~\bibnamefont
  {Neugebauer}}\ and\ \bibinfo {author} {\bibfnamefont {M.}~\bibnamefont
  {Scheffler}},\ }\bibfield  {title} {\bibinfo {title} {{Adsorbate-substrate
  and adsorbate-adsorbate interactions of Na and K adlayers on Al (111)}},\
  }\href {https://journals.aps.org/prb/abstract/10.1103/PhysRevB.46.16067}
  {\bibfield  {journal} {\bibinfo  {journal} {Phys. Rev. B}\ }\textbf {\bibinfo
  {volume} {46}},\ \bibinfo {pages} {16067} (\bibinfo {year}
  {1992})}\BibitemShut {NoStop}%
\bibitem [{\citenamefont {Momma}\ and\ \citenamefont
  {Izumi}(2011)}]{momma2011vesta}%
  \BibitemOpen
  \bibfield  {author} {\bibinfo {author} {\bibfnamefont {K.}~\bibnamefont
  {Momma}}\ and\ \bibinfo {author} {\bibfnamefont {F.}~\bibnamefont {Izumi}},\
  }\bibfield  {title} {\bibinfo {title} {Vesta 3 for three-dimensional
  visualization of crystal, volumetric and morphology data},\ }\href
  {https://doi.org/https://doi.org/10.1107/S0021889811038970} {\bibfield
  {journal} {\bibinfo  {journal} {Journal of applied crystallography}\ }\textbf
  {\bibinfo {volume} {44}},\ \bibinfo {pages} {1272} (\bibinfo {year}
  {2011})}\BibitemShut {NoStop}%
\bibitem [{\citenamefont {Wypych}\ \emph {et~al.}(1998)\citenamefont {Wypych},
  \citenamefont {Weber},\ and\ \citenamefont {Prins}}]{wypych1998scanning}%
  \BibitemOpen
  \bibfield  {author} {\bibinfo {author} {\bibfnamefont {F.}~\bibnamefont
  {Wypych}}, \bibinfo {author} {\bibfnamefont {T.}~\bibnamefont {Weber}},\ and\
  \bibinfo {author} {\bibfnamefont {R.}~\bibnamefont {Prins}},\ }\bibfield
  {title} {\bibinfo {title} {{Scanning tunneling microscopic investigation of
  {1T-MoS$_2$}}},\ }\href {https://doi.org/https://doi.org/10.1021/cm970402e}
  {\bibfield  {journal} {\bibinfo  {journal} {Chemistry of materials}\ }\textbf
  {\bibinfo {volume} {10}},\ \bibinfo {pages} {723} (\bibinfo {year}
  {1998})}\BibitemShut {NoStop}%
\bibitem [{\citenamefont {Tang}\ \emph {et~al.}(2018)\citenamefont {Tang},
  \citenamefont {Li}, \citenamefont {Ma}, \citenamefont {Du}, \citenamefont
  {Liao}, \citenamefont {Gu},\ and\ \citenamefont {Kou}}]{tang2018distorted}%
  \BibitemOpen
  \bibfield  {author} {\bibinfo {author} {\bibfnamefont {X.}~\bibnamefont
  {Tang}}, \bibinfo {author} {\bibfnamefont {S.}~\bibnamefont {Li}}, \bibinfo
  {author} {\bibfnamefont {Y.}~\bibnamefont {Ma}}, \bibinfo {author}
  {\bibfnamefont {A.}~\bibnamefont {Du}}, \bibinfo {author} {\bibfnamefont
  {T.}~\bibnamefont {Liao}}, \bibinfo {author} {\bibfnamefont {Y.}~\bibnamefont
  {Gu}},\ and\ \bibinfo {author} {\bibfnamefont {L.}~\bibnamefont {Kou}},\
  }\bibfield  {title} {\bibinfo {title} {Distorted janus transition metal
  dichalcogenides: Stable two-dimensional materials with sizable band gap and
  ultrahigh carrier mobility},\ }\href
  {https://doi.org/https://doi.org/10.1021/acs.jpcc.8b04161} {\bibfield
  {journal} {\bibinfo  {journal} {The Journal of Physical Chemistry C}\
  }\textbf {\bibinfo {volume} {122}},\ \bibinfo {pages} {19153} (\bibinfo
  {year} {2018})}\BibitemShut {NoStop}%
\bibitem [{\citenamefont {Yang}\ \emph {et~al.}(2019)\citenamefont {Yang},
  \citenamefont {Singh}, \citenamefont {Xu}, \citenamefont {Wang},\ and\
  \citenamefont {Ahuja}}]{yang2019emerging}%
  \BibitemOpen
  \bibfield  {author} {\bibinfo {author} {\bibfnamefont {X.}~\bibnamefont
  {Yang}}, \bibinfo {author} {\bibfnamefont {D.}~\bibnamefont {Singh}},
  \bibinfo {author} {\bibfnamefont {Z.}~\bibnamefont {Xu}}, \bibinfo {author}
  {\bibfnamefont {Z.}~\bibnamefont {Wang}},\ and\ \bibinfo {author}
  {\bibfnamefont {R.}~\bibnamefont {Ahuja}},\ }\bibfield  {title} {\bibinfo
  {title} {{An emerging Janus MoSeTe material for potential applications in
  optoelectronic devices}},\ }\href
  {https://doi.org/https://doi.org/10.1039/C9TC03936H} {\bibfield  {journal}
  {\bibinfo  {journal} {Journal of Materials Chemistry C}\ }\textbf {\bibinfo
  {volume} {7}},\ \bibinfo {pages} {12312} (\bibinfo {year}
  {2019})}\BibitemShut {NoStop}%
\bibitem [{\citenamefont {Hu}\ \emph {et~al.}(2018)\citenamefont {Hu},
  \citenamefont {Jia}, \citenamefont {Zhao}, \citenamefont {Wu}, \citenamefont
  {Stroppa},\ and\ \citenamefont {Ren}}]{hu2018intrinsic}%
  \BibitemOpen
  \bibfield  {author} {\bibinfo {author} {\bibfnamefont {T.}~\bibnamefont
  {Hu}}, \bibinfo {author} {\bibfnamefont {F.}~\bibnamefont {Jia}}, \bibinfo
  {author} {\bibfnamefont {G.}~\bibnamefont {Zhao}}, \bibinfo {author}
  {\bibfnamefont {J.}~\bibnamefont {Wu}}, \bibinfo {author} {\bibfnamefont
  {A.}~\bibnamefont {Stroppa}},\ and\ \bibinfo {author} {\bibfnamefont
  {W.}~\bibnamefont {Ren}},\ }\bibfield  {title} {\bibinfo {title} {{Intrinsic
  and anisotropic Rashba spin splitting in Janus transition-metal
  dichalcogenide monolayers}},\ }\href
  {https://doi.org/10.1103/PhysRevB.97.235404} {\bibfield  {journal} {\bibinfo
  {journal} {Phys. Rev. B}\ }\textbf {\bibinfo {volume} {97}},\ \bibinfo
  {pages} {235404} (\bibinfo {year} {2018})}\BibitemShut {NoStop}%
\bibitem [{\citenamefont {Xia}\ \emph {et~al.}(2018)\citenamefont {Xia},
  \citenamefont {Xiong}, \citenamefont {Du}, \citenamefont {Wang},
  \citenamefont {Peng},\ and\ \citenamefont {Li}}]{xia2018universality}%
  \BibitemOpen
  \bibfield  {author} {\bibinfo {author} {\bibfnamefont {C.}~\bibnamefont
  {Xia}}, \bibinfo {author} {\bibfnamefont {W.}~\bibnamefont {Xiong}}, \bibinfo
  {author} {\bibfnamefont {J.}~\bibnamefont {Du}}, \bibinfo {author}
  {\bibfnamefont {T.}~\bibnamefont {Wang}}, \bibinfo {author} {\bibfnamefont
  {Y.}~\bibnamefont {Peng}},\ and\ \bibinfo {author} {\bibfnamefont
  {J.}~\bibnamefont {Li}},\ }\bibfield  {title} {\bibinfo {title}
  {{Universality of electronic characteristics and photocatalyst applications
  in the two-dimensional Janus transition metal dichalcogenides}},\ }\href
  {https://doi.org/10.1103/PhysRevB.98.165424} {\bibfield  {journal} {\bibinfo
  {journal} {Phys. Rev. B}\ }\textbf {\bibinfo {volume} {98}},\ \bibinfo
  {pages} {165424} (\bibinfo {year} {2018})}\BibitemShut {NoStop}%
\bibitem [{\citenamefont {Wang}\ \emph {et~al.}(2017)\citenamefont {Wang},
  \citenamefont {Wang}, \citenamefont {Guo}, \citenamefont {Zhang},
  \citenamefont {Hu},\ and\ \citenamefont {Chu}}]{wang2017tuning}%
  \BibitemOpen
  \bibfield  {author} {\bibinfo {author} {\bibfnamefont {F.}~\bibnamefont
  {Wang}}, \bibinfo {author} {\bibfnamefont {J.}~\bibnamefont {Wang}}, \bibinfo
  {author} {\bibfnamefont {S.}~\bibnamefont {Guo}}, \bibinfo {author}
  {\bibfnamefont {J.}~\bibnamefont {Zhang}}, \bibinfo {author} {\bibfnamefont
  {Z.}~\bibnamefont {Hu}},\ and\ \bibinfo {author} {\bibfnamefont
  {J.}~\bibnamefont {Chu}},\ }\bibfield  {title} {\bibinfo {title} {{Tuning
  coupling behavior of stacked heterostructures based on {MoS$_2$, WS$_2$, and
  WSe$_2$}}},\ }\href {https://doi.org/https://doi.org/10.1038/srep44712}
  {\bibfield  {journal} {\bibinfo  {journal} {Scientific reports}\ }\textbf
  {\bibinfo {volume} {7}},\ \bibinfo {pages} {1} (\bibinfo {year}
  {2017})}\BibitemShut {NoStop}%
\bibitem [{\citenamefont {Han}\ \emph {et~al.}(2011)\citenamefont {Han},
  \citenamefont {Kwon}, \citenamefont {Kim}, \citenamefont {Ryu}, \citenamefont
  {Yun}, \citenamefont {Kim}, \citenamefont {Hwang}, \citenamefont {Kang},
  \citenamefont {Baik}, \citenamefont {Shin},\ and\ \citenamefont
  {Hong}}]{han2011band}%
  \BibitemOpen
  \bibfield  {author} {\bibinfo {author} {\bibfnamefont {S.~W.}\ \bibnamefont
  {Han}}, \bibinfo {author} {\bibfnamefont {H.}~\bibnamefont {Kwon}}, \bibinfo
  {author} {\bibfnamefont {S.~K.}\ \bibnamefont {Kim}}, \bibinfo {author}
  {\bibfnamefont {S.}~\bibnamefont {Ryu}}, \bibinfo {author} {\bibfnamefont
  {W.~S.}\ \bibnamefont {Yun}}, \bibinfo {author} {\bibfnamefont {D.~H.}\
  \bibnamefont {Kim}}, \bibinfo {author} {\bibfnamefont {J.~H.}\ \bibnamefont
  {Hwang}}, \bibinfo {author} {\bibfnamefont {J.-S.}\ \bibnamefont {Kang}},
  \bibinfo {author} {\bibfnamefont {J.}~\bibnamefont {Baik}}, \bibinfo {author}
  {\bibfnamefont {H.~J.}\ \bibnamefont {Shin}},\ and\ \bibinfo {author}
  {\bibfnamefont {S.~C.}\ \bibnamefont {Hong}},\ }\bibfield  {title} {\bibinfo
  {title} {Band-gap transition induced by interlayer van der waals interaction
  in {MoS$_{2}$}},\ }\href {https://doi.org/10.1103/PhysRevB.84.045409}
  {\bibfield  {journal} {\bibinfo  {journal} {Phys. Rev. B}\ }\textbf {\bibinfo
  {volume} {84}},\ \bibinfo {pages} {045409} (\bibinfo {year}
  {2011})}\BibitemShut {NoStop}%
\bibitem [{\citenamefont {Sahoo}\ \emph {et~al.}(2021)\citenamefont {Sahoo},
  \citenamefont {Sahu}, \citenamefont {Mallik}, \citenamefont {Sharma},
  \citenamefont {Jena}, \citenamefont {Gupta}, \citenamefont {Ahuja},\ and\
  \citenamefont {Sahoo}}]{sahoo2021electric}%
  \BibitemOpen
  \bibfield  {author} {\bibinfo {author} {\bibfnamefont {S.}~\bibnamefont
  {Sahoo}}, \bibinfo {author} {\bibfnamefont {M.~C.}\ \bibnamefont {Sahu}},
  \bibinfo {author} {\bibfnamefont {S.~K.}\ \bibnamefont {Mallik}}, \bibinfo
  {author} {\bibfnamefont {N.~K.}\ \bibnamefont {Sharma}}, \bibinfo {author}
  {\bibfnamefont {A.~K.}\ \bibnamefont {Jena}}, \bibinfo {author}
  {\bibfnamefont {S.~K.}\ \bibnamefont {Gupta}}, \bibinfo {author}
  {\bibfnamefont {R.}~\bibnamefont {Ahuja}},\ and\ \bibinfo {author}
  {\bibfnamefont {S.}~\bibnamefont {Sahoo}},\ }\bibfield  {title} {\bibinfo
  {title} {{Electric Field-Modulated charge transfer in geometrically tailored
  {MoX$_2$/WX$_2$ }(X= S, Se) heterostructures}},\ }\href
  {https://doi.org/https://doi.org/10.1021/acs.jpcc.1c07218} {\bibfield
  {journal} {\bibinfo  {journal} {The Journal of Physical Chemistry C}\
  }\textbf {\bibinfo {volume} {125}},\ \bibinfo {pages} {22360} (\bibinfo
  {year} {2021})}\BibitemShut {NoStop}%
\bibitem [{pat()}]{patel2022rashba}%
  \BibitemOpen
  \bibfield  {title} {\bibinfo {title} {{See Supplemental Material at [URL will
  be inserted by the publisher], which includes detailed information about
  calculations with {HSE06} hybrid functional.}},\ }\href@noop {} {\
  }\BibitemShut {NoStop}%
\bibitem [{\citenamefont {Hong}\ \emph {et~al.}(2014)\citenamefont {Hong},
  \citenamefont {Kim}, \citenamefont {Shi}, \citenamefont {Zhang},
  \citenamefont {Jin}, \citenamefont {Sun}, \citenamefont {Tongay},
  \citenamefont {Wu}, \citenamefont {Zhang},\ and\ \citenamefont
  {Wang}}]{hong2014ultrafast}%
  \BibitemOpen
  \bibfield  {author} {\bibinfo {author} {\bibfnamefont {X.}~\bibnamefont
  {Hong}}, \bibinfo {author} {\bibfnamefont {J.}~\bibnamefont {Kim}}, \bibinfo
  {author} {\bibfnamefont {S.-F.}\ \bibnamefont {Shi}}, \bibinfo {author}
  {\bibfnamefont {Y.}~\bibnamefont {Zhang}}, \bibinfo {author} {\bibfnamefont
  {C.}~\bibnamefont {Jin}}, \bibinfo {author} {\bibfnamefont {Y.}~\bibnamefont
  {Sun}}, \bibinfo {author} {\bibfnamefont {S.}~\bibnamefont {Tongay}},
  \bibinfo {author} {\bibfnamefont {J.}~\bibnamefont {Wu}}, \bibinfo {author}
  {\bibfnamefont {Y.}~\bibnamefont {Zhang}},\ and\ \bibinfo {author}
  {\bibfnamefont {F.}~\bibnamefont {Wang}},\ }\bibfield  {title} {\bibinfo
  {title} {{Ultrafast charge transfer in atomically thin {MoS$_2$/WS$_2$}
  heterostructures}},\ }\href
  {https://doi.org/https://doi.org/10.1038/nnano.2014.167} {\bibfield
  {journal} {\bibinfo  {journal} {Nature nanotechnology}\ }\textbf {\bibinfo
  {volume} {9}},\ \bibinfo {pages} {682} (\bibinfo {year} {2014})}\BibitemShut
  {NoStop}%
\bibitem [{\citenamefont {Yu}\ \emph {et~al.}(2015)\citenamefont {Yu},
  \citenamefont {Cui}, \citenamefont {Xu},\ and\ \citenamefont
  {Yao}}]{yu2015valley}%
  \BibitemOpen
  \bibfield  {author} {\bibinfo {author} {\bibfnamefont {H.}~\bibnamefont
  {Yu}}, \bibinfo {author} {\bibfnamefont {X.}~\bibnamefont {Cui}}, \bibinfo
  {author} {\bibfnamefont {X.}~\bibnamefont {Xu}},\ and\ \bibinfo {author}
  {\bibfnamefont {W.}~\bibnamefont {Yao}},\ }\bibfield  {title} {\bibinfo
  {title} {Valley excitons in two-dimensional semiconductors},\ }\href
  {https://doi.org/https://doi.org/10.1093/nsr/nwu078} {\bibfield  {journal}
  {\bibinfo  {journal} {National Science Review}\ }\textbf {\bibinfo {volume}
  {2}},\ \bibinfo {pages} {57} (\bibinfo {year} {2015})}\BibitemShut {NoStop}%
\bibitem [{\citenamefont {Guan}\ \emph {et~al.}(2018)\citenamefont {Guan},
  \citenamefont {Ni},\ and\ \citenamefont {Hu}}]{guan2018tunable}%
  \BibitemOpen
  \bibfield  {author} {\bibinfo {author} {\bibfnamefont {Z.}~\bibnamefont
  {Guan}}, \bibinfo {author} {\bibfnamefont {S.}~\bibnamefont {Ni}},\ and\
  \bibinfo {author} {\bibfnamefont {S.}~\bibnamefont {Hu}},\ }\bibfield
  {title} {\bibinfo {title} {Tunable electronic and optical properties of
  monolayer and multilayer janus mosse as a photocatalyst for solar water
  splitting: a first-principles study},\ }\href
  {https://doi.org/https://doi.org/10.1021/acs.jpcc.8b00257} {\bibfield
  {journal} {\bibinfo  {journal} {The Journal of Physical Chemistry C}\
  }\textbf {\bibinfo {volume} {122}},\ \bibinfo {pages} {6209} (\bibinfo {year}
  {2018})}\BibitemShut {NoStop}%
\bibitem [{\citenamefont {Yu}\ \emph {et~al.}(2013)\citenamefont {Yu},
  \citenamefont {Liu}, \citenamefont {Zhou}, \citenamefont {Yin}, \citenamefont
  {Li}, \citenamefont {Huang},\ and\ \citenamefont {Duan}}]{yu2013highly}%
  \BibitemOpen
  \bibfield  {author} {\bibinfo {author} {\bibfnamefont {W.~J.}\ \bibnamefont
  {Yu}}, \bibinfo {author} {\bibfnamefont {Y.}~\bibnamefont {Liu}}, \bibinfo
  {author} {\bibfnamefont {H.}~\bibnamefont {Zhou}}, \bibinfo {author}
  {\bibfnamefont {A.}~\bibnamefont {Yin}}, \bibinfo {author} {\bibfnamefont
  {Z.}~\bibnamefont {Li}}, \bibinfo {author} {\bibfnamefont {Y.}~\bibnamefont
  {Huang}},\ and\ \bibinfo {author} {\bibfnamefont {X.}~\bibnamefont {Duan}},\
  }\bibfield  {title} {\bibinfo {title} {Highly efficient gate-tunable
  photocurrent generation in vertical heterostructures of layered materials},\
  }\href {https://doi.org/https://doi.org/10.1038/nnano.2013.219} {\bibfield
  {journal} {\bibinfo  {journal} {Nature nanotechnology}\ }\textbf {\bibinfo
  {volume} {8}},\ \bibinfo {pages} {952} (\bibinfo {year} {2013})}\BibitemShut
  {NoStop}%
\bibitem [{\citenamefont {Xiao}\ \emph {et~al.}(2012)\citenamefont {Xiao},
  \citenamefont {Liu}, \citenamefont {Feng}, \citenamefont {Xu},\ and\
  \citenamefont {Yao}}]{xiao2012coupled}%
  \BibitemOpen
  \bibfield  {author} {\bibinfo {author} {\bibfnamefont {D.}~\bibnamefont
  {Xiao}}, \bibinfo {author} {\bibfnamefont {G.-B.}\ \bibnamefont {Liu}},
  \bibinfo {author} {\bibfnamefont {W.}~\bibnamefont {Feng}}, \bibinfo {author}
  {\bibfnamefont {X.}~\bibnamefont {Xu}},\ and\ \bibinfo {author}
  {\bibfnamefont {W.}~\bibnamefont {Yao}},\ }\bibfield  {title} {\bibinfo
  {title} {{Coupled spin and valley physics in monolayers of $MoS_2$ and other
  group-VI dichalcogenides}},\ }\href
  {https://doi.org/10.1103/PhysRevLett.108.196802} {\bibfield  {journal}
  {\bibinfo  {journal} {Phys. Rev. Lett}\ }\textbf {\bibinfo {volume} {108}},\
  \bibinfo {pages} {196802} (\bibinfo {year} {2012})}\BibitemShut {NoStop}%
\bibitem [{\citenamefont {Ko{\'s}mider}\ \emph {et~al.}(2013)\citenamefont
  {Ko{\'s}mider}, \citenamefont {Gonz{\'a}lez},\ and\ \citenamefont
  {Fern{\'a}ndez-Rossier}}]{kosmider2013large}%
  \BibitemOpen
  \bibfield  {author} {\bibinfo {author} {\bibfnamefont {K.}~\bibnamefont
  {Ko{\'s}mider}}, \bibinfo {author} {\bibfnamefont {J.~W.}\ \bibnamefont
  {Gonz{\'a}lez}},\ and\ \bibinfo {author} {\bibfnamefont {J.}~\bibnamefont
  {Fern{\'a}ndez-Rossier}},\ }\bibfield  {title} {\bibinfo {title} {Large spin
  splitting in the conduction band of transition metal dichalcogenide
  monolayers},\ }\href {https://doi.org/10.1103/PhysRevB.88.245436} {\bibfield
  {journal} {\bibinfo  {journal} {Phys. Rev. B}\ }\textbf {\bibinfo {volume}
  {88}},\ \bibinfo {pages} {245436} (\bibinfo {year} {2013})}\BibitemShut
  {NoStop}%
\bibitem [{\citenamefont {Yuan}\ \emph {et~al.}(2013)\citenamefont {Yuan},
  \citenamefont {Bahramy}, \citenamefont {Morimoto}, \citenamefont {Wu},
  \citenamefont {Nomura}, \citenamefont {Yang}, \citenamefont {Shimotani},
  \citenamefont {Suzuki}, \citenamefont {Toh}, \citenamefont {Kloc} \emph
  {et~al.}}]{yuan2013zeeman}%
  \BibitemOpen
  \bibfield  {author} {\bibinfo {author} {\bibfnamefont {H.}~\bibnamefont
  {Yuan}}, \bibinfo {author} {\bibfnamefont {M.~S.}\ \bibnamefont {Bahramy}},
  \bibinfo {author} {\bibfnamefont {K.}~\bibnamefont {Morimoto}}, \bibinfo
  {author} {\bibfnamefont {S.}~\bibnamefont {Wu}}, \bibinfo {author}
  {\bibfnamefont {K.}~\bibnamefont {Nomura}}, \bibinfo {author} {\bibfnamefont
  {B.-J.}\ \bibnamefont {Yang}}, \bibinfo {author} {\bibfnamefont
  {H.}~\bibnamefont {Shimotani}}, \bibinfo {author} {\bibfnamefont
  {R.}~\bibnamefont {Suzuki}}, \bibinfo {author} {\bibfnamefont
  {M.}~\bibnamefont {Toh}}, \bibinfo {author} {\bibfnamefont {C.}~\bibnamefont
  {Kloc}}, \emph {et~al.},\ }\bibfield  {title} {\bibinfo {title} {Zeeman-type
  spin splitting controlled by an electric field},\ }\href
  {https://doi.org/https://doi.org/10.1038/nphys2691} {\bibfield  {journal}
  {\bibinfo  {journal} {Nature Physics}\ }\textbf {\bibinfo {volume} {9}},\
  \bibinfo {pages} {563} (\bibinfo {year} {2013})}\BibitemShut {NoStop}%
\bibitem [{\citenamefont {Vajna}\ \emph {et~al.}(2012)\citenamefont {Vajna},
  \citenamefont {Simon}, \citenamefont {Szilva}, \citenamefont {Palotas},
  \citenamefont {Ujfalussy},\ and\ \citenamefont {Szunyogh}}]{vajna2012higher}%
  \BibitemOpen
  \bibfield  {author} {\bibinfo {author} {\bibfnamefont {S.}~\bibnamefont
  {Vajna}}, \bibinfo {author} {\bibfnamefont {E.}~\bibnamefont {Simon}},
  \bibinfo {author} {\bibfnamefont {A.}~\bibnamefont {Szilva}}, \bibinfo
  {author} {\bibfnamefont {K.}~\bibnamefont {Palotas}}, \bibinfo {author}
  {\bibfnamefont {B.}~\bibnamefont {Ujfalussy}},\ and\ \bibinfo {author}
  {\bibfnamefont {L.}~\bibnamefont {Szunyogh}},\ }\bibfield  {title} {\bibinfo
  {title} {{Higher-order contributions to the Rashba-Bychkov effect with
  application to the Bi/Ag (111) surface alloy}},\ }\href
  {https://doi.org/10.1103/PhysRevB.85.075404} {\bibfield  {journal} {\bibinfo
  {journal} {Phys. Rev. B}\ }\textbf {\bibinfo {volume} {85}},\ \bibinfo
  {pages} {075404} (\bibinfo {year} {2012})}\BibitemShut {NoStop}%
\bibitem [{\citenamefont {Yao}\ \emph {et~al.}(2017)\citenamefont {Yao},
  \citenamefont {Cai}, \citenamefont {Tong}, \citenamefont {Gong},
  \citenamefont {Wang}, \citenamefont {Wan}, \citenamefont {Duan},\ and\
  \citenamefont {Chu}}]{yao2017manipulation}%
  \BibitemOpen
  \bibfield  {author} {\bibinfo {author} {\bibfnamefont {Q.-F.}\ \bibnamefont
  {Yao}}, \bibinfo {author} {\bibfnamefont {J.}~\bibnamefont {Cai}}, \bibinfo
  {author} {\bibfnamefont {W.-Y.}\ \bibnamefont {Tong}}, \bibinfo {author}
  {\bibfnamefont {S.-J.}\ \bibnamefont {Gong}}, \bibinfo {author}
  {\bibfnamefont {J.-Q.}\ \bibnamefont {Wang}}, \bibinfo {author}
  {\bibfnamefont {X.}~\bibnamefont {Wan}}, \bibinfo {author} {\bibfnamefont
  {C.-G.}\ \bibnamefont {Duan}},\ and\ \bibinfo {author} {\bibfnamefont
  {J.~H.}\ \bibnamefont {Chu}},\ }\bibfield  {title} {\bibinfo {title}
  {Manipulation of the large rashba spin splitting in polar two-dimensional
  transition-metal dichalcogenides},\ }\href
  {https://doi.org/10.1103/PhysRevB.95.165401} {\bibfield  {journal} {\bibinfo
  {journal} {Phys. Rev. B}\ }\textbf {\bibinfo {volume} {95}},\ \bibinfo
  {pages} {165401} (\bibinfo {year} {2017})}\BibitemShut {NoStop}%
\bibitem [{\citenamefont {Korm{\'a}nyos}\ \emph {et~al.}(2014)\citenamefont
  {Korm{\'a}nyos}, \citenamefont {Z{\'o}lyomi}, \citenamefont {Drummond},\ and\
  \citenamefont {Burkard}}]{kormanyos2014spin}%
  \BibitemOpen
  \bibfield  {author} {\bibinfo {author} {\bibfnamefont {A.}~\bibnamefont
  {Korm{\'a}nyos}}, \bibinfo {author} {\bibfnamefont {V.}~\bibnamefont
  {Z{\'o}lyomi}}, \bibinfo {author} {\bibfnamefont {N.~D.}\ \bibnamefont
  {Drummond}},\ and\ \bibinfo {author} {\bibfnamefont {G.}~\bibnamefont
  {Burkard}},\ }\bibfield  {title} {\bibinfo {title} {Spin-orbit coupling,
  quantum dots, and qubits in monolayer transition metal dichalcogenides},\
  }\href {https://doi.org/10.1103/PhysRevX.4.011034} {\bibfield  {journal}
  {\bibinfo  {journal} {Phys. Revi. X}\ }\textbf {\bibinfo {volume} {4}},\
  \bibinfo {pages} {011034} (\bibinfo {year} {2014})}\BibitemShut {NoStop}%
\bibitem [{\citenamefont {Tao}\ and\ \citenamefont
  {Tsymbal}(2018)}]{tao2018persistent}%
  \BibitemOpen
  \bibfield  {author} {\bibinfo {author} {\bibfnamefont {L.}~\bibnamefont
  {Tao}}\ and\ \bibinfo {author} {\bibfnamefont {E.~Y.}\ \bibnamefont
  {Tsymbal}},\ }\bibfield  {title} {\bibinfo {title} {Persistent spin texture
  enforced by symmetry},\ }\href
  {https://doi.org/https://doi.org/10.1038/s41467-018-05137-0} {\bibfield
  {journal} {\bibinfo  {journal} {Nature communications}\ }\textbf {\bibinfo
  {volume} {9}},\ \bibinfo {pages} {1} (\bibinfo {year} {2018})}\BibitemShut
  {NoStop}%
\bibitem [{\citenamefont {Liu}\ \emph {et~al.}(2020{\natexlab{b}})\citenamefont
  {Liu}, \citenamefont {Gao}, \citenamefont {Li}, \citenamefont {Wang},
  \citenamefont {Burton},\ and\ \citenamefont {Ren}}]{liu2020manipulation}%
  \BibitemOpen
  \bibfield  {author} {\bibinfo {author} {\bibfnamefont {C.}~\bibnamefont
  {Liu}}, \bibinfo {author} {\bibfnamefont {H.}~\bibnamefont {Gao}}, \bibinfo
  {author} {\bibfnamefont {Y.}~\bibnamefont {Li}}, \bibinfo {author}
  {\bibfnamefont {K.}~\bibnamefont {Wang}}, \bibinfo {author} {\bibfnamefont
  {L.~A.}\ \bibnamefont {Burton}},\ and\ \bibinfo {author} {\bibfnamefont
  {W.}~\bibnamefont {Ren}},\ }\bibfield  {title} {\bibinfo {title}
  {{Manipulation of the Rashba effect in layered tellurides MTe (M= Ge, Sn,
  Pb)}},\ }\href {https://doi.org/https://doi.org/10.1039/D0TC00003E}
  {\bibfield  {journal} {\bibinfo  {journal} {Journal of Materials Chemistry
  C}\ }\textbf {\bibinfo {volume} {8}},\ \bibinfo {pages} {5143} (\bibinfo
  {year} {2020}{\natexlab{b}})}\BibitemShut {NoStop}%
\bibitem [{\citenamefont {Sharma}\ \emph {et~al.}(2014)\citenamefont {Sharma},
  \citenamefont {Kumar}, \citenamefont {Ahluwalia},\ and\ \citenamefont
  {Pandey}}]{sharma2014strain}%
  \BibitemOpen
  \bibfield  {author} {\bibinfo {author} {\bibfnamefont {M.}~\bibnamefont
  {Sharma}}, \bibinfo {author} {\bibfnamefont {A.}~\bibnamefont {Kumar}},
  \bibinfo {author} {\bibfnamefont {P.}~\bibnamefont {Ahluwalia}},\ and\
  \bibinfo {author} {\bibfnamefont {R.}~\bibnamefont {Pandey}},\ }\bibfield
  {title} {\bibinfo {title} {Strain and electric field induced electronic
  properties of two-dimensional hybrid bilayers of transition-metal
  dichalcogenides},\ }\href {https://doi.org/https://doi.org/10.1063/1.4892798}
  {\bibfield  {journal} {\bibinfo  {journal} {Journal of Applied Physics}\
  }\textbf {\bibinfo {volume} {116}},\ \bibinfo {pages} {063711} (\bibinfo
  {year} {2014})}\BibitemShut {NoStop}%
\end{thebibliography}%
	
\end{document}